\documentclass[apj,twocolumn, twocolappendix]{openjournal}
\usepackage{orcidlink}  

\usepackage{multirow}
\usepackage[parfill]{parskip}
\usepackage{graphicx}
\usepackage{amssymb}
\usepackage{amsmath}
\usepackage{epstopdf}
\usepackage{xcolor}
\usepackage{booktabs}

\usepackage{hyperref}
\hypersetup{
     colorlinks   = true,
     citecolor    = blue,
     linkcolor = blue,
     urlcolor = blue
}

\makeatletter
\long\def\@makecaption#1#2{%
 \noindent\begin{minipage}{0.9999\linewidth}
   \if\csname ftype@\@captype\endcsname 2
   \vskip 2ex\noindent \centering\@table@type@size{\@eapj@cap@font  #1}\par
    #2\par\medskip
   \else
   \vspace*{\abovecaptionskip}\noindent\footnotesize #1 #2\par\vskip \belowcaptionskip
   \fi
 \end{minipage}\par
 }
\def\tablecaption#1{\gdef\@tablecaption{#1}}
\makeatother

\def\apjl{{ApJL}}
\def\apjs{{ ApJS}}

\def\aap{{ A\&A}}

\def\mr{\mathrm}
\def\d{\mr{d}}

\def\mc{\mathcal}

\def\me{m_{\rm e}}
\def\mproton{m_{\rm p}}

\def\kB{k_{\rm B}}
\def\arad{a_{\rm rad}}

\newcommand{\lrb}[1]{\left({#1}\right)}
\newcommand{\lrsb}[1]{\left[{#1}\right]}

\begin{document}


\title{Inflated Supermassive Stars as Little Red Dots and Progenitors of Supermassive Black Holes}

\vspace{-1cm}

\author{Wenbin Lu\,\orcidlink{0000-0002-1568-7461}$^{1}$}

\affiliation{$^1$Department of Astronomy and Theoretical Astrophysics Center, University of California, Berkeley, CA 94720-3411, USA}

\email{wenbinlu@berkeley.edu}

\begin{abstract}

The James Webb Space Telescope has uncovered an abundant population of compact, red sources at high redshifts, termed little red dots (LRDs), whose physical origin remains uncertain. We investigate whether they could be explained by thermally relaxed, hydrogen-burning, metal-enriched supermassive stars (SMSs). We construct one-dimensional models in hydrostatic and thermal equilibrium, including a radiation-pressure-dominated, CNO-burning core and an outer envelope with metallicity-dependent opacity, non-adiabatic convection, and an effective treatment of super-Eddington surface layers. The luminosity remains close to the Thomson-scattering Eddington limit. The Fe-opacity bump drives the formation of a strongly inflated, low-mass envelope, while hydrogen recombination permits the envelope to terminate at a bound photosphere. Across a grid with \(M=10^{4}\)--\(10^{6}\,M_\odot\) and metal mass fractions \(Z=10^{-4}\)--\(10^{-2}\), we find that the most metal-rich models, with \(Z\simeq10^{-2}\), develop highly inflated envelopes and reach \(T_{\rm eff}\sim7000\,{\rm K}\). Since metal cooling disfavors monolithic collapse in star formation, we suggest that such enriched SMSs could instead be assembled through runaway stellar collisions in compact nuclear star clusters, whose dense gas may also favor metal retention. The general-relativistic instability sets a maximum mass and luminosity that depends on the core rotation rate. The nonrotating and rotating models have maximum masses $3\times10^5$ to $3\times 10^6M_\odot$, and maximum luminosity of the order \(10^{44}\,{\rm erg\,s^{-1}}\), comparable to the observed bright-end cutoff of the LRD luminosity function. Core hydrogen depletion can drive initially stable SMSs across the instability threshold after a lifetime of order \(1\,{\rm Myr}\). If each LRD leaves a black hole retaining most of its mass, the observed LRD abundance implies a present-day remnant density of order \(10^{-2}\,{\rm cMpc^{-3}}\), consistent with the local abundance of supermassive black holes. Cool SMSs naturally produce weak X-ray and weak high-ionization lines. Non-LTE effects, particularly Ly$\alpha$ trapping and the resulting enhancement of the hydrogen \(n=2\) population, may significantly modify the Balmer lines, the Balmer break strength, and the continuum opacity. Coupled self-consistently to the envelope structure, these effects may lower $T_{\rm eff}$ below our LTE value of $\sim7000\,{\rm K}$.








\keywords{little red dots -- supermassive stars -- supermassive black holes}

\end{abstract}

\maketitle

\section{Introduction}
\label{sec:introduction}

Observations with the James Webb Space Telescope (JWST) have revealed
an unexpectedly abundant population of compact, red sources at
$z\gtrsim4$, commonly referred to as ``little red dots'' (LRDs)
\citep{2024ApJ...963..129M,2024ApJ...964...39G,
2024ApJ...968...38K,2025ApJ...986..126K}. Many exhibit broad Balmer
emission lines
\citep{2023ApJ...954L...4K,2023ApJ...959...39H} and have therefore been interpreted as
active galactic nuclei (AGNs). Their overall spectral energy
distributions, however, differ substantially from those of ordinary
type-I quasars: the rest-frame optical continuum is extremely red and compact, whereas the rest-frame UV emission is often blue and more spatially extended.
Different selection criteria identify partially overlapping samples, and the
observed population may contain more than one class of objects
\citep{2025ApJ...979..138H}.

Despite possible heterogeneity, several recurring properties
provide a useful set of requirements for physical models \citep{2025arXiv251203130I}. At minimum,
a successful explanation should account for the
following observations.

\begin{enumerate}

\item \emph{A compact, cool, and luminous rest-frame optical
component.}
The red component is generally unresolved or only marginally resolved
by JWST, implying characteristic size limits of order
$10$--$100\,{\rm pc}$, with substantially smaller physical emitting
radii allowed. If this continuum is interpreted as optically thick
emission, typical luminosities $L\sim10^{43}$ to a few $10^{44}\,{\rm erg\,s^{-1}}$ and color temperatures $T_{\rm col}\sim 6000\rm\, K$ imply an emitting scale
\begin{equation}
    R_{\rm BB}
    =
    \left(
    \frac{L}{4\pi\sigma_{\rm SB}T_{\rm col}^{4}}
    \right)^{1/2}
    \simeq
    10^{16}\,{\rm cm}\,
    L_{44}^{1/2}T_{\rm col, 6000}^{-2},
\label{eq:LRD_blackbody_radius}
\end{equation}
which is far smaller than the size constraint from JWST's spatial resolution. Spectroscopic samples further 
show that the blue UV and red optical continua often meet in a
V-shaped turnover located close to the Balmer limit, rather than at an
arbitrary wavelength expected from the crossing of two smooth power
laws \citep{2025ApJ...995..118S}.

\item \emph{Balmer emission, absorption, and continuum opacity.}
Many spectroscopically confirmed LRDs exhibit broad H$\alpha$ or
H$\beta$ emission with widths of order
$10^{3}$--$4\times10^{3}\,{\rm km\,s^{-1}}$
\citep{2023ApJ...959...39H,2024ApJ...963..129M,
2024ApJ...964...39G}. Strong Balmer breaks are also common
\citep{2025ApJ...995..118S}, while Balmer absorption lines have been
detected in individual objects \citep{2025A&A...701A.168D}.
Hydrogen bound-free opacity and Balmer absorption require a substantial
population of hydrogen atoms in the $n=2$ state. Dense, optically thick
gas provides one way of maintaining this population through collisions
and Ly$\alpha$ trapping, whether the gas is part of a stellar atmosphere,
an AGN broad-line region, or a larger reprocessing envelope
\citep{2025ApJ...980L..27I,2025A&A...701A.168D}.

\item \emph{Weak conventional signatures of an unobscured AGN.}
Most LRDs are individually undetected in current X-ray observations,
and stacked samples have lower X-ray luminosities relative to their
optical or broad H$\alpha$ emission than ordinary type-I AGNs
\citep{2024ApJ...974L..26Y,2025ApJ...995...24K}. Their line spectra
also commonly show a deficit of the hard ionizing photons expected
from a standard AGN spectral energy distribution
\citep{2026ApJ..1003...10W}, with a similar deficit reported in broader
samples of high-redshift type-I AGNs \citep{2026A&A...707A..52Z}, indicating that this deficit alone does not disfavor standard-AGN interpretations.
Deep millimeter observations place strong limits on the far-infrared
emission of the broader LRD population \citep{2025ApJ...990L..61C}.
For two of the most luminous LRDs, the combined limits on hot- and
cold-dust emission disfavor the reprocessed luminosity expected from
an intrinsically blue source strongly reddened by dust
\citep{2025ApJ...991L..10S, 2025ApJ...994L..42C}.
Multi-epoch observations find little strong UV-optical variability
in the majority of LRDs, although a minority of variable sources
exists \citep{2025ApJ...983L..26T,2025ApJ...985..119Z, 2026arXiv260521574L}. If the compact red source and the extended blue emission arise from physically distinct regions of different size, variability may be preferentially associated with the red, more compact component.


\item \emph{A high abundance with a luminosity cutoff.}
LRDs have comoving number densities of order
$10^{-5}$--$10^{-4}\,{\rm cMpc^{-3}}$, depending on selection and
luminosity, making them roughly two orders of magnitude more abundant
than UV-selected quasars extrapolated to comparable optical
luminosities
\citep{2024ApJ...963..129M,2024ApJ...968...38K,
2025ApJ...991...37A}. Meanwhile, their luminosity function
declines sharply above
$\lambda L_{5100}\simeq2.5\times10^{44}\,{\rm erg\,s^{-1}}$, well
below the luminosities reached by classical quasars
\citep{2025arXiv250902662M}. Their number density appears to peak near
$z\sim5$ and declines toward lower redshift \citep{2025ApJ...988L..22I}, although LRDs remain
detectable at much lower redshifts:  candidates persist to $z\lesssim2$ \citep{2026arXiv260700084K}, a systematic DESI search has identified 27 LRDs at $z=0.2$--$0.9$ \citep{2026arXiv260521574L}, and several examples have now been reported at $z\simeq0.1$--$0.2$ \citep{2026ApJ...997..364L, 2026MNRAS.545f2235J}.


\item \emph{A compact optical source inside an extended, non-pristine
host.} The red rest-frame optical component that defines an LRD is commonly
unresolved, whereas the rest-frame UV emission is often spatially
extended or morphologically complex
\citep{2025ApJ...983...60C,2025ApJ...992...71R}. This difference
suggests that at least part of the UV light originates from the
surrounding stars rather than from the compact source that
dominates the optical continuum. Narrow nebular lines associated with
the host further show that its ionized gas contains metals: LRD hosts
may be metal-poor relative to present-day massive galaxies, but they
are not primordial
\citep{2026arXiv260631515N,2026MNRAS.550g1220I}. These observations
therefore place the compact red source within an already enriched
galaxy, although they do not directly determine the metallicity of the
unresolved nuclear region.


\end{enumerate}

Most interpretations place an accreting black hole at the center of an
LRD. A dust-obscured AGN naturally supplies broad Balmer lines
\citep{2024A&A...691A..52K}, while
mildly super-Eddington accretion can produce an intrinsically soft,
X-ray-weak spectrum
\citep{2024ApJ...976...96P,2026NatAs..10..868L, 2026ApJ..1007L..13L}.
Super-Eddington disk-atmosphere calculations can also produce red
optical continua and Balmer breaks without requiring absorption by
external gas \citep{2025ApJ...994..113L}. Dense gas surrounding the accretion flow can
absorb the ionizing continuum, enhance the hydrogen $n=2$ population,
and generate Balmer line absorption and a nonstellar Balmer break
\citep{2025ApJ...980L..27I,2026ApJ..1004..153K,
2026arXiv260118864S}. These models can therefore
explain many individual LRD properties. Their challenge is to account
simultaneously for the nearly thermal red continuum, the preferred
V-shaped turnover at the Balmer limit, weak X-ray and dust emission, limited
variability, and the population-level luminosity cutoff. The answer
depends on the geometry, column density, covering fraction, and
radiative transfer of an unusually dense nuclear medium. In this
regime, conventional bolometric corrections and virial black-hole
mass estimates may also be strongly model dependent.

Accretion-powered envelope models take the reprocessing picture
further by embedding an accreting black hole inside optically thick
gas. Classical quasi-stars place a newborn black hole inside a massive,
self-gravitating envelope, whereas recent ``black-hole star'' models
invoke a black-hole-dominated, dense atmosphere or cocoon around a
rapidly accreting supermassive black hole (SMBH) \citep{2008MNRAS.387.1649B,2024ApJ...970..158C, 2025MNRAS.544.3407K, 2025arXiv250316596N,2026ApJ...996...48B,
 2026ApJ...998L...4S,2026OJAp....962505S}.
Both configurations can provide a cool apparent photosphere while
suppressing direct emission from the central engine, but they differ
in their structure, seed requirements, and evolutionary
timescales. We compare them separately with our interpretation in
\S\ref{sec:sms_quasistar_comparison} and
\S\ref{sec:sms_bhstar_comparison}.

Models dominated by an extreme compact stellar population avoid the
need for an active black hole and can associate the Balmer break with
stellar light. The very large velocity dispersion of a sufficiently
dense stellar system might also broaden its nebular lines
\citep{2024ApJ...977L..13B}. However, interpreting the red continuum
as an ordinary stellar population often requires exceptionally large
stellar masses and densities within radii of 10--$100\,{\rm pc}$ --- comparable to or exceeding the densities at which runaway stellar collisions become efficient \citep{2004Natur.428..724P,2006MNRAS.368..121F,2006MNRAS.368..141F}. We in fact invoke such collisional runaway ourselves as a possible formation channel for the SMS progenitor (\S\ref{sec:formation_channels}); the observational distinction is therefore not whether such a dense cluster can exist, but whether the directly visible light comes from the pre-collision cluster itself or from the single merged object it produces, a question we return to in \S\ref{sec:formation_channels}. It is also unclear whether such a stellar systems can reproduce the unresolved, approximately thermal optical component and the broad Balmer line luminosities across the full LRD population. These
difficulties have motivated models in which the luminosity is generated by a single massive central object rather than by a conventional stellar population.

\begin{figure*}[!thb]
    \centering
    \includegraphics[width=0.85\textwidth]{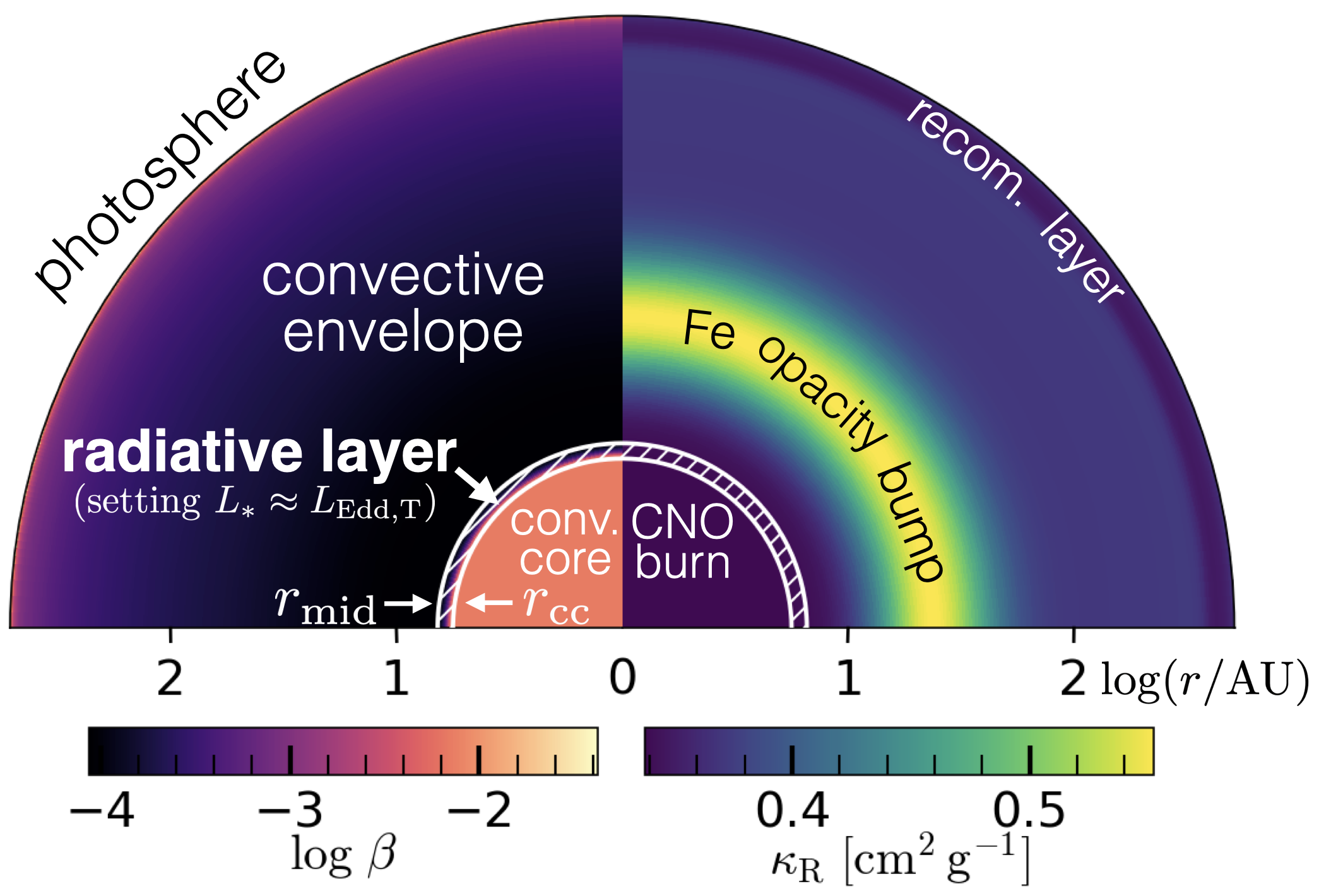}
    \caption{Cross section of a representative inflated SMS solution ($Z=10^{-2}$, $M=10^{6}\,M_{\odot}$,
    $Q_{\rm LIM}=0.1$). The two quadrants show the same radial structure colored by different quantities, both plotted against $\log(r/{\rm AU})$: the gas-pressure fraction $\log\beta$ (left) and the Rosseland-mean opacity $\kappa_{\rm R}$ (right). From the center outward: the convective, CNO-burning core; a geometrically thin radiative layer just outside the convective core, at $r_{\rm cc} < r < r_{\rm mid}$ (\S\ref{sec:sms_model}), which dictates that the total stellar luminosity is almost equal to the Thomson-scattering Eddington luminosity $L_\star\approx (1-\beta)L_{\rm Edd,T}$ for $\beta\ll 1$; the extended convective envelope inflated by the Fe-opacity bump
    (using the effective Eddington limiter parameter $Q_{\rm LIM}=0.1$, \S\ref{sec:gamma_eff_limiter}), where $\beta$ reaches its lowest values
    (darkest region, left) and $\kappa_{\rm R}$ peaks (brightest band,
    right); the outer hydrogen/helium recombination layer, where
    $\kappa_{\rm R}$ falls and $\beta$ rises back up; and the photosphere. The boundaries of the radiative layer --- the convective core boundary radius $r_{\rm cc}$ (\S\ref{sec:sms_model}) and the inner transition-line crossing radius $r_{\rm mid}$ (defined in \S\ref{sec:hse_envelopes} below) --- both lie well inside the Fe-bump-inflated envelope; the base radius $r_{\rm base}$ (no physical importance, not marked) sits just outside $r_{\rm cc}$.
    }
    \label{fig:sms_schematic}
\end{figure*}

In this work, we investigate the hypothesis that at least a subset of
LRDs are thermally relaxed, hydrogen-burning, metal-enriched
supermassive stars (SMSs). SMSs have long been studied as
relativistically unstable stellar configurations
\citep{1963ApJ...138.1090I,1964ApJ...140..417C,
1966ApJ...144..180F}, and were subsequently considered as possible
progenitors of massive black holes in primordial direct-collapse
environments \citep{2003ApJ...596...34B}. Rapidly accreting
primordial SMSs can also develop cool, extended ``supergiant
protostar'' envelopes
\citep{2013ApJ...778..178H,2019PASA...36...27W}, and more recent work
has explored connections between SMS spectra and LRDs
\citep{2026ApJ...998..124N}. The objects considered here occupy a
different physical regime from previous works: they are CNO-burning stars in
thermal and hydrostatic equilibrium, and their envelopes are inflated
by metal opacity rather than by an accretion
flow.

This interpretation separates the luminosity source from the
line-formation problem. Nuclear burning supplies a luminosity close to
the Thomson-scattering Eddington limit,
$L_\star\approx L_{\rm Edd,T}\propto M_\star$, so the observed
luminosity maps directly onto the stellar mass. In the absence
of an accreting compact object, weak X-ray and
high-ionization line emission follow naturally. The central physical
question is whether a metal-enriched SMS can support a
strongly inflated envelope whose photosphere is as cool and extended
as the red component of an LRD.

We construct one-dimensional hydrostatic SMS models containing a
radiation-pressure-dominated, CNO-burning core and an outer envelope
with metallicity-dependent Rosseland-mean opacity and non-adiabatic
convective transport. We find that the Fe-opacity bump drives
substantial envelope inflation at high metallicity, while hydrogen and helium
recombination reduces the opacity in the outermost layers and permits
the inflated structure to terminate at a bound photosphere. Models
with $Z\simeq10^{-2}$ reach $T_{\rm eff}\sim7000\,{\rm K}$ at luminosities characteristic of LRDs,
whereas lower-metallicity models remain hotter and more compact.
The general-relativistic instability predicts a maximum mass and hence a maximum luminosity of order
$10^{44}\,{\rm erg\,s^{-1}}$ (depending on the core rotation rates), which may be comparable to the observed LRD luminosity-function cutoff. Hydrogen depletion can drive initially stable SMSs across this threshold after a lifetime of order $1\,{\rm Myr}$, producing massive black-hole remnants whose cumulative abundance can be compared directly with present-day SMBH demographics.

The cool continuum, maximum luminosity, weak hard-photon emission, and
massive remnants are direct consequences of the SMS model. Explaining the broad
Balmer line emission requires non-LTE atmosphere calculations. We discuss the possible roles of
Ly$\alpha$ trapping, Balmer-continuum reprocessing, and trans-sonic
shocks, but do not treat the observed lines as a prediction of
the present hydrostatic models.

The remainder of the paper is organized as follows.
\S\ref{sec:sms_model} develops the SMS interior and opacity
model. \S\ref{sec:hse_envelopes} presents the hydrostatic
inflated-envelope calculation, and \S\ref{sec:model_results}
describes the metallicity-dependent solutions.
\S\ref{sec:smbh_connection} derives the GR-instability threshold
and compares the remnant population with SMBH demographics.
\S\ref{sec:lrd_predictions} summarizes the principal observable
predictions, and \S\ref{sec:discussion} discusses atmosphere
physics, variability, host environments, formation channels, and
future work. Fig.~\ref{fig:sms_schematic} provides a schematic
overview of the stellar structure and terminology used throughout the
paper. Throughout this work, \(\log\) denotes \(\log_{10}\).

\section{SMS Interior and Opacity Model}
\label{sec:sms_model}

\subsection{Deep-interior structure}
\label{sec:deep_interior}

We compute self-consistent SMS structures for $10^{4}\leq M/M_{\odot}\leq 10^{6}$ and extend to higher masses relevant for GR instability later (\S\ref{sec:smbh_connection}). The composition is taken as uniform, with a fiducial hydrogen mass fraction $X=0.7$, metal mass fraction $Z$, and helium mass fraction $Y=1-X-Z$. We adopt an equation of state consisting of non-degenerate gas and radiation
\begin{equation}
    P=P_{\rm gas}+P_{\rm rad},
    \quad
    P_{\rm gas}=\frac{\rho k_{\rm B}T}{\mu m_{\rm p}},
    \quad
    P_{\rm rad}=\frac{a_{\rm rad}T^{4}}{3},
\label{eq:EOS_P_gas_rad}
\end{equation}
and the gas-pressure fraction is $\beta\equiv P_{\rm gas}/P$.

The deep interior consists of a convective, CNO-burning core surrounded by
a radiative envelope, integrated outward with the constant, fully ionized
Thomson opacity and mean molecular weight,
\begin{equation}
    \kappa_{\rm T}=\frac{(1+X)\sigma_{\rm T}}{2m_{\rm p}},
    \quad
    \mu=\left(2X+\frac{3}{4}Y\right)^{-1},
\label{eq:kappaT_mu_core}
\end{equation}
appropriate above the He$^{2+}$ recombination temperature. The structure of the convective core is governed by mass conservation, hydrostatic equilibrium,
\begin{equation}
    \frac{dm}{dr}=4\pi r^{2}\rho,
    \quad
    \frac{dP}{dr}=-\frac{Gm\rho}{r^{2}},
\label{eq:core_mass_hse}
\end{equation}
energy conservation, and adiabatic temperature gradient
\begin{equation}
    \frac{dL}{dr}=4\pi r^{2}\rho\,\epsilon_{\rm CNO},
    \quad
    \frac{dT}{dr}=\nabla_{\rm ad}\frac{T}{P}\frac{dP}{dr},
\label{eq:core_luminosity_adiabatic}
\end{equation}
with the CNO energy generation rate $\epsilon_{\rm CNO}(\rho,T,Z_{\rm CNO})$
taken from the \citet{2005EPJA...25..455I} analytic fit, the CNO catalyst
abundance,
\begin{equation}
    Z_{\rm CNO}=f_{\rm CNO}Z,
\end{equation}
and the adiabatic gradient,
\begin{equation}
    \nabla_{\rm ad}(\beta)=\frac{2(4-3\beta)}{32-24\beta-3\beta^{2}}.
\label{eq:nabla_ad}
\end{equation}
We fix $f_{\rm CNO}=0.7$ throughout \citep{1998SSRv...85..161G}, since the departure of $f_{\rm CNO}$ from this solar-like value is not strong over the metallicities considered here. The core ends at radius $r_{\rm cc}$, where the radiative and adiabatic temperature gradients coincide (the Schwarzschild criterion),
\begin{equation}
    \nabla_{\rm rad}(r_{\rm cc})=\nabla_{\rm ad}(r_{\rm cc}),
    \quad
    \nabla_{\rm rad}=\frac{3\kappa_{\rm T}PL}{16\pi acGmT^{4}}.
\label{eq:schwarzschild_rcc}
\end{equation}

Beyond $r_{\rm cc}$, the luminosity is fixed at
\begin{equation}
L(r > r_{\rm cc})=L_{\star}\equiv L(r_{\rm cc})
\label{eq:L_beyond_rcc}
\end{equation}
since CNO burning is negligible outside the core. We integrate the radiative envelope using $\ln P$ as the independent variable, which remains
well behaved even as the envelope becomes very extended (unlike $r$):
\begin{equation}
    \frac{dr}{d\ln P}=-\frac{k_{\rm B}Tr^{2}}{Gm\mu m_{\rm p}\beta},
    \quad
    \frac{d\ln T}{d\ln P}=\frac{\Gamma}{4(1-\beta)},
\label{eq:envelope_dr_dlnT_dlnP}
\end{equation}
with
\begin{equation}
    \beta=1-\frac{P_{\rm rad}}{P},
    \quad
    \rho=\frac{P\beta\mu m_{\rm p}}{k_{\rm B}T},
    \quad
    \Gamma=\frac{\kappa_{\rm T}L_{\star}}{4\pi cGm},
\label{eq:beta_rho_def}
\end{equation}
and we keep track of the enclosed mass $m(r)$ by integrating $dm/d\ln P=4\pi r^{2}\rho\,(dr/d\ln P)$. In a radiative layer, we combine hydrostatic equilibrium, $dP/dr=-Gm\rho/r^{2}$, with the radiative-diffusion form of $dT/dr$ (which gives $dP_{\rm rad}/dr=-\Gamma\,Gm\rho/r^{2}$), and hence obtain the following general identity in any radiative zone
\begin{equation}
    \frac{dP_{\rm rad}}{dP}=\Gamma.
\label{eq:dPrad_dP_Gamma}
\end{equation}
This system is integrated from $r_{\rm cc}$ down to $T_{\rm base}\equiv10^{6}\,$K, well above He$^{2+}$ recombination (so a constant opacity and mean molecular weight apply).
We denote the corresponding radius $r_{\rm base}\equiv r(T_{\rm base})$. This value of $T_{\rm base}$ is chosen to be low enough that the exterior mass above $r_{\rm base}$ is negligible and close to the Fe-opacity-bump temperature, yet high enough that the opacity there is given by electron scattering; as we show below, the resulting boundary condition is then essentially exact regardless of the precise value of $T_{\rm base}$, since the optical depth at $r_{\rm base}$ is large.

An important consequence of the Schwarzschild condition at the convective-core boundary is that it \textit{determines} the stellar
luminosity $L_\star$. Taylor-expanding $\nabla_{\rm ad}(\beta)$ for $\beta\ll 1$ gives
$\nabla_{\rm ad}\approx\tfrac14(1+3\beta^{2}/32)$, so the Schwarzschild
condition $\nabla_{\rm rad}=\nabla_{\rm ad}$, together with $\nabla_{\rm rad}=\tfrac14(P/P_{\rm rad})(L/L_{\rm Edd})$,
gives
\begin{equation}
    \frac{L(r_{\rm cc})}{L_{\rm Edd}(r_{\rm cc})} = \Gamma(r_{\rm cc})
    \approx1-\beta(r_{\rm cc})+\frac{3}{32} \beta^{2}(r_{\rm cc}).
\label{eq:L_Ledd_rcc_Gamma}
\end{equation}
Differentiating $\beta=1-P_{\rm rad}/P$ and using
$\d P_{\rm rad}/\d P=\Gamma$, we find the $\beta$ stratification in any radiative layer
\begin{equation}
\frac{\d\beta}{\d\ln P} = 1-\Gamma-\beta,
\label{eq:dbeta_dlnP}
\end{equation}
and hence
\begin{equation}
\frac{\d\beta}{\d\ln P} (r_{\rm cc})\approx - \frac{3}{32} \beta^{2}(r_{\rm cc}),
\label{eq:dbeta_dlnP_rcc}
\end{equation}
This means that $\beta$ changes only by a small, second-order amount across the many decades in pressure between $r_{\rm cc}$ and $r_{\rm base}$ (as confirmed by our numerical integration).

\begin{figure*}
    \centering
    \includegraphics[width=0.8\textwidth]{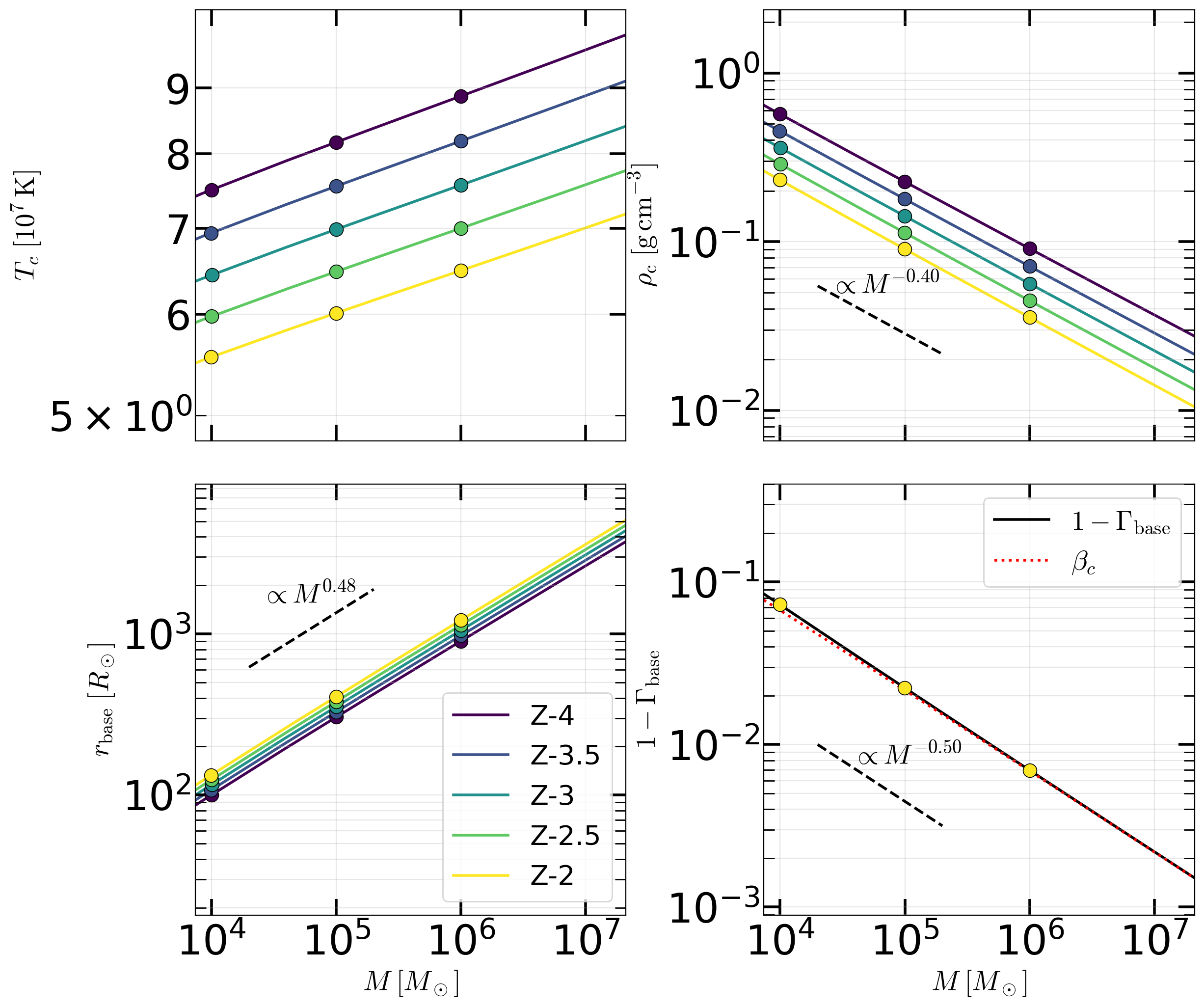}
    \caption{Deep-interior sequences of base models (\S\ref{sec:sms_model}),
    integrated with constant Thomson-scattering opacity $\kappa_{\rm T}$
    and $X=0.7$, $f_{\rm CNO}=0.7$, for five metallicities
    ($Z=10^{-4}$ to $10^{-2}$, denoted as `Z-4' to `Z-2').
    Solid curves show the results from a dense $T_c$ grid for each metallicity;
    filled circles mark the three mass-targeted models at
    $M=10^{4},10^{5},10^{6}\,M_{\odot}$ used throughout this work.
    Clockwise from top left: central temperature $T_c$; central density
    $\rho_c$; Eddington factor $1-\Gamma_{\rm base}\approx \beta_c$ (metallicity-independent); and the base radius $r_{\rm base}$ (where $T=T_{\rm base}\equiv 10^{6}\,{\rm K}$). At higher metallicities, the base models at a fixed mass are slightly more extended and the threshold mass for GR instability is higher (\S\ref{sec:gr_instability_mass}).
    }
    \label{fig:sms_base}
\end{figure*}

The specific entropy of radiation pressure-dominated gas scales as $s\propto T^3/\rho\propto \beta^{-1}$, so the convective core has a nearly constant $\beta\approx \beta_c\approx \beta(r_{\rm cc})$, where $\beta_c$ is the central gas-pressure fraction. 
The near constancy of $\beta$ in the radiative layer further means that
\begin{equation}
\beta_c \approx \beta(r_{\rm cc}) \approx \beta(r_{\rm base})\approx 1-\Gamma(r_{\rm base}).
\label{eq:betac_approx}
\end{equation}
This also means that the Eddington factor stays nearly constant $\Gamma(r_{\rm cc})\approx \Gamma(r_{\rm base})$ to within a factor of $\mc{O}(\beta^2)$. For fully ionized gas with a constant Thomson scattering opacity $\kappa_{\rm T}$, this shows that the radiative layer only contains a mass of $\mc{O}(\beta^2 M)$, so that
\begin{equation}
m(r_{\rm cc})\approx m(r_{\rm base})\approx M.
\label{eq:mass_rcc_approx_M}
\end{equation}
Since $L_{\rm Edd}(r_{\rm cc})= L_{\rm Edd,T}$, the total luminosity of the star is then given by
\begin{equation}
L_\star \approx (1-\beta_c) L_{\rm Edd,T}, \quad L_{\rm Edd,T}\equiv \frac{4\pi cGM}{\kappa_{\rm T}}.
\label{eq:Lstar_LeddT}
\end{equation}
For the SMS mass range $10^4\leq M/M_\odot \leq 10^6$ considered in this work, since $\beta_c\ll 1$, we find that every SMS radiates close to, and only slightly below, the Thomson-scattering Eddington luminosity evaluated at its total mass, essentially independent of the detailed opacity structure of the outer envelope.


Next, we derive the boundary condition at $r_{\rm base}$ under the assumption of a constant opacity --- the effects of variable opacity in the cooler regions of the outer envelope will be considered as a small perturbation to the boundary condition at $r_{\rm base}$ in \S \ref{sec:hse_envelopes}.

For a nearly constant $\Gamma$ in the radiative layer, $\d P_{\rm rad}/\d P=\Gamma$ integrates trivially to $P_{\rm rad}=\Gamma P+C$ for some constant $C$. Using the optical depth $\tau(r)=\int_r^\infty\rho\kappa_{\rm T}\,dr'$, the temperature profile under the plane-parallel Eddington-approximation is $T^4(\tau) = (3/4) T_{\rm eff}^4(\tau + 2/3)$. Using $T_{\rm eff}^4 = \Gamma L_{\rm Edd,T}/(4\pi \sigma_{\rm SB} R^2)$ where $R$ is the photospheric radius\footnote{The photospheric radius $R$ here is only defined based on a hypothetical star with constant opacity $\kappa_{\rm T}$, for which $R$ is located slightly above $r_{\rm base}$.} where $\tau=2/3$, we then find the radiation pressure profile
\begin{equation}
P_{\rm rad}(\tau) = a_{\rm rad} T^4(\tau)/3 = \Gamma(GM/\kappa_{\rm T}R^{2})(\tau+2/3).
\label{eq:Prad_tau}
\end{equation}
On the other hand, we can also obtain the gas pressure profile by integrating the equation of hydrostatic equilibrium $\d P_{\rm gas}/\d \tau = (1-\Gamma)GM/(\kappa_{\rm T}R^2)$ with the boundary condition $P_{\rm gas}(\tau=0) = 0$, and this gives
\begin{equation}
P_{\rm gas}(\tau) = {(1-\Gamma) GM\over \kappa_{\rm T}R^2} \tau.
\label{eq:Pgas_tau}
\end{equation}
Combining $P_{\rm rad}(\tau)$ and $P_{\rm gas}(\tau)$, we first obtain the gas-pressure fraction at the photosphere $\tau=2/3$,
\begin{equation}\label{eq:beta_ph_const_kap}
    \beta(\tau=2/3) = {P_{\rm gas}(2/3)\over P_{\rm rad}(2/3)+ P_{\rm gas}(2/3)} = {1-\Gamma\over 1+\Gamma},
\end{equation}
and then the integration constant
\begin{equation}
    C = P_{\rm rad}(\tau)-\Gamma P(\tau) = \frac{2}{3}\Gamma(1-\Gamma)\frac{GM}{\kappa_{\rm T}R^{2}}.
\label{eq:C_const}
\end{equation}
Since the total pressure $P(\tau)$ grows linearly with $\tau$ at large optical depths while $C$ stays fixed, and since
$T_{\rm base}=10^{6}\,$K lies at $\tau(r_{\rm base})\gg1$, we find
\begin{equation}
{C\over P(r_{\rm base})} \sim {1-\Gamma\over \tau(r_{\rm base})} \ll 1-\Gamma.
\label{eq:C_over_P_negligible}
\end{equation}
This shows that the boundary condition at $r_{\rm base}$ can be taken as $C\approx 0$ to high precision as long as the optical depth across a local scale height near $r_{\rm base}$ is much greater than unity. Later in \S \ref{sec:hse_envelopes}, we will consider how an opacity-driven inflated envelope exerts a small perturbation on the condition of $C\approx 0$ at $r_{\rm base}$. Nevertheless, the small perturbation has a negligible effect on the core structure, which is largely determined by the convective core region within $r_{\rm cc}$.


The inner boundary condition is a near-center Taylor expansion of
$(m,P,L,T)$ about $r=0$ for fixed central temperature $T_c$ and trial
central density $\rho_c$. The outer boundary condition at $r_{\rm base}$ is taken to be $P_{\rm rad} = \Gamma P$ at $r_{\rm base}$, which gives the following residual function 
\begin{equation}
    f(\rho_c)\equiv\Gamma_{\rm base}-\frac{P_{\rm rad}(r_{\rm base})}{P(r_{\rm base})}=0,
\label{eq:f_rhoc_residual}
\end{equation}
For each $T_c$, $\rho_c$ is adjusted by a fine-tuned shooting method
(Appendix~\ref{app:shooting}) until $|f(\rho_c)|\lesssim10^{-9}$. This
procedure yields a converged mass $M = m(r_{\rm base})$ for each central temperature $T_c$ and metallicity $Z$. For a given $Z$, it is then possible to find the appropriate $T_c$ that reproduces a given mass $M$.

In our models with realistic opacity in the outer envelope, we first calculate the interior structure of a star as specified by $(M, Z)$, and then the corresponding base state $(r_{\rm base},P_{\rm base},T_{\rm base},\Gamma_{\rm base},L_{\star})$ serves as the inner boundary condition for the envelope calculation in
\S\ref{sec:hse_envelopes}.

Fig.~\ref{fig:sms_base} shows the resulting base-model sequences across
five metallicities $10^{-4}\leq Z\leq 10^{-2}$. Radiation pressure dominates
throughout ($\beta_c\ll1$). The solutions follow closely the well-known scalings of the standard Eddington model (for a constant $\beta$) with a nearly fixed central temperature due to the thermostat effect of CNO burning. From $P_c\propto T_c^4 \propto M^2/r_{\rm base}^4$ and $T_c\approx \mr{const}$, we find $r_{\rm base}\propto M^{1/2}$ and $\rho_c\propto M/r_{\rm base}^3\propto M^{-1/2}$. Our numerically obtained slopes, $\rho_c\propto M^{-0.40}$ and $r_{\rm base}\propto M^{0.48}$, deviate mildly from these ideal values because the central temperature is not exactly constant (as seen from the top left panel of Fig.~\ref{fig:sms_base}).

Note that $\beta_c$ and $\Gamma_{\rm base}$ are essentially independent of metallicity, because our base models are well approximated by a polytropic model with polytropic index $n=3$. As shown in Appendix \ref{app:constant_beta_core}, in the limit of $\beta_c\ll 1$, the stellar mass is uniquely determined by the polytropic constant $K_\beta =P/\rho^{1 + 1/n} \propto M^{2/3}$, and this leads to $\beta_c\approx 1-\Gamma_{\rm base}\propto \rho_c/T_c^3 \propto M^{-1/2}$, which agrees with our numerical solutions.

The small gas-pressure contribution also determines the net binding energy and hence the thermal timescale of the otherwise marginally bound radiation-dominated core. As shown in Appendix~\ref{app:constant_beta_core}, the constant-\(\beta\), \(n=3\) polytropic model gives a net binding energy
\(\lvert E_{\rm bind}\rvert=(3\beta_c/4)GM^2/r_{\rm base}\) (eq. \ref{eq:constant_beta_binding_energy}). The corresponding core Kelvin--Helmholtz timescale is
\begin{align}
    t_{\rm KH,c}
    &\equiv
    \frac{|E_{\rm bind}|}{L_\star}
    \simeq
    \frac{3\beta_c}{4}
    \frac{GM^2}{r_{\rm base}L_{\rm Edd,T}}
    \nonumber\\
    &\simeq 4\times10^3\,{\rm yr} \frac{\kappa_{\rm T}}{0.34\,{\rm cm^2\,g^{-1}}},
\label{eq:core_KH_timescale}
\end{align}
where we have taken $L_*\approx L_{\rm Edd,T}$ and an approximate mass-radius relation of $r_{\rm base}\simeq 1\times 10^3 R_\odot (M/10^6M_\odot)^{1/2}$ (to within $\sim$20\%, for our base models shown in Fig.~\ref{fig:sms_base}). Across our model grid, \(t_{\rm KH,c}\) is a few thousand years, nearly independent of the stellar mass and weakly dependent on metallicity (at the $\sim$20\% level). This is much shorter than the Myr-scale hydrogen-burning lifetime (see eq. \ref{eq:tau_SMS_GR} later), so the SMS core has sufficient time to become thermally relaxed and evolve in quasi-thermal equilibrium.


\subsection{Outer-envelope opacity and mean molecular weight}
\label{sec:opacity}

\begin{figure}
    \centering
    \includegraphics[width=\linewidth]{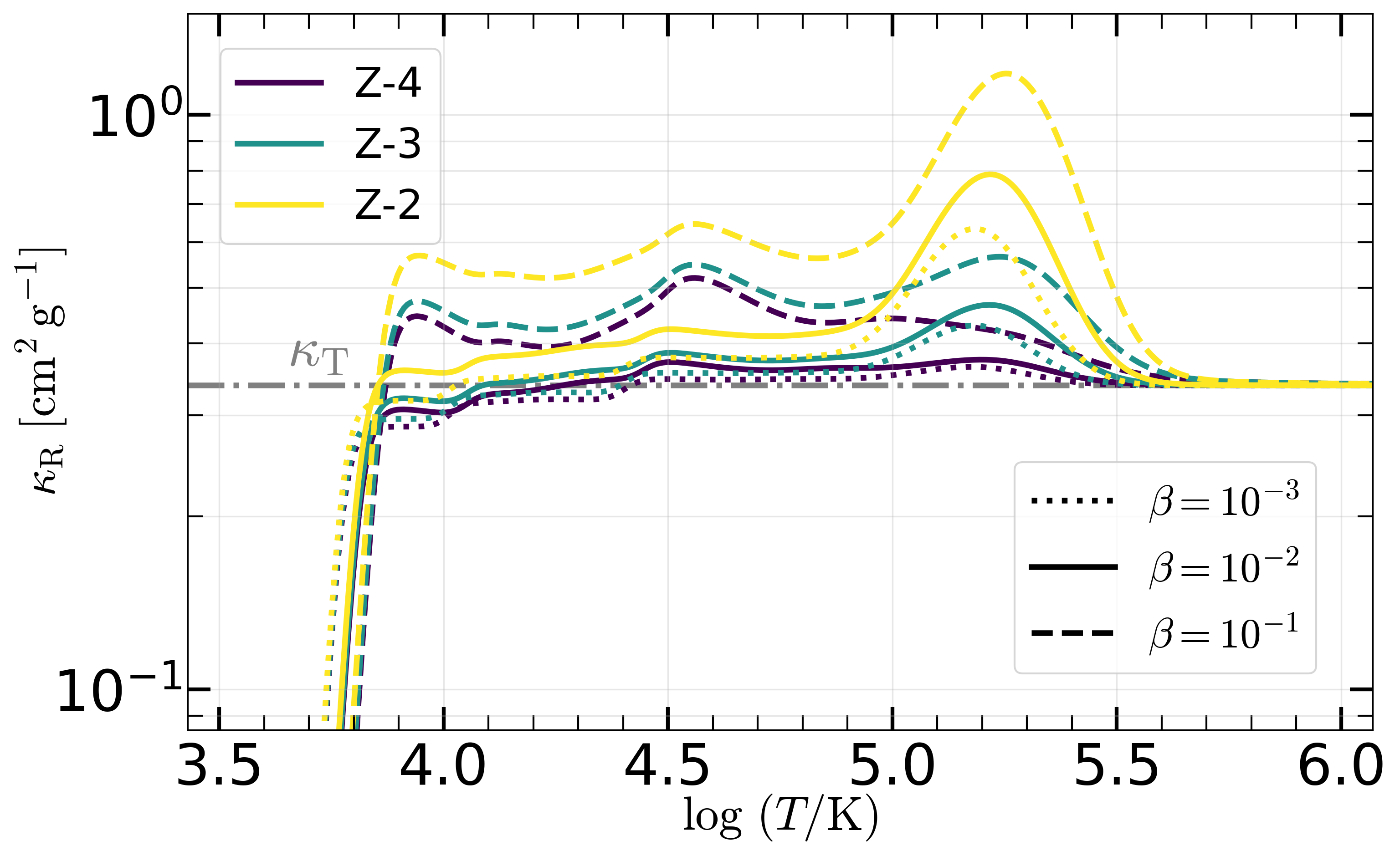}
    \caption{Total opacity model $\kappa_{\rm R}(T,\beta,Z)$ for the solar
    Fe-group abundance pattern, evaluated at three representative
    metallicities ($Z=10^{-4},10^{-3},10^{-2}$) and three
    gas-pressure fractions ($\beta=10^{-3},10^{-2},10^{-1}$),
    compared with the constant Thomson-scattering opacity $\kappa_{\rm T}$
    (gray dash-dotted). The steep rise near $\log T\approx3.8$--$4.0$ marks
    hydrogen recombination; the shoulder near $\log T\approx4.5$ arises from
    helium bound-free opacity; and the prominent, strongly $Z$- and
    $\beta$-dependent peak near $\log T\approx5.0$--$5.3$ is the iron-group
    opacity bump that drives the envelope inflation. At $T\gtrsim 10^{5.5}\rm\, K$ all curves
    converge to $\kappa_{\rm T}$ as the gas gets fully ionized.}
    \label{fig:kap_sms_tot}
\end{figure}

In the outer layers where $T<T_{\rm base}$, the constant, fully ionized $\kappa_{\rm T}$ of
\S\ref{sec:deep_interior} is no longer adequate: the iron-group opacity bump becomes important near $\log T\sim 5.2$ and at lower temperatures hydrogen and helium recombine. We therefore construct a model for the Rosseland-mean opacity $\kappa_{\rm R}(T,\beta,Z)$ under local thermodynamic equilibrium (LTE) decomposed as
\begin{equation}
    \kappa_{\rm R}(T,\beta,Z)
    =
    \kappa_{\rm H+He}(T,\beta)
    +
    \kappa_{\rm metal}(T,\beta,Z)\,y_{\rm H}(T,\beta),
\label{eq:kappaR_decomp}
\end{equation}
where $y_{\rm H}$ is the hydrogen ionization fraction.

The hydrogen and helium term $\kappa_{\rm H+He}$ is metallicity-independent
and includes electron scattering, bound-free and free-free absorption
(H, He$^0$, He$^+$, and H$^-$), each computed via the Saha equation for
the three ionization stages H$\to$H$^+$, He$^0\to$He$^+$, and
He$^+\to$He$^{2+}$ (Appendix~\ref{app:opacity_model}). The monochromatic
opacities are summed before the Rosseland harmonic-mean integral; bound-bound (line) opacity is neglected throughout, justified by the low densities characteristic of the SMS envelopes
considered here. This term alone reproduces the steep opacity rise near
$\log T\approx3.8$--$4.0$ (hydrogen recombination) and the shoulder near
$\log T\approx4.5$ (helium bound-free opacity) seen in
Fig.~\ref{fig:kap_sms_tot}.

\begin{figure*}
    \centering
    \includegraphics[width=0.85\textwidth]{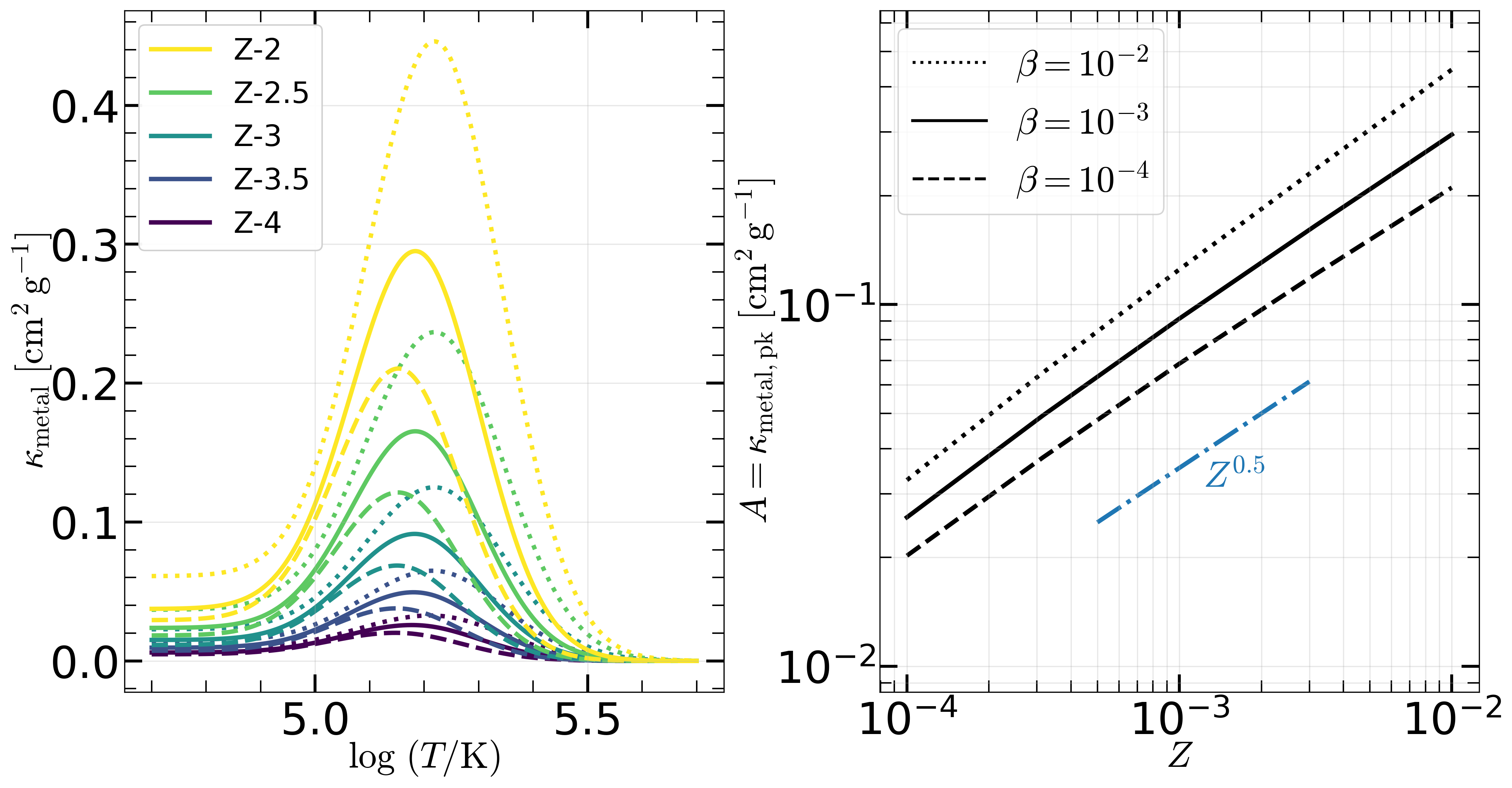}
    \caption{Shape (left) and peak amplitude (right) of the metal contribution to the Rosseland-mean opacity $\kappa_{\rm metal}(T,\beta, Z)$ for solar abundance pattern across five metallicities (``Z-4'' to ``Z-2'', colors) and three representative gas-pressure fractions ($\beta=10^{-4},10^{-3},10^{-2}$, linestyles). The peak amplitude increases monotonically with both $Z$ and $\beta$ and is well described locally by $\kappa_{\rm metal,pk}\propto Z^{\approx 0.5}$ with a mild positive dependence of the exponent on $\beta$, while the bump's temperature location ($\log T\approx5.2$) and width are very weakly dependent on both $Z$ and $\beta$.
    }
    \label{fig:kap_metal}
\end{figure*}

The metal contribution $\kappa_{\rm metal}(T,\beta,Z)$ is obtained from
the LANL TOPS opacity tables \citep{2016ApJ...817..116C} by subtracting a metal-free run from a
metal-enriched run at the same $(T,\rho)$, then fit at each $\beta$ to a
Gaussian + plateau model,
\begin{equation}
    \kappa_{\rm metal}=A\,G+\kappa_{\rm p}(1-G)H,
\label{eq:kappa_metal_fit}
\end{equation}
$$
    G=e^{-\frac12\left(\frac{\log T-\log T_{0}}{\sigma}\right)^{2}},
    \ 
    H=\left[1+e^{4(\log T-\log T_{0})/\sigma}\right]^{-1},
$$
with the four parameters $(A, T_{0},\sigma \kappa_{\rm p})$. This fit reproduces the raw TOPS-derived $\kappa_{\rm metal}$ to with an RMS residual of $\lesssim10\%$ across our full $(Z,\beta)$ grid (as shown Fig.~\ref {fig:kap_metal_fit}). These four parameters are then fitted by degree-4 (or 5) polynomials in $x = \log\lrsb{\beta/(1-\beta)}$. This isolates the iron-group opacity bump (Fig.~\ref{fig:kap_metal}), whose peak amplitude increases monotonically with both $Z$ and $\beta$. The peak amplitude scales as $A=\kappa_{\rm metal,pk}\propto Z^{\approx0.5}$, and this is because Fe ions contribute to the Rosseland-mean opacity in the form of a large number of collisionally broadened lines and the contribution from each line is determined by the part of the Lorentzian wings above the Thomson scattering opacity. The peak temperature location ($\log T\approx5.2$) and width of the Fe bump remain nearly independent of $Z$ and $\beta$ (see Fig. \ref{fig:kap_metal_params}). We additionally multiply the metal contribution $\kappa_{\rm metal}$ by the hydrogen ionization fraction to roughly account for the suppression of metal opacity below $\sim10^4\rm\,K$ --- at such low temperatures, the opacity $\kappa_{\rm R}$ is dominated by hydrogen, so our results are not affected by the exact recombination temperatures of metals.

Reducing the Fe-group abundance from the solar pattern \citep{1998SSRv...85..161G} by a factor of $f_{\rm Fe}=10^{-0.5}$ rescales the bump amplitude by a comparable factor without changing its shape or location
(Appendix~\ref{app:opacity_model}); we adopt this reduced pattern alongside
the solar one as a check for the sensitivity of our results to
the uncertain relative abundance of iron-group elements at low metallicity.

The two terms are combined into a single tabulated model, $\kappa_{\rm
R}(T,\beta,Z)$: $\kappa_{\rm H+He}$ is precomputed once on a
$(\log T,\log\beta)$ grid and the metal term added per metallicity, with a
bicubic spline on $\log\kappa_{\rm R}$ giving fast evaluation. The resulting total opacity, shown in Fig.~\ref{fig:kap_sms_tot}, has strong, $\beta$- and $Z$-dependent Fe-opacity bump, which ultimately drives the envelope inflation discussed in \S\ref{sec:hse_envelopes}. Hydrogen recombination
at low $T$ then causes $\kappa_{\rm R}$ to decline sharply, which
regulates where the inflated envelope can terminate at a cool photosphere.

The mean molecular weight is obtained self-consistently from the same
three Saha ionization fractions,
\begin{equation}
    \mu(T,\beta)^{-1}=X(1+y_{\rm H})+\frac{Y}{4}\Big[1+y_{\rm He^{0}}(1+y_{\rm He^{+}})\Big].
\label{eq:mu_inverse_envelope}
\end{equation}
At high temperatures $T\gg 3\times10^4\rm\, K$, the above expression matches the constant $\mu=(2X+3Y/4)^{-1}$ used in \S\ref{sec:deep_interior}. Because $\mu(\rho,T)$ is more naturally computed at fixed $\rho$
than at fixed $\beta$, evaluating it at a given $(T,\beta)$ requires
inverting $P_{\rm gas}=\rho k_{\rm B}T/(\mu m_{\rm p})$ against
$\beta=P_{\rm gas}/(P_{\rm gas}+P_{\rm rad})$ via a one-dimensional root
find in $\rho$; $\mu$ is tabulated and spline-interpolated in the same way
as $\kappa_{\rm R}$ for fast repeated evaluation.

This model is validated over $4000\,{\rm K}\leq T\leq10^{6}\,{\rm K}$
and $10^{-5}\leq\beta\leq0.9$, for five metallicities from $Z=10^{-4}$ to
$10^{-2}$, spanning the full range of conditions encountered in the
envelope calculations of \S\ref{sec:hse_envelopes}. Its principal
limitations are the neglect of bound-bound line opacity and the LTE assumption. Further validation details, including the full Saha ionization treatment and a comparison against the raw TOPS tables, are given in Appendix~\ref{app:opacity_model}.

\section{Hydrostatic Inflated Envelopes}
\label{sec:hse_envelopes}

\begin{figure*}
    \centering
    \includegraphics[width=0.8\linewidth]{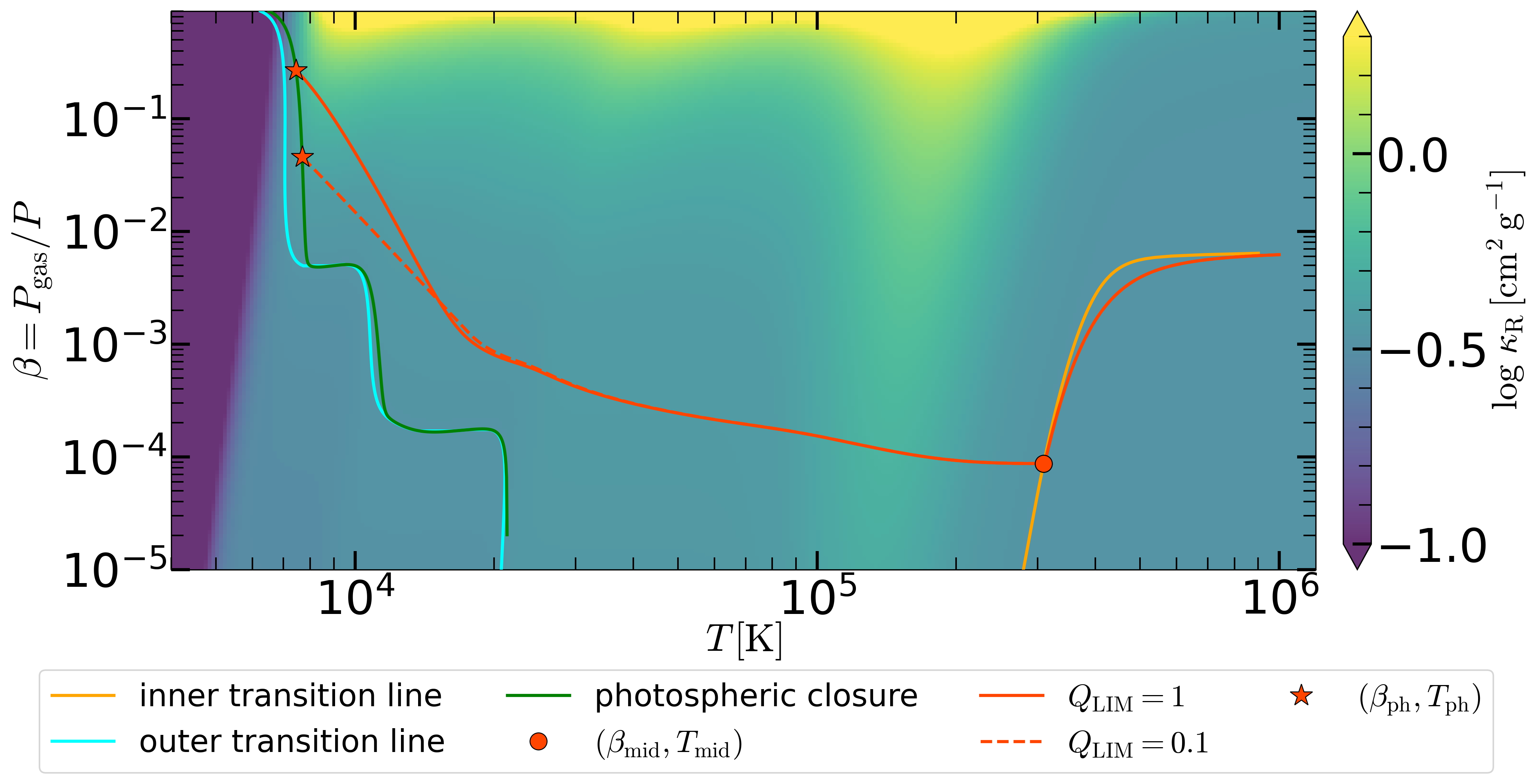}
    \caption{Envelope trajectories in the $(T,\beta)$ plane for the
    representative $Z=10^{-2}$, $M=10^{6}\,M_{\odot}$ models, overlaid on the background opacity map
    $\log\kappa_{\rm R}(T,\beta)$. The red circle marks
    $(\beta_{\rm mid},T_{\rm mid})$, where the profile crosses the inner
    transition line (orange curve, \S\ref{sec:hse_envelopes}) that anchors
    the two-point shooting method; from there the profile is integrated
    outward through the Fe-opacity-bump convective envelope to the
    photosphere, shown for two choices of effective-Eddington limiters $Q_{\rm LIM}=1$ (solid)
    and $0.1$ (dashed). Stars mark the resulting
    $(\beta_{\rm ph},T_{\rm ph})$ photospheric states, set by the
    intersection with the photospheric closure curve (green,
    \S\ref{sec:hse_envelopes}) near the outer transition line (cyan).
    The modest separation between the $Q_{\rm LIM}=1$ and $0.1$
    photospheric points along the temperature direction shows that $T_{\rm eff}$ is insensitive to the
    effective-Eddington limiter discussed in \S\ref{sec:model_results}.
    }
    \label{fig:envelope_solutions}
\end{figure*}

\subsection{Convective instability and the transition line}
\label{sec:transition_line}

Below $T_{\rm base}$ the opacity is given by
$\kappa_{\rm R}(T,\beta,Z)$ of \S\ref{sec:opacity}, so the local Eddington luminosity is
\begin{equation}
L_{\rm Edd}=\frac{4\pi cGM}{\kappa_{\rm R}(T,\beta,Z)}.
\label{eq:Ledd_kappaR}
\end{equation}
We use the constant total mass $M=m(r_{\rm base})$ throughout the envelope, as the region above
$r_{\rm base}$ contains negligible mass despite its large radial extent.
We also take the total stellar luminosity $L=L_\star=\mr{const}$ throughout the envelope and neglect mechanical luminosity carried by an outflow. This approximation requires any mass loss to remain dynamically unimportant.

We define a \textit{maximum} Eddington factor as
\begin{equation}
\Gamma_{\star}(T,\beta)\equiv\frac{L_{\star}}{L_{\rm Edd}},
\label{eq:Gammastar_def}
\end{equation}
which assumes that the total stellar luminosity $L_\star$ is transported by radiative diffusion. Realistically, it is possible that only part of the stellar luminosity $L_{\rm rad}$ is transported by radiative diffusion and the rest $L_{\rm conv}$ by convection in the cool envelope, so we have the following general decomposition
\begin{equation}
L_\star = L_{\rm rad} + L_{\rm conv},
\label{eq:Lstar_decomp}
\end{equation}
and the local Eddington factor is given by the radiative luminosity (which couples to pressure gradient and hence resists gravity)
\begin{equation}
\Gamma \equiv {L_{\rm rad}\over L_{\rm Edd}}\leq \Gamma_*.
\label{eq:Gamma_def}
\end{equation}
In the following, we discuss the convective boundary in the outer envelope.

A layer is convectively unstable when the radiative gradient, evaluated as
if all of $L_\star$ were carried radiatively, $\nabla_{\rm rad}=\Gamma_{\star}/[4(1-\beta)]$, exceeds the adiabatic gradient $\nabla_{\rm ad}(\beta)$ (eq.~\ref{eq:nabla_ad}). This condition is equivalent to
\begin{equation}
    \Gamma_{\star}>\Gamma_{\rm conv}
    \equiv
    \frac{1-7\beta/4+3\beta^{2}/4}{1-3\beta/4-3\beta^{2}/32}
    \approx1-\beta+\frac{3\beta^{2}}{32},
\label{eq:convective_condition}
\end{equation}
and since $\beta < 10^{-2}$ near the convective transition boundary for all cases considered in this work, $\Gamma_{\rm conv}\approx1-\beta$ to good approximation.

\paragraph{The transition line.} In \S\ref{sec:deep_interior} we showed that
$d\beta/d\ln P=1-\Gamma-\beta$ holds generally in any radiative layer. Evaluating the right-hand side in this purely radiative limit ($\Gamma\to\Gamma_{\star}$) defines the transition line,
\begin{equation}
    g(\beta,T)\equiv1-\Gamma_{\star}(\beta,T)-\beta=0,
\label{eq:transition_line_def}
\end{equation}
which is almost identical to the convective boundary (eq. \ref{eq:convective_condition}). Thus, the transition line can be effectively considered as the convective boundary itself, although our numerical treatment keeps track of the full $\Gamma_{\rm conv}$. Note that $g$ is the sign-changing quantity in $d\beta/d\ln P$, because $\beta$ decreases moving outward (decreasing $P$) through a radiative layer with $g>0$ and $\beta$ increases outward if $g<0$. Our solutions with inflated envelopes indeed show that $g=0$ marks the minimum $\beta$, which is usually much smaller than $\beta(r_{\rm base})\approx \beta_c$ (see Fig. \ref{fig:envelope_solutions}).

Two possible crossings exist in a typical envelope profile: an \textit{inner
transition line}, just above the Fe-bump peak temperature, marking the onset of
convection; and an \textit{outer transition line}, in the hydrogen and helium  recombination region, marking where
convection ends. We are primarily interested in the inner transition line,
since the gas-pressure fraction on the transition line $\beta_{\rm mid}$
largely controls how inflated the resulting convective envelope becomes
(\S\ref{sec:two_point_shooting}). The outer transition line is
comparatively unimportant, since it frequently lies beyond the
photospheric radius itself.

\subsection{Non-adiabatic mixing-length treatment}
\label{sec:mlt}

In the convective regions of the outer envelope, we adopt a non-adiabatic mixing-length theory (MLT) similar to that of \citet{1965ApJ...142..841H} to explicitly resolve the split of the stellar luminosity between radiative and convective parts as in eq.~\eqref{eq:Lstar_decomp}. The non-adiabatic part of MLT sets the local convective efficiency by comparing the radiative-diffusion time across a rising blob to the time over which it mixes with its surroundings; the full derivation, including the convective velocity $v_c$ and the superadiabaticity $\nabla-\nabla_{\rm ad}$ is given in Appendix~\ref{app:mlt}.

\begin{figure}
    \centering
    \includegraphics[width=\linewidth]{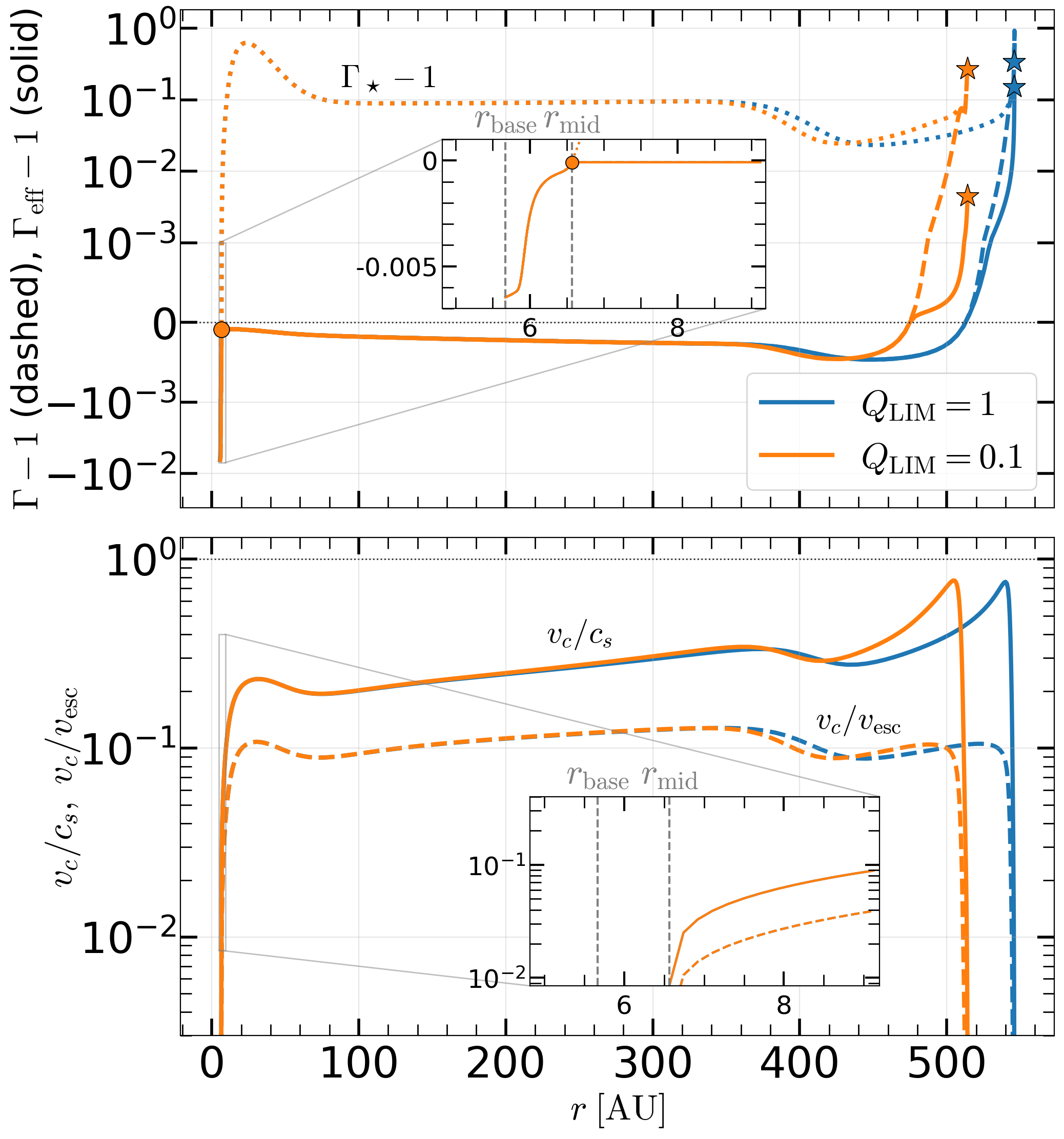}
    \caption{MLT diagnostics through the envelope of the
    $Z=10^{-2}$, $M=10^{6}\,M_{\odot}$ solution, from $r_{\rm base}$ to the
    photosphere $r_{\rm ph}$, for $Q_{\rm LIM}=1$ (blue) and $0.1$ (orange).
    \textit{Upper panel:} $\Gamma-1$ and $\Gamma_{\rm eff}-1$. The $\Gamma\approx1$ attractor
    (\S\ref{sec:mlt}) holds $|\Gamma-1|\lesssim10^{-3}$ across the deep
    convective envelope and only breaks down near the photosphere, where convection fails and $\Gamma-1$ rises sharply; $\Gamma_{\rm eff}-1$ is visibly suppressed below the $\Gamma-1$
    there, more strongly for the smaller $Q_{\rm LIM}=0.1$.
    \textit{Lower panel:} the convective velocity relative to the local
    sound speed $v_c/c_s$ and to the local escape speed
    $v_c/v_{\rm esc}$. Note that, near the photosphere, $v_c/c_s$ approaches unity, suggesting that the fractional density fluctuation $\Delta \rho/\rho$ also reaches order unity and we expect the porosity effect to become important.
    }
    \label{fig:limiter_envelope}
\end{figure}

Over most of the envelope, MLT drives convection toward a $\Gamma\approx1$
\textit{attractor} \citep{2012A&A...538A..40G}: wherever the layer is
formally super-Eddington ($\Gamma_\star>1$), convection carries just enough
of the flux to pull the actual radiative Eddington factor $\Gamma$ back down
just below unity; $\nabla$ tracks $\nabla_{\rm ad}$ closely and the convective blob
behaves adiabatically. Fig.~\ref{fig:limiter_envelope} (upper panel)
shows this explicitly for two representative $Z=10^{-2}$,
$M=10^{6}\,M_{\odot}$ solutions ($Q_{\rm LIM}=0.1, 1$): throughout the deep convective envelope,
$|\Gamma-1|$ is suppressed to below $10^{-3}$, orders of magnitude below the super-Eddington excess of a purely radiative layer ($\Gamma_\star-1$). This attractor breaks down, however, in a layer close to the
photosphere where the opacity is still high enough that a purely radiative
layer would formally be super-Eddington, yet radiative diffusion outpaces
the mixing of a buoyant blob before it can release its heat, so convection
is suppressed instead of relieving the excess flux --- visible in
Fig.~\ref{fig:limiter_envelope} as the sharp rise of $\Gamma-1$ away from
the attractor near the outer edge of both curves. This genuinely
non-adiabatic but super-Eddington regime is what motivates the Eddington
limiter model introduced below.

Left untreated, this super-Eddington, non-convective layer is problematic. With $\Gamma\to\Gamma_\star>1$, eq.~\eqref{eq:dbeta_dlnP} shows that $d\beta/d\ln P=1-\Gamma-\beta$ is dominated by $1-\Gamma$ (as $\beta\ll \Gamma-1$ in our solutions), so a small further increase in $\ln P$ drives a large fractional change in
$\beta$ and hence a sharp density spike, growing by orders of
magnitude over a narrow range in temperature or pressure. Such a spike is
known to be dynamically unstable to multi-dimensional perturbations
\citep{2003ApJ...596..509B}, and local radiation-hydrodynamic simulations of
massive-star envelopes at the Fe-opacity bump \citep{2015ApJ...813...74J}
show explicitly the outcome of this instability: it saturates into strong,
near-sonic turbulence whose large density fluctuations open low-opacity
channels for radiation to escape, lowering the effective (mass-weighted)
opacity and pulling the effective Eddington factor back toward unity rather
than letting $\beta$ spike indefinitely.

We do not attempt to reproduce this 3D turbulent regulation
from first principles. Instead, guided by its qualitative outcome, we adopt
in \S\ref{sec:gamma_eff_limiter} a 1D effective-opacity
(``porosity'') limiter that caps $\Gamma$ near unity wherever MLT predicts
convective suppression, limiting (but not eliminating) the density inversion while
remaining agnostic about the detailed turbulent structure. As
Fig.~\ref{fig:limiter_envelope} shows, the MLT convective
velocity $v_c$ rises close to the sound speed $c_s$ even in
this breakdown layer for our converged solutions, consistent
with the subsonic-breakdown criterion of Appendix~\ref{app:mlt}.

\subsection{Effective Eddington limiter}
\label{sec:gamma_eff_limiter}

We absorb the effect of 3D turbulent regulation in the super-Eddington radiative layers into an effective 1D model on the Eddington factor,
\begin{equation}
    \Gamma_{\rm eff}
    =
    1+
    \frac{\Gamma-1}
    {1+(\Gamma-1)/(Q_{\rm LIM}\beta)}
    \quad
    (\Gamma>1),
\label{eq:Gamma_eff_def}
\end{equation}
applied in place of $\Gamma$ in the envelope equations (\S\ref{sec:envelope_equations}) whenever MLT predicts $\Gamma>1$; for
$\Gamma\leq1$, $\Gamma_{\rm eff}=\Gamma$ identically. Eq.
\eqref{eq:Gamma_eff_def} has the following limits: as $\Gamma\to1^+$, $\Gamma_{\rm eff}\to\Gamma$
continuously; and when $\Gamma-1\gg Q_{\rm LIM}\beta$, $\Gamma_{\rm eff}\to1+Q_{\rm LIM}\beta$. Capping the effective Eddington factor excess $\Gamma_{\rm eff}-1$ proportional to $\beta$ itself, rather than at a nearly fixed value in the limit of $\Gamma-1\gg \beta$, is what prevents the runaway growth in $\beta$ according to $\d \beta/\d\ln P = 1-\Gamma-\beta$ (eq.~\ref{eq:dbeta_dlnP}). Nevertheless, this functional form is a prescription for the porosity effect, mimicking how the low-opacity channels of the turbulent state allows radiation to escape.

We treat $Q_{\rm LIM}$ as a model parameter. The local simulations of
\citet{2015ApJ...813...74J} give $\Gamma-1\simeq0.1$ and $\beta\simeq0.1$
based on the volume-averaged opacity in their most strongly
super-Eddington (\texttt{StarTop}) case, together with an effective,
mass-weighted $\Gamma_{\rm eff}-1\simeq0.05$ once the turbulent density
fluctuations are accounted for. In their model, $\Gamma_{\rm eff}-1$ is indeed suppressed to a value somewhat below $\beta$. Given the order-of-magnitude nature of this calibration with a single simulated case (where $\Gamma_{\rm eff}-1$ itself is uncertain at the order unity level), we consider two limiter values of $Q_{\rm LIM}=0.1$ and $1$ in this work. Fig.~\ref{fig:limiter_envelope} compares the results of the two directly, and \S\ref{sec:model_results} compares their effect on the full grid of converged solutions. We find that the photospheric gas conditions (e.g., $\beta_{\rm ph}$, $\rho_{\rm ph}$) are strongly affected by our choices of $Q_{\rm LIM}$, but the effective temperatures of our solutions are only weakly affected, at the level $\lesssim 10\%$.

\subsection{Envelope equations}
\label{sec:envelope_equations}

We integrate the outer envelope, from $r_{\rm base}$ out to the photosphere,
using the same $\ln P$-coordinate system as in \S\ref{sec:deep_interior} for the deep radiative layer,
\begin{equation}
    \frac{dr}{d\ln P}
    =
    -\frac{k_{\rm B}Tr^{2}}
    {GM\mu m_{\rm p}\beta},
\label{eq:outer_envelope_dr_dlnP}
\end{equation}
\begin{equation}
    \frac{d\beta}{d\ln P}
    =
    1-\Gamma_{\rm eff}-\beta,
\label{eq:outer_envelope_dbeta_dlnP}
\end{equation}
with two substitutions relative to \S\ref{sec:deep_interior}: the constant
$\kappa_{\rm T}$ and $\mu$ are replaced by the self-consistent
$\kappa_{\rm R}(T,\beta,Z)$ and $\mu(T,\beta)$ of \S\ref{sec:opacity}, and
the purely radiative $\Gamma$ is replaced by its MLT-derived,
limiter-capped counterpart $\Gamma_{\rm eff}$ of \S\ref{sec:gamma_eff_limiter}. At each step, $(T,\rho)$
follow algebraically from the local $(P,\beta)$, which in turn gives $\kappa_{\rm R}$, the local
Eddington luminosity (eq.~\ref{eq:Ledd_kappaR}), and hence $\Gamma$ and
$\Gamma_{\rm eff}$.

Integrated outward from $r_{\rm base}$, the solution passes through a radiative layer up to the transition line, and then switches to a convective zone driven by the Fe-opacity bump, with a $\Gamma\approx1$ attractor regulating the growth of $\beta$, and finally reaches the hydrogen/helium recombination zone leading to a cool photosphere.


\subsection{Two-point shooting}
\label{sec:two_point_shooting}

The envelope integration starts at $r_{\rm base}$ with $\beta(r_{\rm base})$ perturbed slightly away from its base value $\beta_{\rm base,0}\approx 1-\Gamma_{\rm base}$, which is obtained for a hypothetical star with a constant opacity $\kappa_{\rm T}$ (\S\ref{sec:deep_interior}). Realistic opacity leads to an inflated outer envelope, which would perturb the gas conditions at $r_{\rm base}$ (defined as where $T= T_{\rm base}=10^6\rm\, K$). Thus, we take
\begin{equation}\label{eq:f_beta_perturbation}
    \beta(r_{\rm base})=\beta_{\rm base,0}(1-f_{\beta}),
\end{equation}
with the fractional perturbation $f_{\beta}$ the free shooting parameter that is determined by the photospheric boundary condition (\S\ref{sec:photospheric_closure}). Shooting directly
from $r_{\rm base}$ to the photosphere in a single pass, however, is
numerically intractable: the trajectory bifurcates so sharply that
changing $f_{\beta}$ by one part in $10^{15}$ swings the solution between unphysical branches, so it is difficult to use a single root-find method to determine which $f_{\beta}$ satisfies the photospheric boundary condition.

We overcome this numerical limitation by anchoring the shooting on the inner transition line
(\S\ref{sec:transition_line}) in
two stages. First, for a grid of trial $f_{\beta}$, we integrate outward
from $r_{\rm base}$ until the trajectory crosses the inner transition
line,
\begin{equation}
    1-\Gamma_{\star}(T_{\rm mid},\beta_{\rm mid})
    -\beta_{\rm mid}=0,
\label{eq:inner_transition_condition}
\end{equation}
recording the crossing point $(r_{\rm mid},T_{\rm mid},\beta_{\rm mid})$; this defines a map $f_{\beta}\to\beta_{\rm mid}$. Empirically, the crossing
radius $r_{\rm mid}(\beta_{\rm mid})$ is a smooth,
monotonically decreasing function of $\beta_{\rm mid}$ over the trial
grid. Tabulating and interpolating this smooth relation lets us read off
$(r_{\rm mid},T_{\rm mid})$ for any desired $\beta_{\rm mid}$
without re-integrating the numerically fragile inner segment for that
specific value. Second, starting from $(r_{\rm mid},T_{\rm mid},
\beta_{\rm mid})$, we integrate the same envelope equations
(\S\ref{sec:envelope_equations}) outward to the photosphere and evaluate
the photospheric-closure residual (\S\ref{sec:photospheric_closure}); this
outer segment can itself bifurcate, so we again scan over $\beta_{\rm mid}$
within the tabulated range from the first stage. This decouples the fragile inner segment from the outer segment; the full procedure underlying both scans is given in Appendix~\ref{app:shooting}.

The resulting $f_{\beta}$ is remarkably uniform across the grid: solving
for the value that satisfies the photospheric closure gives
$f_{\beta}\approx0.09$--$0.11$ for every converged
$(M,Z)$ combination, essentially independent of mass and metallicity. A
$\sim10\%$ perturbation to $\beta(r_{\rm base})$ is, in principle, large
enough to ask whether it feeds back on the deep-interior structure of
\S\ref{sec:deep_interior}, which is solved with the unperturbed boundary
condition $\beta_{\rm base,0}$ (i.e., $f_{\beta}=0$) and then held fixed
throughout the envelope calculation. We tested this directly by
re-solving the core with the perturbed boundary condition
$\beta(r_{\rm base})=\beta_{\rm base,0}(1-f_{\beta})$ at $f_{\beta}=0.1$
and comparing to the unperturbed sequence at fixed stellar mass: $T_c$ and
$\rho_c$ shift by a fractional amount of $\sim10^{-6}$. The base radius
$r_{\rm base}$ itself, where the perturbation is directly applied, is the most sensitive quantity, but even so shifts by a fraction of at most $\sim10^{-3}$ across the grid. The corresponding change in
GR-instability threshold mass $M_{\rm th}$ (\S\ref{sec:gr_instability_mass}) is also negligible. We therefore treat the core sequences of \S\ref{sec:deep_interior} as independent of $f_{\beta}$
throughout this work, without iterating the core and envelope solutions
to convergence in $f_{\beta}$.

\subsection{Photospheric closure}
\label{sec:photospheric_closure}

We define the photosphere radius $r_{\rm ph}$ as where the local gas
temperature matches the effective temperature implied by $L_\star$,
\begin{equation}
    T_{\rm ph}
    = T_{\rm eff}=
    \left(
    \frac{L_{\star}}
    {4\pi r_{\rm ph}^{2}\sigma_{\rm SB}}
    \right)^{1/4},
\label{eq:Tph_def}
\end{equation}
equivalently the depth at which the exterior optical depth reaches
$\tau_{\rm ph}=2/3$. Near the photosphere $\Gamma_\star<1$, so the layer is
hydrostatic and radiative diffusion dominates transport ($\Gamma\approx
\Gamma_\star$); together with plane-parallel Eddington-approximation
$T^4(\tau) = \tfrac34T_{\rm ph}^4(\tau+\tfrac23)$. Using
$dP/d\tau=GM/(\kappa_{\rm R}r_{\rm ph}^2)$, we find
\begin{equation}
    \tau(P) = \frac{r_{\rm ph}^2}{GM}\int_{P_0}^{P}\kappa_{\rm R}\big(\beta(P'),P'\big)\,dP',
\label{eq:tau_of_P}
\end{equation}
where $P_0\equiv P(\tau=0)$. Adopting $P_{\rm gas}(\tau=0)=0$ (assuming negligible wind), we find $P_0= P_{\rm rad}(\tau=0) = aT_{\rm ph}^4/6$. Let us denote the pressure at $\tau_{\rm ph}=2/3$ as $P_{\rm ph} = aT_{\rm ph}^4/[3(1-\beta_{\rm ph})]$. Evaluating
eq.~\eqref{eq:tau_of_P} at $\tau_{\rm ph}=2/3$ gives the following closure
condition
\begin{equation}
    \frac{2}{3}\frac{GM}{r_{\rm ph}^2}
    =
    \int_{P_0}^{P_{\rm ph}}\kappa_{\rm R}\big(\beta(P),P\big)\,dP,
\label{eq:photospheric_closure_exact}
\end{equation}
which requires the full $\beta(P)$ profile between $P_0$ and $P_{\rm ph}$
and is seemingly not local to $(\beta_{\rm ph}, T_{\rm ph})$ alone. 

However, since the $\beta(P)$ profile is locally determined by eq.~\eqref{eq:outer_envelope_dbeta_dlnP}, this allows us to obtain a local closure relation $\beta_{\rm ph}(T_{\rm ph})$ that is independent of the detailed profile leading up to $r_{\rm ph}$. The numerical solver for the photospheric closure is presented in Appendix \ref{app:boundary_curves}.

To gain a better analytical understanding of the photospheric closure, we Taylor-expand $\kappa_{\rm R}$ to first order in $\ln P$ about the photosphere along the local trajectory, $\kappa_{\rm R}(P)\approx\kappa_{\rm ph}\left[1+\xi (\ln P-\ln P_{\rm ph})
\right]$, where the local logarithmic slope $\xi \equiv(d\ln\kappa_{\rm R}/d\ln P)|_{\rm ph}= (d\ln\kappa_{\rm R}/d\ln\beta)|_{\rm ph}\,(1-\Gamma_{\star,\rm ph}- \beta_{\rm ph})/\beta_{\rm ph}$ follows from the chain rule and
eq.~\eqref{eq:outer_envelope_dbeta_dlnP} evaluated at the photosphere
(with $\Gamma_{\rm eff}\to\Gamma_\star$, since $\Gamma_\star<1$ there).
Carrying out the resulting elementary integral in eq.~\eqref{eq:photospheric_closure_exact} gives an approximate local closure
relation
\begin{multline}
    \left.\frac{d\ln\kappa_{\rm R}}{d\ln\beta}\right|_{\rm ph}
    \frac{1-\Gamma_{\star,\rm ph}-\beta_{\rm ph}}{\beta_{\rm ph}}
    \Big[1+\beta_{\rm ph}+ \\
    (1-\beta_{\rm ph})\ln\!\left(\frac{1-\beta_{\rm ph}}{2}\right)\Big]
    \approx
    (1+\beta_{\rm ph})-\frac{1-\beta_{\rm ph}}{\Gamma_{\star,\rm ph}},
\label{eq:photospheric_closure_local}
\end{multline}
which depends only on local photospheric quantities and can be solved for $\beta_{\rm ph}(T_{\rm ph})$.

It turns out that the photospheric closure curve lies close to the outer transition line
(\S\ref{sec:transition_line}),
\begin{equation}
    1-\Gamma_{\star}-\beta=0.
\label{eq:outer_transition_condition}
\end{equation}
For $\beta_{\rm ph}\ll1$ and
$(d\ln\kappa_{\rm R}/d\ln\beta)|_{\rm ph}\sim\mathcal{O}(1)$, the
left-hand side of eq.~\eqref{eq:photospheric_closure_local} is of order
$(1-\Gamma_{\star,\rm ph}-\beta_{\rm ph})/\beta_{\rm ph}$, which diverges
as $\beta_{\rm ph}\to0$ unless simultaneously
$1-\Gamma_{\star,\rm ph}-\beta_{\rm ph}\to0$, while the right-hand side
stays of order unity throughout. This argument is only approximate: the expansion parameter
$\ln(P_{\rm ph}/P_0)=\ln2\approx0.7$ is not particularly small, and the
approximation degrades wherever $(d\ln\kappa_{\rm R}/d\ln\beta)|_{\rm ph}$
is itself very small.

\begin{figure*}[!htb]
    \centering
    \includegraphics[width=0.85\textwidth]{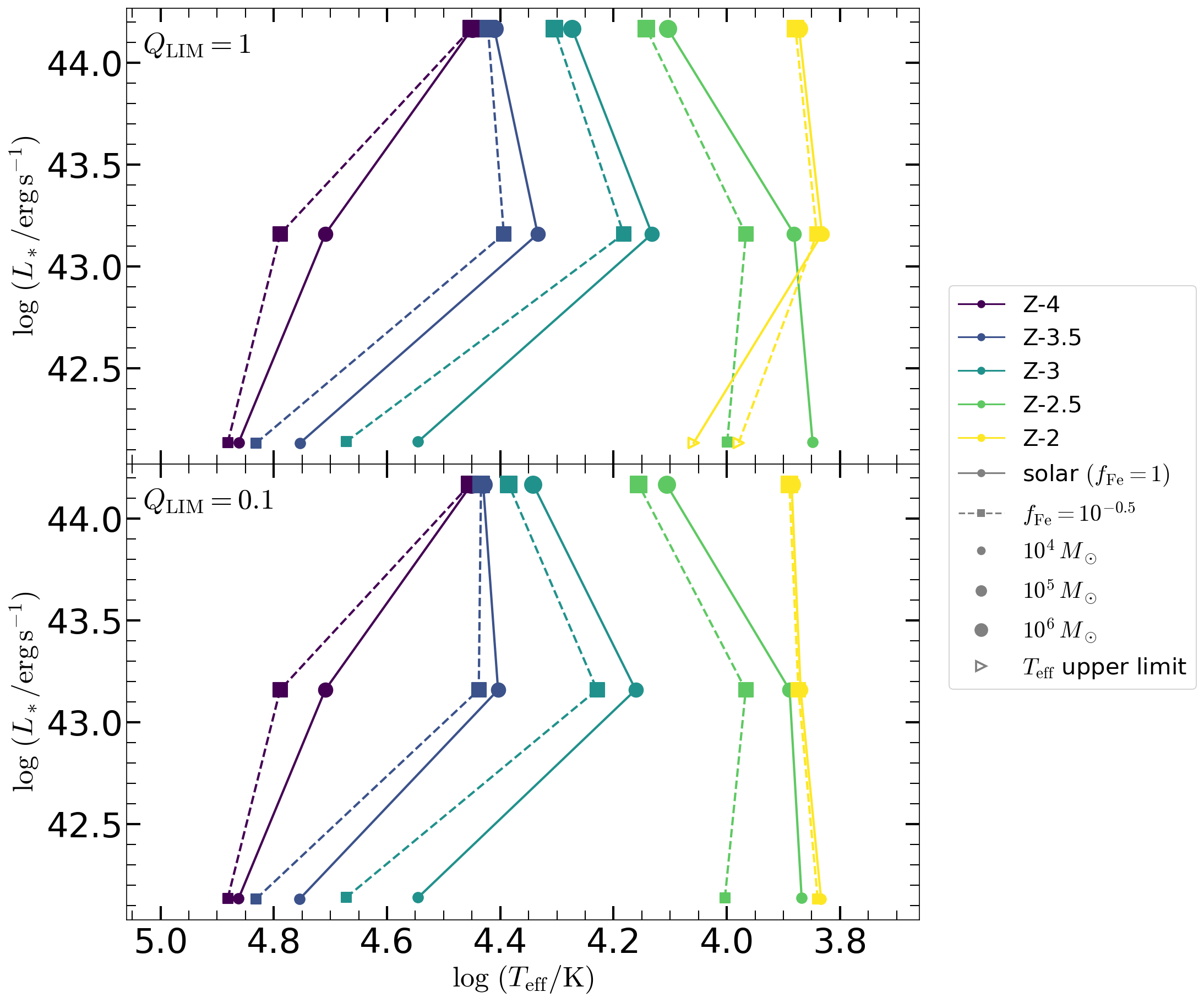}
    \caption{HR diagram of all the solutions.
    Within each $(Z,\, f_{\rm Fe})$ sequence, $L_{\star}$ tracks the Eddington luminosity ($L_{\star}\approx L_{\rm Edd,T}\propto M$) and $T_{\rm eff}$ depends non-monotonically on mass. Increasing metallicity shifts models from hot, compact solutions at low $Z$ to progressively cooler, more inflated ones, with the coolest models ($Z\sim10^{-2}$) reaching $T_{\rm eff}\sim7000\,{\rm K}$ and entering the LRD-like
    temperature range. Reducing the Fe-group abundance to
    $f_{\rm Fe}=10^{-0.5}$ (dashed) shifts each sequence to modestly hotter
    $T_{\rm eff}$ at fixed $Z$ without affecting the high-metallicity cool
    branch. The two open triangles mark the cases
    for which only an upper limit on $T_{\rm eff}$ was obtained because the solution requires $\beta_{\rm ph}$ beyond the tabulated opacity domain ($\beta_{\rm max}=0.9$).
    Comparing the top and bottom panels shows
    that $T_{\rm eff}$ is robust (to within $\lesssim 10\%$) to the $Q_{\rm LIM}=1$ vs.\ $0.1$ choice.
    }
    \label{fig:sms_hr_combined}
\end{figure*}


A useful check of our photospheric closure is the special case of constant opacity, for which $(d\ln\kappa_{\rm R}/d\ln\beta)=0$ and eq.~\eqref{eq:photospheric_closure_local} directly recovers eq.~\eqref{eq:beta_ph_const_kap}.

What ultimately permits a bound photosphere to form at all is hydrogen recombination: it is the associated drop in $\kappa_{\rm R}$ that lets $\Gamma_\star$ fall below unity and the envelope settle into hydrostatic equilibrium close to the photosphere, rather than continuing to inflate indefinitely.






\section{Metallicity-Dependent Solutions}
\label{sec:model_results}

\subsection{Model grid}

We solve the full envelope problem of \S\ref{sec:hse_envelopes} over a grid
of three masses and five metallicities,
\begin{equation} \label{eq:grid_masses_metallicities}
\begin{split}
    M/M_{\odot}&\in\{10^{4},10^{5},10^{6}\},\\
    \log\,Z&\in\{-4,-3.5,-3,-2.5,-2\},
\end{split}
\end{equation}
each at both the solar and reduced-Fe ($f_{\rm Fe}=10^{-0.5}$) abundance patterns and both limiter values $Q_{\rm LIM}=1$ and $0.1$, for $60$ models in total.

The Hertzsprung-Russell (HR) diagram containing all cases is shown in Fig.~\ref{fig:sms_hr_combined}. Most of the grid converges to a root of the shooting procedure (\S\ref{sec:two_point_shooting}). Two cases, both at
$Q_{\rm LIM}=1$, $Z=10^{-2}$, $M=10^{4}\,M_{\odot}$ (solar and reduced-Fe),
do not reach a photosphere at all and the solution demands $\beta_{\rm ph}$
beyond the tabulated opacity domain's $\beta_{\rm max}=0.9$. Thus, only upper limits on $T_{\rm eff}$ are
available for these two cases, marked by open triangles in Fig.~\ref{fig:sms_hr_combined}.

\subsection{Full stellar profiles}

\begin{figure}
    \centering
    \includegraphics[width=\linewidth]{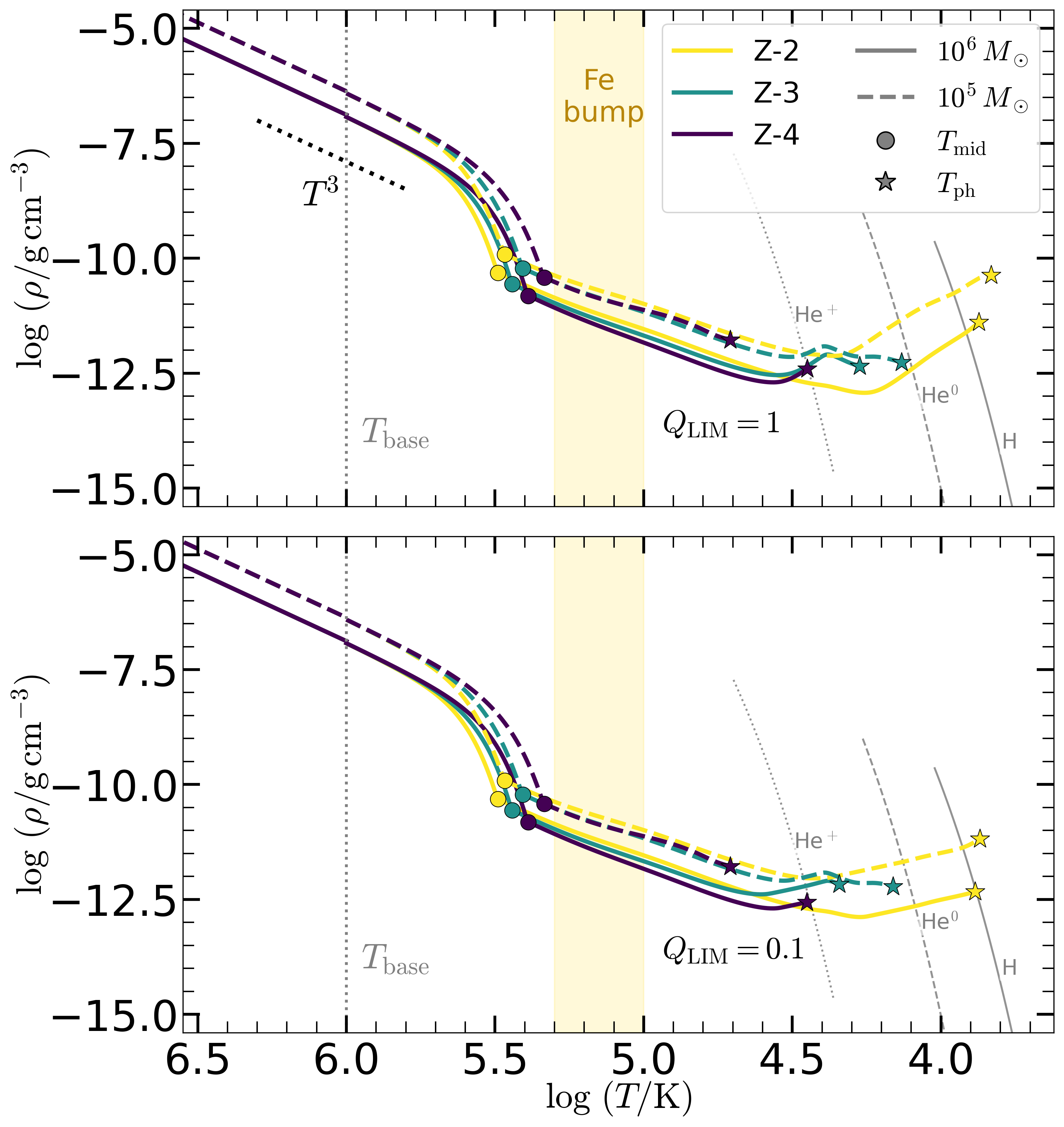}
    \caption{Structure of converged SMS envelope solutions, from the radiative layer (upper left) through $T_{\rm base}=10^{6}\,{\rm K}$ (vertical dotted line) into the Fe-bump-driven convective envelope and out to the photosphere (lower right), for $Z=10^{-2},10^{-3},10^{-4}$ (color) at $M=10^{6}\,M_{\odot}$ (solid) and $10^{5}\,M_{\odot}$ (dashed). Circles mark the inner transition-line crossing $(T_{\rm mid},\beta_{\rm mid})$; stars mark the photospheric point $T_{\rm ph}$. The rapid density drop between $T_{\rm base}$ and the transition line $T_{\rm mid}$ makes it possible for the development of a low-density convective envelope that extends across the Fe-opacity bump all the way to the recombination zone. The thin gray lines show the three main recombination fronts for He$^+$ (dotted), He$^0$ (dashed), and H (solid), each for an ionization fraction of $y=0.9$. The density profile flattens and turns up near the photosphere because the surface layers become radiative and super-Eddington (rapid photon diffusion suppresses convection) --- this region is modeled using an effective Eddington limiter $Q_{\rm LIM}=1$ (upper panel) and $0.1$ (lower panel). The density inversion is weaker for $Q_{\rm LIM}=0.1$ (lower Eddington factor).
    }
    \label{fig:sms_profile}
\end{figure}

Fig.~\ref{fig:sms_profile} shows full $(T,\rho)$ profiles from the
convective core through $T_{\rm base}$ and out to the photosphere, for
representative $Z=10^{-2},10^{-3},10^{-4}$ cases at $M=10^{5},10^{6}\,M_{\odot}$. The dense core from the center through $r_{\rm base}$ is described by the constant-opacity convective-radiative solution with a roughly constant $\beta$ or $\rho\propto T^3$. As the system transitions to an envelope with variable $\kappa_{\rm R}(T,\beta,Z)$, the density rapidly drops between $r_{\rm base}$ and $r_{\rm mid}$ due to decreasing $\beta$ at the rise of the Fe-opacity bump. After the inner transition line ($r_{\rm mid}$) is crossed, convection sets in and gives rise to a power-law $\rho\propto T^{3}$ profile again. The power-law profile holds until the recombination zone where the surface layer returns to radiative, super-Eddington transport, and $\rho$ turns back up again when approaching the photosphere.


Across all masses considered, the degree of inflation, $r_{\rm ph}/r_{\rm base}$, grows sharply with
metallicity as the Fe-opacity bump gets stronger:
from $r_{\rm ph}/r_{\rm base} \sim \mathcal{O}(1\text{--}10)$ at $Z=10^{-4}$ to $\gtrsim10^{2}$ at
$Z=10^{-2}$. Despite spanning up to two orders of magnitude in radius, this entire envelope contains negligible mass: integrating from $r_{\rm base}$ to $r_{\rm ph}$ for the most inflated case in the grid
($Z=10^{-2}$, $M=10^{6}\,M_{\odot}$) gives $\Delta m_{\rm env}/M\sim\mc{O}(10^{-4})$, and correspondingly smaller fractions for the less-inflated cases. This confirms the assumption of
constant $M=m(r_{\rm base})$ adopted throughout the envelope.

This $Z$-dependent inflation directly sets the photospheric temperature
across the grid (Fig.~\ref{fig:sms_hr_combined}): low-metallicity models
remain relatively compact and hot, while higher $Z$ increases the strength of the Fe-opacity bump and produces progressively cooler, more inflated envelopes. Our models reach LRD-like effective temperatures for $Z\gtrsim 10^{-2}$ where the
$10^{5}$--$10^{6}\,M_{\odot}$ models reach
\begin{equation}
    T_{\rm eff}\sim7000\,\mathrm{K},
    \quad
    {r_{\rm ph}/r_{\rm base}}\sim10^{2}.
\label{eq:Teff_rph_range}
\end{equation}
As such cool effective temperatures are the direct result of hydrogen recombination, we expect this to be the case for even higher metallicities.

\subsection{Dependence on Fe abundance}

In the high-redshift universe, e.g., $z\sim5$ where the inferred LRD
formation rate peaks, chemical enrichment may still be dominated by
core-collapse supernovae, which preferentially produce $\alpha$-elements
with a comparatively low iron yield relative to the delayed contribution
from Type Ia supernovae \citep[e.g.,][]{2013ARA&A..51..457N, 2017ApJ...848...25M}. Observational evidence for such enrichment includes the super-solar
$\alpha$/Fe (i.e., Fe-poor) abundance ratio measured in star-forming
galaxies at $z\simeq3.4$ \citep{2021MNRAS.505..903C}. However, abundance
patterns at higher redshifts may be diverse: \citet{2026MNRAS.547ag123I}
inferred super-solar Fe/O in the $z=4$--7 galaxy subsamples. Motivated by the possibility of Fe-poor enrichment, while recognizing this observational uncertainty, we also consider models with
the Fe-group abundance reduced by $f_{\rm Fe}=10^{-0.5}$ relative to the
solar pattern.

Our models with the solar Fe-group pattern (solid) and a reduced pattern $f_{\rm Fe}=10^{-0.5}$ (dashed) are compared on the HR diagram in Fig. \ref{fig:sms_hr_combined}. Because $f_{\rm Fe}$ rescales only the Fe-opacity bump amplitude (\S\ref{sec:opacity}) and leaves the CNO catalyst abundance roughly unchanged, our deep-interior structure assumes a fixed $f_{\rm CNO}=0.7$. The effect of reduced Fe abundance is confined entirely to how strongly the envelope inflates. We find that this shift is modest and metallicity dependent: at $Z=10^{-2}$, where the bump is strongest and models already sit on the cool branch, reducing $f_{\rm Fe}$ shifts $T_{\rm eff}$ by at most a few percent, whereas at $Z\lesssim10^{-3}$, where the bump is weaker and models are more compact and the same Fe abundance reduction can shift $T_{\rm eff}$ by about 10\%.

The high-metallicity cool branch itself is therefore robust to this abundance-pattern uncertainty: even at reduced
$f_{\rm Fe}=10^{-0.5}$, the Fe bump at $Z=10^{-2}$ remains strong enough to drive
substantial inflation, so the qualitative conclusion that LRD-like
temperatures require high metallicity does not depend sensitively on the
exact Fe-group abundance.

This weak sensitivity to $f_{\rm Fe}$ can be understood from the shape of
$\kappa_{\rm metal}(T,\beta,Z)$ itself. The condition for a locally
super-Eddington, convective layer is roughly $\Gamma_\star(\beta,T)>1-\beta$
(eq.~\ref{eq:convective_condition}). Near the peak of the Fe-group opacity bump
($T_0\sim1.5\times10^5$--$2\times10^5\,{\rm K}$), reducing $f_{\rm Fe}$ lowers the
bump amplitude nearly in proportion, weakening the local super-Eddington
excess there. However, once the local temperature drops below
$\sim10^5\,{\rm K}$, the Fe-group bump no longer contributes, and the
opacity is instead set by the plateau arising from the non-Fe metals, which we find to be nearly 
independent of $f_{\rm Fe}$ (see Fig.~\ref{fig:kap_metal_fem} in Appendix \ref{app:metal_opacity}). Consequently, whether the inflated convective envelope survives all the way down to the
H-recombination temperatures -- and hence whether the star lands on
the cool, inflated branch -- is controlled mainly by whether the
plateau opacity keeps $\Gamma_\star>1-\beta$ satisfied, a condition set
essentially by $Z$ and $\beta$ rather than by $f_{\rm Fe}$. Reducing
$f_{\rm Fe}$ can therefore only shrink the margin locally at the bump,
without altering the plateau-controlled outcome at lower temperatures,
which is why the resulting shift in $T_{\rm eff}$ is modest.

Independent support for this plateau-controlled picture comes from
comparing across mass at fixed $Z$. Since $\beta\propto M^{-1/2}$
(\S\ref{sec:deep_interior}), lower-mass models have systematically smaller $\beta$ and hence a larger gap of $1-\beta$ for the plateau opacity to overcome. For $M=10^3\,M_\odot$ at $Z=10^{-2}$, the
Fe-bump peak is still comfortably super-Eddington, yet the envelope
fails to inflate down to H-recombination temperatures (see Fig. \ref{fig:sms_hr_M1e3_check} later), because the smaller $\beta$ at this mass raises the threshold that the plateau opacity must clear. This confirms that
it is the plateau opacity relative to $\beta$ that determines whether inflation reaches the
H-recombination region, consistent with the weak, only mildly
metallicity-dependent effect of $f_{\rm Fe}$ found here.







\section{GR Collapse and Remnant Demographics}
\label{sec:smbh_connection}

\subsection{GR-instability mass threshold}
\label{sec:gr_instability_mass}

The core of a radiation-pressure-dominated SMS has an adiabatic index only slightly larger than $4/3$ \citep{1984ApJ...280..825B}:
\begin{equation}\label{eq:adiabatic_index}
    \gamma_{\rm ad}(\beta) - {4/3}= \beta/6 + {\beta^2/48} + \mc{O}(\beta^3).
\end{equation}
In Newtonian gravity, an exact $\gamma_{\rm ad}=4/3$ configuration is marginally unstable to radial perturbations. The small gas-pressure contribution raises the adiabatic index and stabilizes the star, whereas the first post-Newtonian correction to gravity has the opposite effect and eventually drives a radial instability
\citep{1964ApJ...140..417C,1966ApJ...144..180F, 1986ApJ...307..675F}. We characterize the magnitude of the
relativistic correction by
\begin{equation}
    \epsilon_{\rm GR}
    \equiv 2.6324\,P_c/(\rho_c c^2),
\label{eq:epsilonGR_def}
\end{equation}
where the numerical prefactor comes from \citet{1964ApJ...140..417C}; $\epsilon_{\rm GR}$ is of the order $GM/(Rc^2)$ and measures the importance of the relativistic terms in the equation of hydrostatic equilibrium. We also define the rotational energy parameter
\begin{equation}
    \eta
    \equiv
    T_{\rm rot}/|W|,
\label{eq:eta_def}
\end{equation}
where $T_{\rm rot}$ and $W$ are the rotational kinetic and
gravitational binding energies, respectively.

To first order in the gas-pressure, relativistic, and rotational
corrections, the marginal-stability condition can be written as (see Appendix \ref{app:gr_instability})
\begin{equation}
    \epsilon_{\rm GR}
    \approx
    \frac{\beta_c}{6}
    +
    \frac{2}{3}\eta.
\label{eq:epsilonGR_marginal_stability}
\end{equation}
For the models considered in this work, \(\beta_c,\epsilon_{\rm GR},\eta\lesssim10^{-2}\), so the second-order terms are generally \(\lesssim10^{-4}\). The star is stable when $\epsilon_{\rm GR}$ lies below the right-hand
side of eq.~\eqref{eq:epsilonGR_marginal_stability} and becomes
unstable when it exceeds it. The gas-pressure term therefore
stabilizes the star, while rotation supplies additional support and
allows the star to reach a larger compactness before collapse
\citep{1999ApJ...526..941B,2016ApJ...818..157S,
2025ApJ...978...58S}. We consider three representative rotation
rates: $\eta=0$, $\eta=\eta_{\rm max}/2$, and
$\eta=\eta_{\rm max}$, where
$\eta_{\rm max}\simeq0.009$ is the approximate mass-shedding limit
for a rigidly rotating, radiation-dominated SMS core
\citep{1999ApJ...526..941B}. This value should not be interpreted as
a universal upper limit for differentially rotating stars.

\begin{figure}
    \centering
    \includegraphics[width=\linewidth]
    {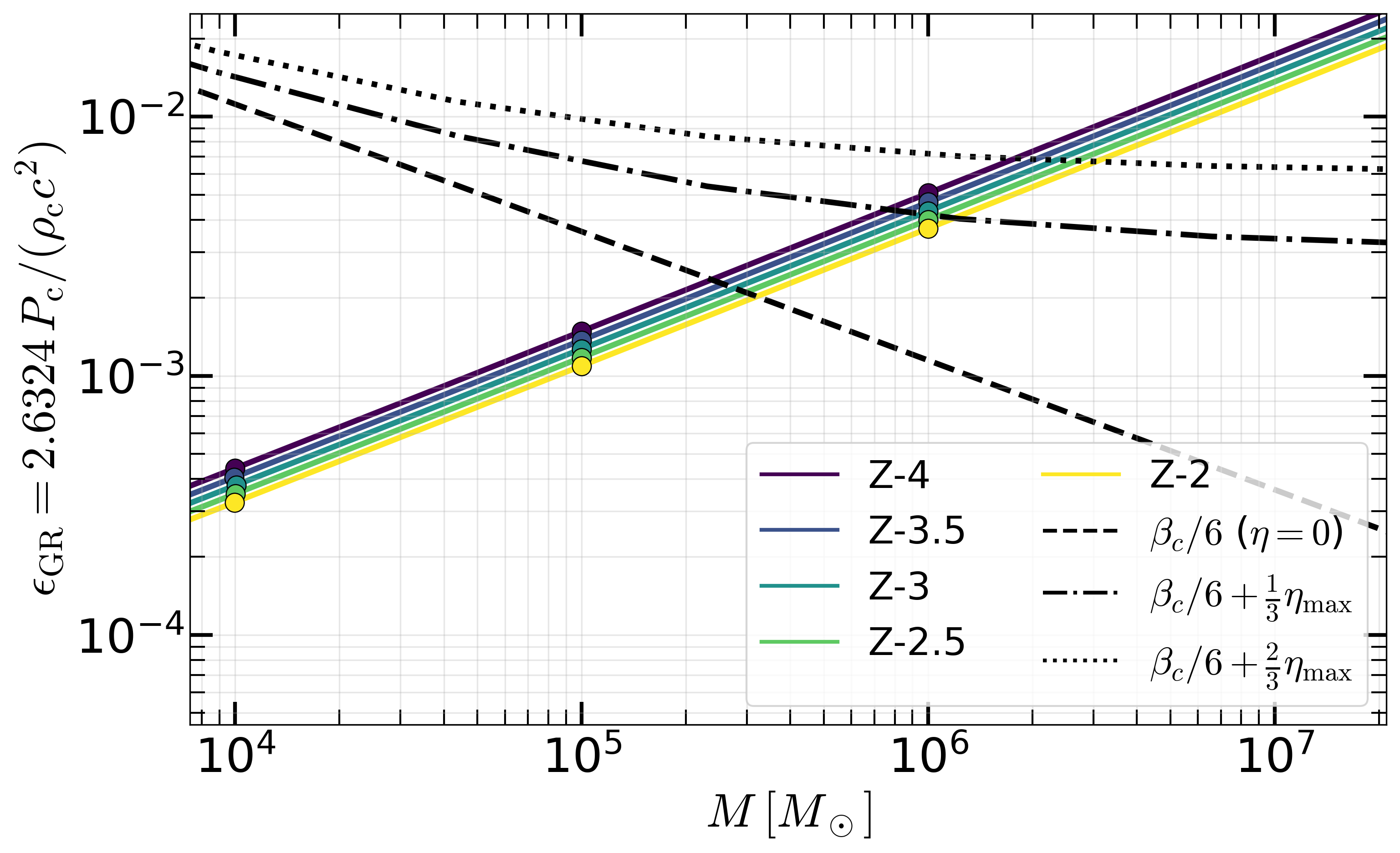}
    \caption{The relativistic instability parameter
    $\epsilon_{\rm GR}$ (eq.~\ref{eq:epsilonGR_def}) as a function of mass for
    the $X=0.7$ base-model sequences of
    Fig.~\ref{fig:sms_base}. Black curves show the approximately
    metallicity-independent marginal-stability boundary
    $\beta_c/6+\tfrac{2}{3}\eta$ for three rotation rates:
    $\eta=0$ (dashed), $\eta=\eta_{\rm max}/2$ (dash-dotted), and
    $\eta=\eta_{\rm max}$ (dotted). Each
    intersection of a colored model sequence with a black stability
    curve defines the threshold mass $M_{\rm th}(Z,\eta)$. Higher
    metallicity and faster rotation both raise $M_{\rm th}$, although
    the rotational dependence is considerably stronger.}
    \label{fig:sms_gr_instability}
\end{figure}

For each composition and rotation rate, we evaluate
$\epsilon_{\rm GR}(M)$ and $\beta_c(M)$ from the deep-interior model
sequences and identify the mass at which
eq.~\eqref{eq:epsilonGR_marginal_stability} is first satisfied.
Fig.~\ref{fig:sms_gr_instability} illustrates this construction for
the unevolved composition $X=0.7$. At fixed rotation rate ($\eta$), increasing the
metallicity shifts the intersection to a slightly higher mass.
Rotation has a much larger effect: even modest rotational support
substantially increases the mass required to reach the instability.

The threshold also depends on the hydrogen abundance and the CNO
catalyst abundance. The physical dependence enters through three
quantities: the CNO-burning rate, the Thomson-scattering Eddington
luminosity, and the mean molecular weight. We adopt the reference
composition
\begin{equation}
    X_0=0.7,
    \quad
    Z_0=10^{-2},
\end{equation}
and define the normalized composition factors
\begin{equation}
    \begin{split}
    \mathcal{A}(X,Z)
    &\equiv
    \frac{Z_{\rm CNO}}{Z_{{\rm CNO},0}}
    \frac{X(1+X)}{X_0(1+X_0)}
    \approx
    \frac{Z}{10^{-2}}
    \frac{X(1+X)}{1.19},\\
    \mathcal{B}(X)
    &\equiv
    \frac{\mu^{-1}(X)}{\mu^{-1}(X_0)}
    =
    \frac{0.75+1.25X}{1.625}.
    \end{split}
\label{eq:AB_def_main}
\end{equation}
Here $Z_{\rm CNO}$ is the total C+N+O catalyst abundance. The
approximate expression for $\mathcal{A}$ assumes that the
CNO-to-total-metal mass fraction $f_{\rm CNO}$ remains fixed. Although the relative
C, N, and O abundances change during CNO-cycle burning, their total
catalyst abundance is approximately conserved. The full homology
derivation of these composition factors is given in
Appendix~\ref{app:gr_instability}.

The numerical model sequences and the analytical composition
scaling can be summarized by
\begin{equation}
    M_{\rm th}(X,Z,\eta)
    \approx
    \mathcal{A}^{p}
    \mathcal{B}^{q} M_0,
\label{eq:Mth_composition_fit}
\end{equation}
where the fitted coefficients are listed in
Table~\ref{tab:Mth_fit}.

\begin{table}
    \centering
    \caption{Coefficients in the composition-dependent
    GR-instability threshold,
    eq.~\eqref{eq:Mth_composition_fit}.}
    \label{tab:Mth_fit}
    \begin{tabular}{lccc}
        \hline
        Rotation rate
        & $M_0$
        & $p$
        & $q$\\
        & $(M_\odot)$ & & \\
        \hline
        $\eta=0$
        & $3.20\times10^{5}$
        & $0.065$
        & $0.967$\\
        $\eta=\eta_{\rm max}/2$
        & $1.19\times10^{6}$
        & $0.098$
        & $0.377$\\
        $\eta=\eta_{\rm max}$
        & $3.06\times10^{6}$
        & $0.116$
        & $0.170$\\
        \hline
    \end{tabular}
\end{table}

For the unevolved abundance $X=0.7$ and the metallicity interval
$10^{-4}\leq Z\leq10^{-2}$, eq.~\eqref{eq:Mth_composition_fit}
gives
$M_{\rm th}\simeq(2.4$--$3.2)\times10^{5}\,M_\odot$ for a
nonrotating SMS,
$(7.6\times10^{5}$--$1.2\times10^{6})\,M_\odot$ for
$\eta=\eta_{\rm max}/2$, and
$(1.8$--$3.1)\times10^{6}\,M_\odot$ at the
rigid-rotation mass-shedding limit\footnote{Another possible mechanism to increase the maximum mass of stable SMSs is on-going accretion \citep[e.g.,][]{2012ApJ...750...66J}, which may prevent the SMS core from settling into thermal equilibrium but is not considered in this work.}. The metallicity dependence is
weak because the high temperature sensitivity of CNO burning acts
as a thermostat. A larger CNO abundance allows the required
luminosity to be generated at a slightly lower central temperature
and compactness, thereby raising the mass at which the relativistic
correction overcomes gas-pressure support
\citep{2025ApJ...978...58S}.

\begin{figure}
    \centering
    \includegraphics[width=0.48\textwidth]
    {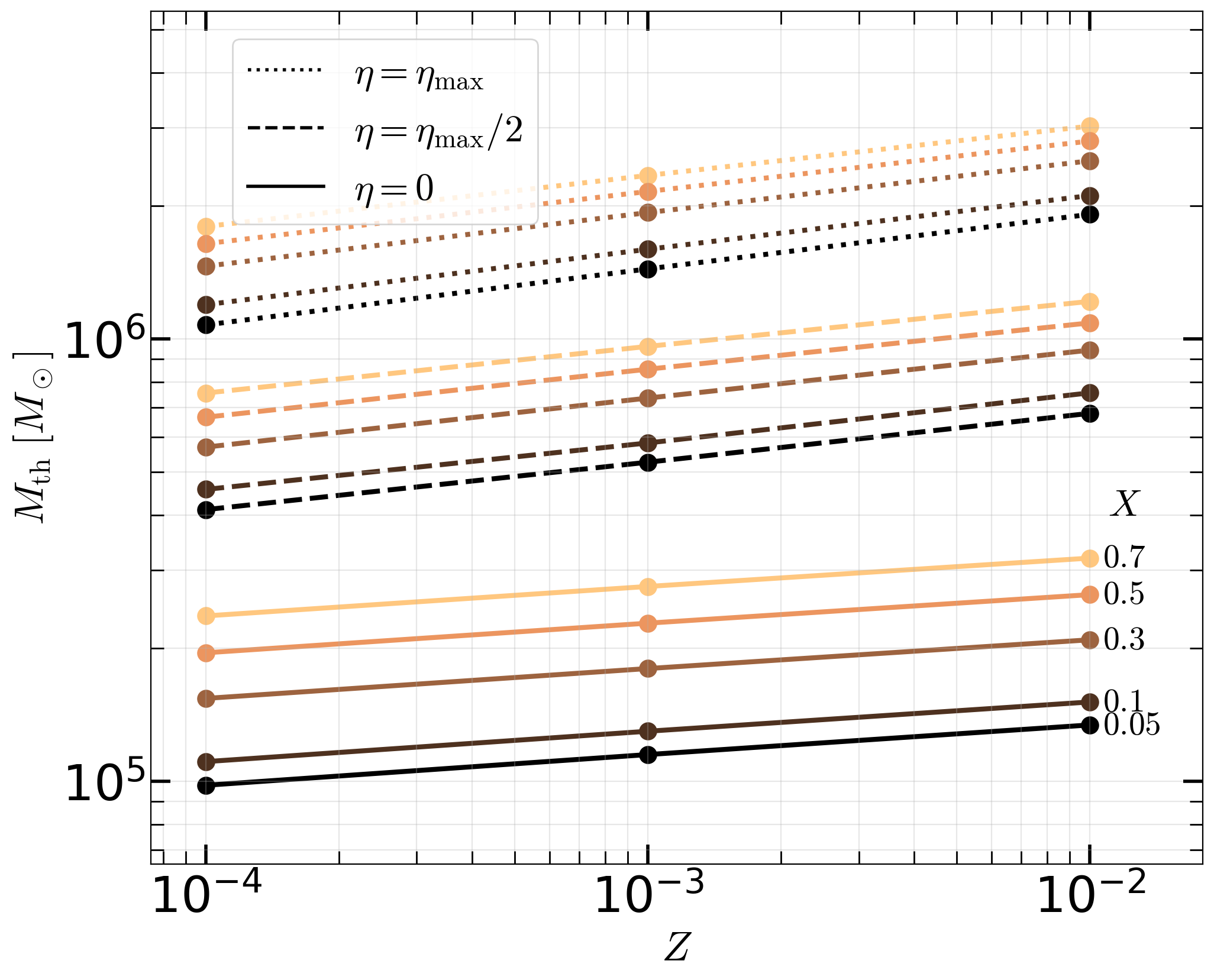}
    \caption{GR-instability threshold mass
    $M_{\rm th}(X,Z,\eta)$ obtained by solving eq. (\ref{eq:epsilonGR_marginal_stability})
    for base-model sequences computed at
    $X\in\{0.7,0.5,0.3,0.1,0.05\}$ and
    $Z\in\{10^{-4},10^{-3},10^{-2}\}$. The three line styles
    denote $\eta=0$ (solid), $\eta=\eta_{\rm max}/2$
    (dashed), and $\eta=\eta_{\rm max}$ (dotted).
    At fixed $Z$ and $\eta$, the threshold mass decreases as
    hydrogen is depleted. An initially stable SMS can therefore
    cross the instability boundary during core hydrogen burning.
    At fixed $X$ and $\eta$, the weaker upward trend with $Z$
    reflects the CNO-abundance dependence discussed in the text.}
    \label{fig:mth_xz}
\end{figure}

Hydrogen depletion has a stronger evolutionary effect. As $X$ decreases, the combined changes in the CNO-burning rate, Thomson-scattering opacity, and mean molecular weight reduce $M_{\rm th}$. Thus, an SMS that is stable at zero-age main-sequence can cross the relativistic instability boundary later in its evolution, even if its total mass and rotation rate
remain fixed. This behavior is consistent with previous evolutionary and stability calculations of nuclear-burning SMSs \citep{1984ApJ...280..825B, 1986ApJ...307..675F, 2020MNRAS.494.2236W, 2025ApJ...978...58S}. For a star of mass $M_\star$, we define the collapse abundance $X_{\rm crit}$ implicitly by
\begin{equation}
    M_\star
    =
    M_{\rm th}\!\left(X_{\rm crit},Z,\eta\right).
\label{eq:Xcrit_def}
\end{equation}
If $X_{\rm crit}>0$, the GR instability truncates the observable
SMS phase before the complete exhaustion of core hydrogen.
Fig.~\ref{fig:mth_xz} shows this evolutionary crossing over the
composition and rotation grid.

The hydrogen abundance at GR collapse also determines the duration of the observable
SMS phase. Between the initial abundance $X_0$ and the onset of the
GR instability at $X_{\rm crit}$, the nuclear energy released by
hydrogen burning is approximately
\begin{equation}
    E_{\rm nuc}
    \simeq
    \epsilon_{\rm H}\Delta X M_\star c^2,
    \quad
    \Delta X\equiv X_0-X_{\rm crit},
\label{eq:Enuc_before_GR}
\end{equation}
where $\epsilon_{\rm H}\simeq0.007$ is the mass-energy conversion
efficiency of hydrogen burning. The corresponding SMS lifetime is
therefore
\begin{equation}
    \begin{split}
    \tau_{\rm SMS}
    &\approx
    \frac{\epsilon_{\rm H}\Delta X M_\star c^2}{L_\star}
    \approx
    \frac{\epsilon_{\rm H}\Delta X\kappa_{\rm T}c}
         {4\pi G(1-\beta_c)}\\
    &\approx
    1.35\,{\rm Myr}\,
    \frac{\epsilon_{\rm H}}{0.007}
    \frac{\Delta X}{0.5}
    \frac{\kappa_{\rm T}}
               {0.34\,{\rm cm^2\,g^{-1}}}.
    \end{split}
\label{eq:tau_SMS_GR}
\end{equation}
Here we have used $L_\star\approx(1-\beta_c)L_{\rm Edd,T}\approx L_{\rm Edd,T}$. The explicit
dependence on $M_\star$ cancels because both the available nuclear
fuel and the Eddington luminosity scale linearly with stellar mass.
Thus, within the chemically homogeneous approximation, the lifetime
is controlled primarily by the amount of hydrogen consumed before
the star crosses the GR-instability boundary. For stars that remain
GR-stable throughout core hydrogen burning, one may set
$X_{\rm crit}\approx 0$, giving the full hydrogen-burning lifetime. The
resulting Myr-scale duration is consistent with detailed evolutionary
calculations of thermally relaxed SMSs
\citep{1986ApJ...307..675F,2020MNRAS.494.2236W}.

The composition scalings above assume a chemically homogeneous burning region, a fixed total CNO catalyst abundance, and a fixed rotation parameter. Mass loss changes both sides of the instability condition by reducing $M_\star$, while angular-momentum transport or rotational mass loss can change $\eta$ during the stellar lifetime. The initial spin distribution and its subsequent evolution therefore broaden both the maximum stable SMS mass and the range of hydrogen abundances at which collapse occurs. The highly inflated outer envelope is not expected to modify the threshold appreciably, because it contains little mass and contributes negligibly to the compactness that controls $\epsilon_{\rm GR}$; the instability is set primarily by the deep interior.

Published threshold masses span a broad range because the result
depends not only on composition and rotation but also on the entropy
profile, convective-core mass, accretion history, and adopted
stability criterion. In particular, rapidly accreting primordial
SMSs can have strongly entropy-stratified structures that differ from
the chemically homogeneous, thermally relaxed models considered
here \citep{2018MNRAS.474.2757H,2021A&A...650A.204H,
2023MNRAS.521..463H,2024A&A...689A.169S}. Our nonrotating
thresholds are toward the upper end of commonly quoted SMS collapse
masses because the comparatively large CNO abundance lowers the
central temperature and compactness at a given mass. The strong
increase at $\eta=\eta_{\rm max}$ should be interpreted more
cautiously: it follows from applying the first-order condition in
eq.~\eqref{eq:epsilonGR_marginal_stability} up to the
mass-shedding limit, where fully relativistic rotating-star
calculations become preferable
\citep{2025ApJ...978...58S}.

Finally, because the stellar luminosity is close to the
Thomson-scattering Eddington luminosity, the GR instability mass threshold maps
directly onto a characteristic upper luminosity,
\begin{equation}
\begin{split}
    L_{\rm th}
    &\approx
    \frac{4\pi Gc}{\kappa_{\rm T}}M_{\rm th}\\
    &\approx
    1.5\times10^{44}\,\mr{{\rm erg\,s^{-1}}}\,
    \frac{M_{\rm th}}{10^{6}\,M_\odot}
    \frac{0.34\,{\rm cm^2\,g^{-1}}}{\kappa_{\rm T}}.
\end{split}
\label{eq:Lth_GR}
\end{equation}
The GR instability can therefore provide an upper luminosity scale for SMS.
However, the predicted cutoff is not a single luminosity, as it is
broadened by the distributions of rotation, hydrogen abundance, and
metallicity.

\subsection{Collapse outcomes and black-hole remnant masses}
\label{sec:bh_remnants}

Crossing the GR-instability boundary does not by itself
determine the final fate of the SMS. Once the fundamental radial
mode becomes unstable, the core contracts on its
dynamical timescale, but the subsequent evolution depends on the
competition between relativistic collapse, nuclear energy release,
and centrifugal support. Three outcomes are possible: prompt
formation of a black hole, one or more thermonuclear pulsations
followed by renewed contraction, or complete thermonuclear
disruption with no black-hole remnant
\citep{1986ApJ...307..675F,2012ApJ...749...37M,
2023MNRAS.523.1629N,2024PhRvD.110c1301N,
2025ApJ...978...58S}.

In the black-hole-forming branch, the early collapse is expected to
be nearly homologous. The central density and compactness increase
until an apparent horizon forms near the center, after which the
remaining stellar material either falls directly into the black hole
or, if it has sufficient angular momentum, circularizes into a
surrounding disk. In the most extreme case, the collapse of a
uniformly rotating $n=3$ polytropic SMS at the mass-shedding limit produces a
prompt black hole containing approximately $90\%$ of the initial
stellar mass, with dimensionless spin $\chi_{\rm BH}\simeq0.75$,
while the remaining $\sim10\%$ forms a rotationally supported disk
\citep{2002ApJ...572L..39S}.


Rotation can therefore produce an energetic transient without
preventing the formation of a massive black hole. Relativistic
collapse calculations find that a disk formed around the nascent
black hole can lead to accretion shocks which eject mass even when nuclear
burning is dynamically subdominant. For rapidly rotating models,
the unbound mass may be of order $1\%$ of the initial core
mass, with characteristic velocities of $\sim0.1c$ and
enormous kinetic energies 
\citep{2017PhRvD..96h3016U,2025ApJ...981..119F}. A larger
fraction, up to $\sim10\%$, may temporarily remain
in the disk, and subsequent accretion and disk outflows may further increase the energy output.

The main alternative picture to black hole formation is that nuclear burning reverses the
collapse before an event horizon forms. Because an SMS near the GR
instability is only weakly bound, a comparatively small energy
release can be dynamically important. A \textit{necessary}, although not
sufficient, condition for reversing the collapse is approximately
\begin{equation}
    \Delta E_{\rm nuc}
    \gtrsim
    |E_{\rm bind}|
    \sim
    \beta_c {GM_\star^2\over R} \sim 10^{-5} M_\star c^2,
\label{eq:nuclear_reversal_condition}
\end{equation}
where $|E_{\rm bind}|$ is the net binding energy of the stellar core (eq. \ref{eq:constant_beta_binding_energy}). However, once the collapse has already proceeded for a fraction of the dynamical timescale, the amount of energy needed to reverse the collapse will be of the order of the kinetic energy of the infalling gas, which may be much larger than $|E_{\rm bind}|$.

\begin{figure*}
    \centering
    \includegraphics[width=0.95\textwidth]
    {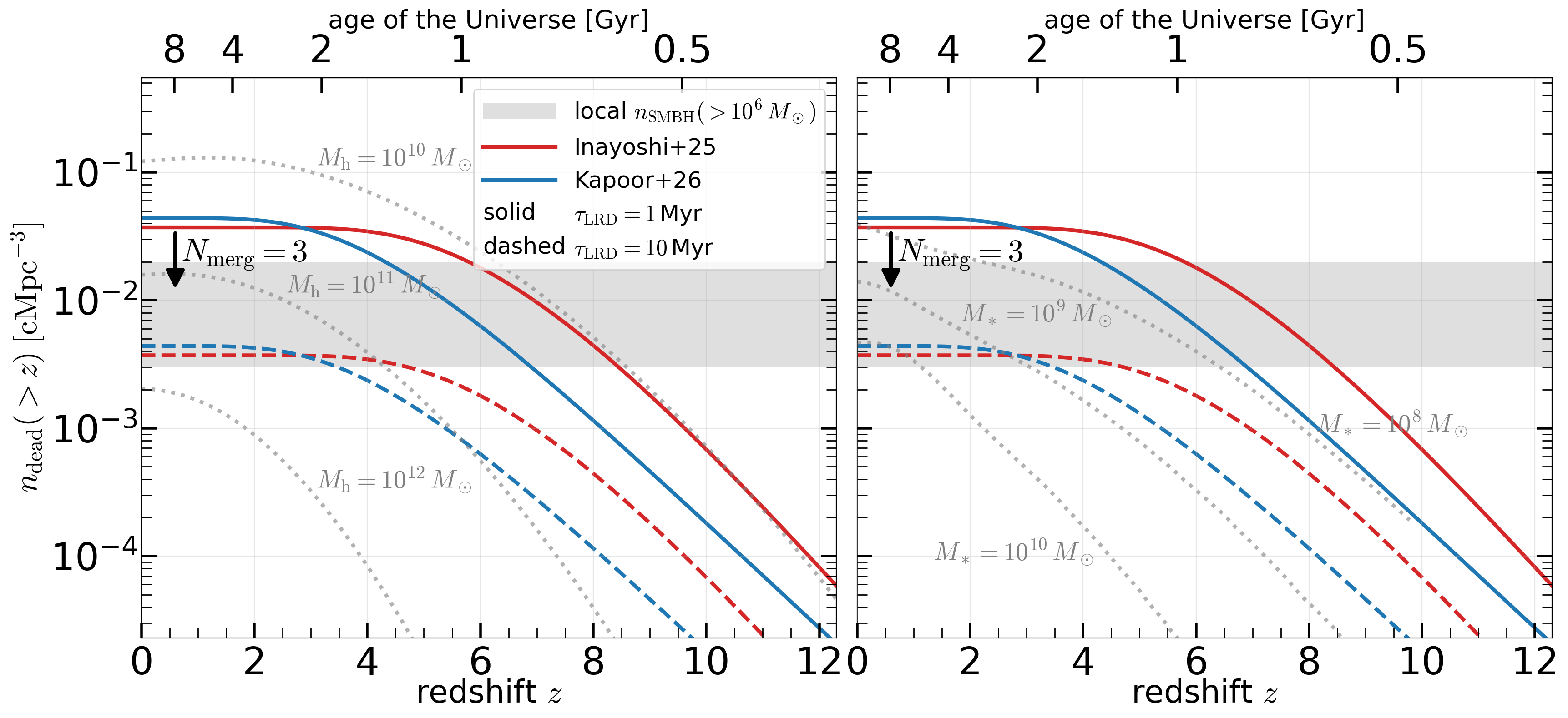}
    \caption{
    Cumulative comoving number density of LRD remnants formed at redshifts $>z$. The red and blue curves use the LRD number-density histories of \citet{2025ApJ...988L..22I} and \citet{2026arXiv260700084K}, respectively. Solid and dashed curves correspond to SMS lifetimes of $\tau_{\rm SMS}=1\,{\rm Myr}$ and $10\,{\rm Myr}$; the
    cumulative density scales as $n_{\rm dead}\propto\tau_{\rm SMS}^{-1}$. The gray band shows the uncertain range of local number density of SMBHs above $10^{6}\,M_\odot$. The black arrow illustrates an effective reduction by $N_{\rm merg}=3$ due to mergers among LRD descendants. After this correction, a lifetime of order
    $1\,{\rm Myr}$ produces a remnant density comparable to the present-day SMBH abundance, whereas a lifetime of
    $10\,{\rm Myr}$ falls below the adopted local range. \textit{Left}: Gray curves show the cumulative abundances of dark-matter halos above $10^{10}$, $10^{11}$, and $10^{12}\,M_\odot$, calculated using the halo mass function
    of \citet{2008ApJ...688..709T}. \textit{Right}: Gray curves show the cumulative abundances of galaxies above stellar masses of $10^{8}$, $10^9$, $10^{10}\,M_\odot$, following the UniverseMachine model of \citet{2019MNRAS.488.3143B}. Abundance matching against these gray curves (\S\ref{sec:host_halos_descendants}) predicts a characteristic host halo mass of $M_{\rm h}\sim10^{10}$--$6\times10^{10}\,M_\odot$ and stellar mass of $M_\star\sim10^{8}\,M_\odot$ for LRDs near $z\sim5$.
    }
    \label{fig:n_dead_lrd}
\end{figure*}

Relativistic calculations found that CNO burning may halt the collapse of some
$M_\star\sim5\times10^{5}\,M_\odot$ models at
abundances comparable to our $Z=10^{-2}$ case, with rotation modifying the critical composition
\citep{2012ApJ...749...37M}. Other calculations of metal-enriched
or rapidly accreting SMSs likewise find thermonuclear pulsations or
complete disruption over parts of parameter space
\citep{2023MNRAS.523.1629N,2024PhRvD.110c1301N}. In contrast, recent
stability estimates argue that hydrogen-burning SMS cores with
plausible metallicities are likely to form black holes because nuclear
burning does not release sufficient energy to reverse the collapse
\citep{2025ApJ...978...58S}. Fully relativistic collapse simulations
at primordial CNO abundance likewise find nuclear burning to be
dynamically subdominant, resulting in black-hole formation
\citep{2025ApJ...981..119F}.

For the demographic calculation below, we assume that every SMS
eventually forms a black hole which retains nearly all the mass.
The remnant-abundance estimates in the following subsections are
thus conditional on the SMS interpretation of LRDs and on the
fiducial black-hole-forming fraction.

\subsection{From LRD remnants to present-day SMBHs}
\label{sec:remnant_demographics}

A central prediction of the SMS interpretation is the cumulative
abundance of massive-black-hole remnants. Let
$\Phi_{\rm LRD}(z)$ denote the instantaneous comoving number density
of active LRDs. If each object remains visible for a time
$\tau_{\rm LRD}\simeq\tau_{\rm SMS}$, and this time is short compared
with the cosmological timescale over which $\Phi_{\rm LRD}$ evolves,
the implied formation rate of SMSs is
\begin{equation}
    \dot n_{\rm SMS}(z)
    \approx
    \frac{\Phi_{\rm LRD}(z)}{\tau_{\rm SMS}}.
\label{eq:SMS_formation_rate}
\end{equation}
Under the fiducial assumptions that one LRD corresponds to one SMS and
that each SMS produces a massive black hole, the cumulative
comoving density of remnants formed at redshifts $z'>z$ is therefore
\begin{equation}
    n_{\rm dead}(z)
    =
    \int_{\max(z,z_{\rm min})}^{z_{\rm max}}
    \frac{\Phi_{\rm LRD}(z')}{\tau_{\rm SMS}}
    \left|\frac{dt}{dz'}\right|\,dz'.
\label{eq:n_dead_integral}
\end{equation}
We adopt $z_{\rm min}=0.5$ for the calculation shown below and
evaluate eq.~(\ref{eq:n_dead_integral}) using the analytical redshift
distributions\footnote{This only includes LRDs brighter than absolute magnitude $M\approx -20$ at rest-frame wavelength $\lambda=5100\,\AA$. The faint end slope of the luminosity function of LRDs is fairly flat \citep{2025arXiv250902662M}, so the uncertainty associated this magnitude cut is modest.} $\Phi_{\rm LRD}(z)$ inferred by \citet{2025ApJ...988L..22I} and
\citet{2026arXiv260700084K}. The two prescriptions differ in the
detailed redshift distribution of LRD formation, but give similar
total remnant abundances and the contribution from below $z_{\rm min}=0.5$ is negligible for our purpose here. For example, the latter gives
\begin{equation}
    n_{\rm dead}(z=0)
    \simeq
    4.4\times10^{-2}\,\mathrm{cMpc^{-3}}
    \frac{1\,\mathrm{Myr}}{\tau_{\rm SMS}},
\label{eq:n_dead_fiducial}
\end{equation}
whose normalization has order-unity uncertainty due to small-number statistics in $\Phi_{\rm LRD}(z)$ for some redshift bins. The inverse dependence on $\tau_{\rm SMS}$ means that, at fixed observed LRD abundance, a shorter lifetime requires a larger formation rate and hence produces more remnants.

The appropriate observational comparison is the present-day
abundance of massive black holes. This quantity remains uncertain
near and below $M_{\rm BH}\sim10^{6}\,M_\odot$, where dynamical measurements
are incomplete and the inferred black-hole mass function depends on
the adopted host scaling relations, their intrinsic
scatter and selection biases, and the black-hole occupation fraction
\citep{2004MNRAS.351..169M,2007MNRAS.378..198G,
2009MNRAS.400.1451V,2016MNRAS.460.3119S,
2019ApJ...883L..18G,2025ApJ...992..176Z}.
We therefore adopt a deliberately broad interval
\begin{equation}
    3\times10^{-3}
    \lesssim
    {n_{\rm SMBH,0}(>10^{6}\,M_\odot) \over \mathrm{cMpc^{-3}}}
    \lesssim
    2\times10^{-2},
\label{eq:local_SMBH_density}
\end{equation}
which is intended to bracket the dominant systematic uncertainties. This estimate does not include lower-mass objects with $M_{\rm BH}<10^6M_\odot$ or wandering SMBHs that are not located in galactic nuclei. The strongest constraint on the lower end of the SMBH mass function comes from observations of tidal disruption events, and \citet{2023ApJ...955L...6Y} showed that the SMBH mass function is nearly flat ($\d n/\d \log M\simeq \mr{\,const}$) in the mass range of $\sim 10^5$ to $\sim 10^7M_\odot$, so including SMBHs down to e.g., $10^5M_\odot$ will only weakly affect our results.

Mergers between LRD descendants can reduce the number of distinct
black holes surviving to the present day. We describe this effect
with an effective reduction factor $N_{\rm merg}$.
For the illustrative choice $N_{\rm merg}=3$, motivated by the
order-unity number of mergers expected along massive-halo assembly
histories \citep{2010MNRAS.406.2267F}, we obtain the number density of present-day black holes contributed by LRDs
\begin{equation}
    n_{\rm BH,LRD}(z=0)
    \simeq
    1.5\times10^{-2}\,\mathrm{cMpc^{-3}}
    \frac{1\,\mathrm{Myr}}{\tau_{\rm SMS}}
    {3\over N_{\rm merg}}.
\label{eq:LRD_local_BH_density}
\end{equation}
As shown in Fig.~\ref{fig:n_dead_lrd}, this prediction lies within
the estimated local SMBH abundance for
$\tau_{\rm SMS}\sim1\,\mathrm{Myr}$. In contrast, a lifetime of
$10\,\mathrm{Myr}$ gives a merger-corrected density of only
$\sim10^{-3}\,\mathrm{cMpc^{-3}}$, below the local SMBH abundance interval.
The comparison therefore provides a nontrivial demographic test of
the model: an SMS lifetime of order $1\,\mathrm{Myr}$ naturally
produces enough remnants to account for a substantial, and
potentially dominant, fraction of present-day SMBHs above
$10^{6}\,M_\odot$. This agreement is one of the strongest quantitative predictions of
the SMS interpretation.

The gray curves on the cumulative dark matter halo number densities (left panel) and the galaxy number densities (right panel) in Fig.~\ref{fig:n_dead_lrd} are used here as abundance landmarks; their implications for LRD host halos and cosmological descendants are discussed in \S\ref{sec:host_halos_descendants}.

\subsection{Comparison with quasi-star models}
\label{sec:sms_quasistar_comparison}

Quasi-stars were originally proposed to overcome the difficulty of
growing SMBHs from stellar-remnant seeds within the
limited time available at high redshift. In these models, a newborn
black hole is embedded within a much more massive, quasi-hydrostatic
gaseous envelope. The accretion luminosity is regulated near the
Eddington luminosity of the entire system rather than that of
the black hole alone, permitting accretion that is highly
super-Eddington relative to the central black-hole mass
\citep{2008MNRAS.387.1649B,2010MNRAS.402..673B,
2024ApJ...970..158C}. A quasi-star can therefore amplify an initially
modest seed much more rapidly than ordinary Eddington-limited
accretion.

The SMS channel considered here addresses the seed problem differently.
On the black-hole-forming branch, the
$10^{5}$--$10^{6}\,M_\odot$ star itself collapses to form a massive
seed while retaining most of its mass
(\S\ref{sec:bh_remnants}). An intermediate phase in which a small
black hole must grow by consuming a much more massive envelope is
therefore not required. In this sense, the original motivation for a
quasi-star phase is weakened if LRDs are already nuclear-burning SMSs,
although quasi-stars remain a distinct possible interpretation of the
LRD phenomenon.

The two models also make different assumptions about the state of the
envelope after black-hole formation. A quasi-star
constructs a massive envelope in quasi-hydrostatic equilibrium around
an accreting black hole, with the central accretion luminosity
supporting the overlying gas
\citep{2008MNRAS.387.1649B,2024ApJ...970..158C}. Our calculations,
by contrast, describe the star only before black-hole formation and
terminate at the onset of a global GR instability. Furthermore, the inflated region above $r_{\rm base}$ that produces the cool LRD-like photosphere contains only $\Delta m_{\rm env}/M_\star\sim10^{-4}$ of the stellar mass. It is therefore not by itself the massive fuel reservoir assumed in a quasi-star. Relativistic collapse calculations generally find that most of the SMS enters the black hole promptly, with only a rotation-dependent fraction remaining in an accretion disk \citep{2025ApJ...981..119F}. Thus, the formation and the long-term survival of a quasi-hydrostatic envelope are not implied by the pre-collapse hydrostatic solutions presented here.

The closest well-studied analogues of a black hole embedded inside a
pre-existing stellar envelope are collapsars and failed supernovae.
In these systems, the outer layers initially preserve approximately
their pre-collapse structure because they have not yet responded to
the loss of central pressure support. As the collapse propagates
outward, a sufficiently powerful accretion flow can launch jets or
disk wind and unbind the star, whereas a weaker engine allows most of
the envelope to fall into the black hole
\citep{1999ApJ...524..262M,2006ARA&A..44..507W,
2015PhR...561....1K,2018MNRAS.476.2366F,
2019MNRAS.485L..83Q,2022MNRAS.511..176A}. In neither case does the
original envelope remain intact for much longer than its dynamical or
fallback time. Collapsars and failed supernovae are not direct models
of quasi-stars, but their rapid evolution suggests that a long-lived quasi-star envelope likely requires the gas to re-establish a regulated quasi-hydrostatic configuration.


This distinction is particularly important for LRD demographics. For
an LRD-like configuration with
$R_{\rm env}\sim10^{16}\,{\rm cm}$ and
$M\sim10^{6}\,M_\odot$, the global dynamical time is only a few years.
If the observed LRD phase were simply the transient response of a
pre-existing envelope to central collapse, adopting
$\tau_{\rm LRD}\sim3\,{\rm yr}$ in
eq.~\eqref{eq:LRD_local_BH_density} would give
$n_{\rm BH,LRD}\sim 5\times 10^{3}\,{\rm cMpc^{-3}}$ (taking $N_{\rm merg}\sim 3$), five to six orders of magnitude above the local SMBH abundance. An accretion-powered envelope can therefore represent
the bulk of the LRD population only if it is a true quasi-star rather
than a collapsar-like configuration.

Once a quasi-hydrostatic configuration has been established, its
fuel-consumption time can instead be substantially longer. Let
$\eta_{\rm acc}$ be the energy efficiency (including both electromagnetic and outflow kinetic power) of black-hole accretion near the innermost regions of the accretion flow
and let $f_{\rm acc}$ be the fraction of the envelope mass processed
through the black hole while the source remains in its quasi-star
state. If the luminosity is regulated near the Thomson-scattering
Eddington luminosity of the total quasi-star mass $M_{\rm QS}$, then
\begin{equation}
\begin{split}
    \tau_{\rm QS}
    &\simeq
    \frac{\eta_{\rm acc}f_{\rm acc}M_{\rm env}c^2}
         {L_{\rm Edd,T}(M_{\rm QS})} =
    \eta_{\rm acc}f_{\rm acc}
    \frac{M_{\rm env}}{M_{\rm QS}}
    \frac{\kappa_{\rm T}c}{4\pi G}\\
    &\simeq
    20\,{\rm Myr}\,
    \frac{\eta_{\rm acc}}{0.1}
    \frac{f_{\rm acc}}{0.5}
    \frac{M_{\rm env}}{M_{\rm QS}}
    \frac{\kappa_{\rm T}}
    {0.34\,{\rm cm^2\,g^{-1}}}.
\end{split}
\label{eq:quasistar_lifetime}
\end{equation}
The final estimate adopts the envelope-dominated limit
$M_{\rm env}\simeq M_{\rm QS}$. It is possible that winds and structural transitions cause an
LRD-like phase to occupy only part of the evolution and shorten the
observable duration. Nevertheless, modern quasi-star calculations
indeed find lifetimes of order tens of Myr
\citep{2024ApJ...970..158C,2026ApJ...996...48B,
2026ApJ...998L...4S}.

The demographic calculation in
\S\ref{sec:remnant_demographics} constrains the duration for which
each independent remnant appears as an LRD, irrespective of whether
its luminosity is supplied by nuclear burning or black-hole
accretion. At fixed observed LRD abundance,
$n_{\rm BH,LRD}\propto\tau_{\rm LRD}^{-1}$: an extremely short
collapsar-like phase would overproduce remnants, whereas a much longer
quasi-star phase would produce too few. For example,
$\tau_{\rm LRD}=20\,{\rm Myr}$ gives
$n_{\rm BH,LRD}\simeq7.5\times10^{-4}\,{\rm cMpc^{-3}}$ for
$N_{\rm merg}=3$, below the adopted local abundance of SMBHs. The comparison empirically favors an
LRD-visible lifetime of order 1 Myr, naturally consistent with the
hydrogen-burning lifetime of the SMS models.

This argument does not exclude quasi-stars as individual LRDs, provided that they only show an LRD-like phase lasting for $\sim 1\rm\, Myr$.
The relation between LRD counts and
local SMBH demographics therefore provides a common empirical
lifetime test for both nuclear-burning SMSs and accretion-powered
quasi-stars.

\subsection{Comparison with black-hole-star models}
\label{sec:sms_bhstar_comparison}

The term ``black-hole star'' (BH$\star$) has recently been used for
LRD models in which a rapidly accreting black hole is surrounded by a
dense and nearly thermalizing gaseous atmosphere or cocoon
\citep{2025arXiv250316596N,2025arXiv251121820D,
2026OJAp....962505S,2026ApJ..1005L..37T}. The terminology is not yet
uniform, and some proposed configurations may interpolate between
BH$\star$'s and quasi-stars. Here we use BH$\star$ specifically for
models in which the black hole dominates the gravitational potential
and the surrounding gas primarily reprocesses the output of a
super-Eddington accretion flow. This differs from the quasi-star
models of \S\ref{sec:sms_quasistar_comparison}, in which a massive
self-gravitating envelope regulates the luminosity through the
Eddington limit of the entire system.

BH$\star$ models have several important attractions. A central
accretion flow naturally supplies the ionizing photons needed to power
the Balmer lines, while high-density gas can absorb and thermalize the
emergent continuum, produce a strong Balmer break, and broaden the
lines through scattering or gas motions. The freedom to vary the
accretion rate, density profile, column density, and viewing geometry
allows such models to reproduce a wide range of individual LRD
spectra. This spectral flexibility is an advantage over the present
SMS calculations, which predict the cool continuum but do not yet
provide a quantitative model for the broad Balmer lines.

The evolutionary starting point is nevertheless different. A
BH$\star$ model begins with a massive black hole and describes how it
accretes from (and is concealed by) its surrounding gas, but the
model does not by itself specify how the initial seed black hole formed and grew over time. In the
SMS interpretation, the luminous LRD is instead the progenitor of the
seed: the same $10^{5}$--$10^{6}\,M_\odot$ object that produces the
observed LRD emission subsequently undergoes GR collapse and leaves a
black hole of comparable mass. The SMS model therefore connects the
observable phase directly to the formation of a massive seed, whereas
the seed mass and its prior growth history are assumptions in
the BH$\star$ model.

The two pictures also differ in how the LRD lifetime is determined.
For an SMS, the available hydrogen fuel and its nearly Eddington
luminosity give a nuclear-burning lifetime of order 1 Myr
(eq.~\ref{eq:tau_SMS_GR}). In a BH$\star$, the observable lifetime
depends on the gas-supply history and on how often the accretion flow
is surrounded by a sufficiently dense and optically thick cocoon.
More generally, the cumulative duration of the BH$\star$ phase can be written schematically as $\tau_{\rm BH\star} = \int f_{\rm duty}(t)\,dt$, where $f_{\rm duty}$ includes the duty-cycle for the LRD-like phase. Population-based BH$\star$ interpretations have inferred duty cycles
of order $1\%$ and cumulative lifetimes of order $10\,{\rm Myr}$
\citep[e.g.,][]{2026ApJ..1000...90I, 2026OJAp....962505S, 2026arXiv260110573A}. By contrast, the remnant-demographic comparison in \S\ref{sec:remnant_demographics} favors a cumulative
LRD-visible lifetime of order $1\,{\rm Myr}$ if LRD remnants are to
provide a substantial fraction of present-day SMBHs. A BH$\star$
model can accommodate this constraint through a shorter or more
intermittent gas-enshrouded phase, but the duty cycle depends on the highly uncertain accretion and gas-replenishment history.

A related distinction concerns the upper luminosity cutoff \citep{2025arXiv250902662M}. For an accreting black hole $L_{\rm BH\star} = \lambda_{\rm Edd} L_{\rm Edd,T}$,
where the Eddington ratio $\lambda_{\rm Edd}$ is set by the accretion rate and its
radiative efficiency. The BH$\star$ framework therefore does not by
itself select a maximum LRD luminosity: reproducing the observed
bright-end cutoff requires a cutoff in the black-hole mass function,
a restriction on $\lambda_{\rm Edd}$, or a characteristic upper limit
to the available gas supply. Recent atmosphere models inferred black hole masses of $\sim10^{3}$--$10^{5},M_\odot$ and Eddington ratios of $\lambda_{\rm Edd}\sim$a few to 300 from individual LRDs and stacked spectra \citep{2026arXiv260909265C,2026arXiv260909274S}. These estimates rely on the fitted surface gravity but the structure of these strongly accelerated envelopes have not been self-consistently determined. In the SMS model, as $L_\star\approx L_{\rm Edd,T}\propto M_\star$, the maximum mass from the GR-instability threshold mass maps directly onto the maximum LRD luminosity.

Finally, a BH$\star$ must reconcile an intrinsically ionizing central
engine with the weak X-rays and high-ionization lines observed from
most LRDs. Super-Eddington accretion may intrinsically soften the
spectrum by weakening the hot corona and reprocessing the emission from the inner accretion flow by disk winds \citep{2024ApJ...976...96P}. Even so, if a substantial hard
ionizing component is produced, the surrounding gas must intercept
and reprocess nearly all of it before it either reaches the observer
or illuminates lower-density gas in the environment. This motivates a geometrically
thick cocoon with a covering fraction approaching unity
\citep{2026Natur.649..574R,2026arXiv260118864S}. Otherwise,
low-column sightlines should reveal luminous X-rays, while escaping
ionizing radiation should generate stronger He\,II,
C\,IV, [Ne\,V] lines. The covering fraction, column density,
ionization structure, and reprocessing efficiency represent 
additional uncertainties in the BH$\star$ model.

These considerations do not rule out BH$\star$'s as LRDs.
The flexibility in the BH$\star$ model makes it less predictive at the population
level: the seed mass, accretion history, duty cycle, luminosity
cutoff, and reprocessing geometry must be specified separately. The
SMS interpretation instead provides a more tightly connected
evolutionary sequence.
The agreement between the Myr-scale nuclear lifetime coupled with the
LRD abundance and local SMBH demographics is therefore a distinctive
advantage of the SMS model.

\section{Predictions and Observational Tests}
\label{sec:lrd_predictions}

The SMS interpretation makes predictions on several fronts: direct observables of the SMS photosphere itself
(\S\ref{sec:cool_photospheres}--\ref{sec:host_halos_descendants},
\S\ref{sec:spectral_environment}--\ref{sec:little_blue_dots}), the
terminal transient produced when the core reaches the GR instability
(\S\ref{sec:collapse_transient}), and long-lived signatures left
behind in its descendants (\S\ref{sec:host_halos_descendants},
\S\ref{sec:chemical_fossil_record}). We order these predictions roughly from the most robust to the more speculative. Table~\ref{tab:sms_observation_comparison} summarizes the these predictions and observational tests.

\subsection{Cool photospheres}
\label{sec:cool_photospheres}

The most direct observational prediction of our models is that
metal-enriched SMSs can develop intrinsically cool photospheres while
remaining close to the Thomson-scattering Eddington luminosity. As
shown in Fig.~\ref{fig:sms_hr_combined}, the
$M_\star=10^{4}$--$10^{6}\,M_\odot$ models at
$Z\simeq10^{-2}$ converge toward $T_{\rm eff}\sim 7000\,{\rm K}$.
The red rest-frame optical continua observed in LRDs
\citep{2024ApJ...963..129M,2024ApJ...964...39G}, when interpreted as
optically thick emission, imply characteristic color temperatures in the range of $3000$--$9000\rm\,K$ \citep{2025ApJ...994..113L, 2025arXiv251121820D}, close to (but many lower than) the
temperatures predicted by our models.

The cool temperature is a consequence of opacity-regulated envelope
inflation. At low metallicity, the Fe-opacity bump and the opacity plateau at lower temperatures is too weak to sustain a highly inflated envelope, and the SMS envelope remains compact and hot. At $Z\simeq10^{-2}$, the stronger opacity bump produces a low-density envelope extending far beyond the
stellar core. The envelope continues to expand until hydrogen
recombination causes the Rosseland-mean opacity to decrease sharply,
allowing the radiative acceleration to fall below gravity and a bound
photosphere to form. The photospheric temperature is therefore set
primarily by hydrogen recombination, providing a thermostat that
naturally places the cool branch near $T_{\rm eff}\sim7000\,{\rm K}$.

This prediction is relatively insensitive to the main
uncertainties explored in our envelope calculation. Changing the
effective Eddington limiter from $Q_{\rm LIM}=1$ to $0.1$ has negligible effect on 
$T_{\rm eff}$ of our cool branch. At $Z=10^{-2}$, reducing the Fe-group abundance by
$f_{\rm Fe}=10^{-0.5}$ shifts the predicted temperature by less than a few
percent, because the opacity at lower temperatures below the Fe opacity bump remains sufficiently high to place
the star on the inflated cool branch. The nearly vertical
high-metallicity sequences in Fig.~\ref{fig:sms_hr_combined} also
show that the photospheric temperature depends only weakly on stellar
mass over the luminosity range relevant to LRDs.

\begin{figure}
    \centering
    \includegraphics[width=0.48\textwidth]{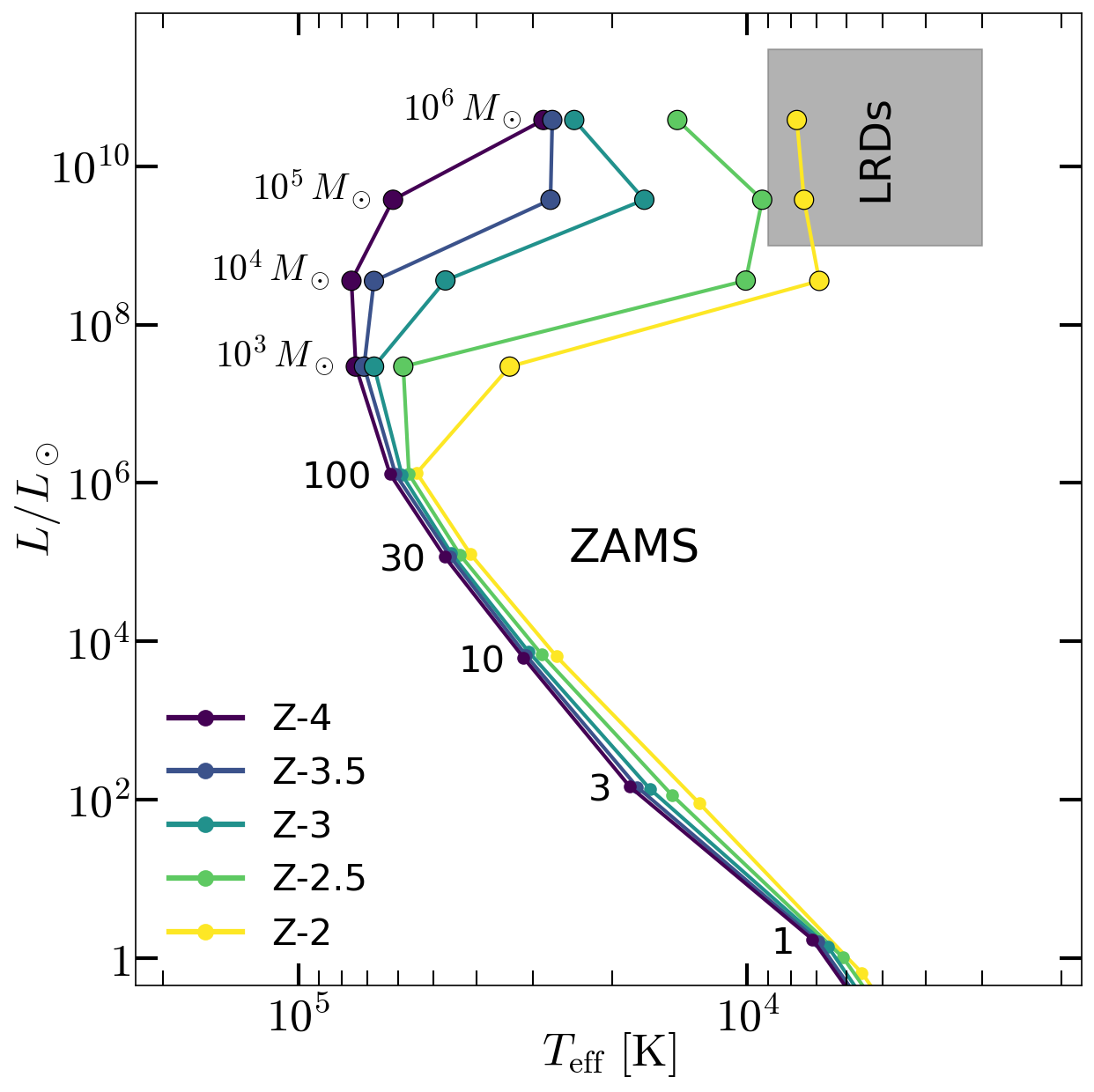}
    \caption{The HR diagram for our models with $M=10^3, 10^4, 10^5, 10^6\,M_\odot$ with $Q_{\rm LIM}=0.1$, reduced-Fe abundance $f_{\rm Fe}=10^{-0.5}$ for each metallicity in our grid (indicated by colors), together with the zero-age main sequence (ZAMS) from MIST models \citep{2026ApJS..283...64D} for ordinary stars with $M\leq 100M_\odot$. The $M=10^{3}\,M_\odot$ models it sits well off the cool branch, remaining compact and hot even for $Z=10^{-2}$. The gray shaded region marks the
    observationally inferred (but model-dependent) ranges of effective temperature ($3000$--$9000\rm\, K$) and luminosity ($10^9$--$3\times 10^{11}L_\odot$) by \citet{2025arXiv251121820D}.
    }
    \label{fig:sms_hr_M1e3_check}
\end{figure}

However, we do not expect this cool branch to persist to arbitrarily low masses. As a
supplementary check, we solved the models at $M=10^{3}\,M_\odot$ across the same metallicity grid. Fig.~\ref{fig:sms_hr_M1e3_check} shows the models with reduced-Fe abundance pattern ($f_{\rm Fe}=10^{-0.5}$, the more conservative case for envelope inflation), together with MIST models for ordinary zero-age main-sequence stars\footnote{Since the MIST grid does not contain $f_{\rm Fe}=10^{-0.5}$, we use $f_{\rm Fe}=10^{-0.4}$ instead, but the difference in the resulting positions on the HR diagram is negligible for our purpose here. } for $M\leq 10^2M_\odot$ at the same metallicities \citep{2016ApJS..222....8D, 2016ApJ...823..102C, 2026ApJS..283...64D}. We find that the $M=10^3M_\odot$ models sit well off the cool branch even for $Z=10^{-2}$. This is because the stars at $M\lesssim10^{3}\,M_\odot$ are further from the Thomson-scattering Eddington luminosity, $1-\Gamma \approx \beta_c \simeq 0.2 (M/10^{3}\,M_\odot)^{-1/2}$ (eq. \ref{eq:betac_mass_analytic}), so the outer envelope transitions to the sub-Eddington regime at temperatures well above the hydrogen-recombination temperature --- the hydrogen-recombination thermostat mechanism described above never engages. Only at $M\gtrsim 10^4M_\odot$, our high metallicity ($Z=10^{-2}$) solutions reach the cool branch.

We caution that the comparison between our predicted effective temperatures $T_{\rm eff}$ and the observationally inferred LRD color temperatures should be made only at the limited precision in the present theoretical and observational treatments. Our models adopt LTE Rosseland-mean opacities and an approximate photospheric closure; although sufficient for calculating the global interior structure, these assumptions do not predict the frequency-dependent emergent spectrum or the relation between effective and color temperatures. The cool-branch endpoint near $T_{\rm eff}\sim 7000\,{\rm K}$ should therefore not be regarded as a strict temperature floor. Possible non-LTE modifications of $T_{\rm eff}$ and the formation of the Balmer features are discussed in \S\ref{sec:discussion}. Conversely, the observationally inferred temperatures are themselves model dependent. For example, \citet{2025arXiv251121820D} fitted the rest-frame $0.42$--$1\,\micron$ spectra of LRDs with a three-parameter modified blackbody, and they inferred color temperatures of 3000--9000 K and luminosities of $10^{9}$--$3\times 10^{11}\,L_\odot$. Although the resulting continuum peak near $\lambda_{\rm peak}\simeq0.6$--$0.7\,\micron$ is relatively robust, the fitted color temperature is strongly degenerate with the assumed power-law modification to the blackbody. Moreover, longer-wavelength MIRI observations generally require additional emission components \citep[e.g., dust or nebular continuum emission,][]{2026arXiv260220247P}, so the modified blackbody should only be regarded as an empirical description.

Finally, we note that the emission from cool SMS photosphere differs from the alternative picture of gas/dust enshrouded star formation in that (1) the latter is spatially extended with an intrinsic size of the order pc or larger (depending on the compactness of the stellar/gas distribution) and that (2) the latter does not predict observable variability.



\subsection{A rotation-dependent maximum luminosity}
\label{sec:upper_luminosity}

A second direct prediction of the SMS interpretation is a maximum
luminosity set by the onset of the GR instability. As an SMS
radiates close to the Thomson-scattering Eddington luminosity,
$L_\star\approx L_{\rm Edd,T}\propto M_\star$, its luminosity provides
a direct measure of its mass. An SMS cannot remain on the
hydrogen-burning branch once its mass exceeds the instability
threshold $M_{\rm th}(X,Z,\eta)$. The threshold mass therefore
maps directly onto the limiting luminosity given by
eq.~(\ref{eq:Lth_GR}).

For the initial composition\footnote{The maximum luminosity depends very weakly on the evolution on the main-sequence because, as $X$ decreases, both $M_{\rm th}$ and $\kappa_{\rm T}$ decrease at the same time, and these two factors nearly cancel such that $\d \ln L_{\rm th}/\d \ln X \approx 0.2, -0.1, -0.2$ for $\eta=0, \eta_{\rm max}/2, \eta_{\rm max}$.} $X=0.7$ and the metallicity
$Z=10^{-2}$ relevant to the cool models, we obtain
\begin{equation}
    L_{\rm max}
    \simeq
    \begin{cases}
        4.8\times10^{43}\,{\rm erg\,s^{-1}},
        & \eta=0,\\[3pt]
        1.8\times10^{44}\,{\rm erg\,s^{-1}},
        & \eta=\eta_{\rm max}/2,
    \end{cases}
\label{eq:Lmax_rotation}
\end{equation}
where the second row is for an SMS whose core rotates at $1/\sqrt{2}$ of the break-up angular speed. Rotation raises the luminosity ceiling because it stabilizes the star against relativistic collapse and allows it to
reach a larger mass. We therefore consider
\begin{equation}
    L_{\rm max}\sim5\times10^{43}\text{ to }2\times10^{44}\,
    {\rm erg\,s^{-1}}
\label{eq:robust_Lmax_range}
\end{equation}
as the robust upper limit for nonrotating through moderately rotating SMSs.

Formally extending the first-order stability condition to
$\eta=\eta_{\rm max}$ gives a larger maximum luminosity,
$L_{\rm max}\sim4.6\times10^{44}\,{\rm erg\,s^{-1}}$ at $X=0.7$ and
$Z=10^{-2}$. This value is less secure because the
perturbative stability criterion is being applied at the
mass-shedding limit, where fully relativistic rotating-star
calculations are preferable
\citep{1999ApJ...526..941B,2016ApJ...818..157S}. Gravity darkening, differential
rotation and angular-momentum loss introduce additional uncertainties.

The current LRD population shows a sharp decline in the luminosity
function above\footnote{From an LRD sample at $4.5 < z < 4.9$, \citet{2025arXiv250902662M} inferred a characteristic luminosity of $\lambda L_{5100}^*\simeq 1.2\times10^{44}\rm\, erg\,s^{-1}$ based on a Schechter function fit and a break luminosity of $\lambda L_{5100}^*\simeq 3.2\times10^{44}\rm\, erg\,s^{-1}$ based on a broken power-law fit.} $\lambda L_{5100}\simeq10^{43}\mbox{--}3\times10^{44}\,{\rm erg\,s^{-1}}$ \citep{2025arXiv250902662M}. Matching this scale therefore requires SMS cores rotating close to the mass-shedding limit ($\eta\sim \eta_{\rm max}/2$); whereas the nonrotating model prediction of $5\times10^{43}\rm\, erg\,s^{-1}$ falls short by a factor of a few below the
observed cutoff. The existence of a cutoff on a comparable luminosity scale is qualitatively consistent with the
GR-instability interpretation, but the comparison depends sensitively on the (uncertain) distribution of SMS core rotation rates. A precise comparison is not yet possible because the
observed cutoff is defined using a monochromatic optical luminosity,
whereas eq.~(\ref{eq:Lmax_rotation}) refers to the total
luminosity. Although recent multiwavelength measurements favor
smaller bolometric corrections for LRDs than for ordinary quasars,
the correction remains sensitive to the continuum model and host galaxy subtraction \citep{2026ApJ...996..129G}. A direct comparison will ultimately require synthetic spectra for the cool SMS
atmospheres.

The predicted cutoff should not be abrupt. A distribution
of rotation rates broadens the maximum stable mass, while hydrogen
depletion, metallicity, angular-momentum evolution, and mass loss
introduce additional scatter. Nevertheless, the SMS model predicts
that the abundance of cool LRDs should decline rapidly above a
characteristic luminosity of $5\times 10^{43}$--a few $\times10^{44}\,{\rm erg\,s^{-1}}$. Establishing the location and shape of this bright-end decline would therefore provide a population-level test of
our model.

\subsection{Host halos and cosmological descendants}
\label{sec:host_halos_descendants}

The demographic comparison in
\S\ref{sec:remnant_demographics} also provides a conditional estimate
of the halo masses associated with LRDs and their descendants. At
$z=0$, the adopted abundance of SMBHs above $10^{6}\,M_\odot$ is
comparable to the cumulative abundance of dark-matter halos with
masses between $10^{11}$ and $10^{12}\,M_\odot$ (Fig.~\ref{fig:n_dead_lrd}, left panel). Note that this correspondence is only an abundance-based estimate, as the SMBH occupation fraction, wandering SMBHs, and merger histories remain uncertain.

To connect these descendants to the LRD epoch, we integrate the mean halo accretion rate of
\citet{2010MNRAS.406.2267F} to obtain the mean halo mass track
\begin{equation}\label{eq:Fakhouri_mass_track}
\begin{split}
    &\frac{M_{\rm h}(z)}{10^{12}\,M_\odot}
    =
    \Big[
    \left(
    \frac{M_{\rm h,0}}{10^{12}\,M_\odot}
    \right)^{-0.1}
    \\
    &+
    0.062
    \left\{
    1.11z-0.11\ln(1+z)
    \right\}
    \Big]^{-10},
\end{split}
\end{equation}
where $M_{\rm h,0}$ is the halo mass at $z=0$ and we have adopted
the Millennium cosmology, with
$H_0=73\,{\rm km\,s^{-1}\,Mpc^{-1}}$. The above equation gives
\begin{equation}
    M_{\rm h}(z=5)
    \simeq
    \begin{cases}
        1\times10^{10}\,M_\odot,
        & M_{\rm h,0}=10^{11}\,M_\odot,\\
        6\times10^{10}\,M_\odot,
        & M_{\rm h,0}=10^{12}\,M_\odot.
    \end{cases}
\label{eq:halo_mass_mapping_z5}
\end{equation}
Thus, if LRDs are SMSs that seed a substantial fraction of the local
SMBH population, their characteristic host halos at $z\simeq5$ should
have masses of $10^{10}$--$6\times10^{10}\,M_\odot$,
with the abundance comparison favoring values toward the lower end of
this range. These estimates describe mean main-progenitor tracks;
individual halos have substantial scatter in their assembly histories,
and some LRD remnants may reside in satellite haloes or become wandering
black holes. The predicted host-mass scale can be tested through LRD
clustering, gravitational lensing, and independent dynamical
constraints on their host galaxies.

On the other hand, we can also compare the number density of LRD remnants for $\tau_{\rm LRD}\simeq 1\rm\,Myr$ with the galaxy stellar mass function model (Fig. \ref{fig:n_dead_lrd}, right panel) at epochs with abundant LRD formation ($z \gtrsim 4$). Based on this comparison, our SMS model predicts that the typical stellar mass of LRD host galaxies at high redshifts is $\sim 10^8\rm\,M_\odot$ or slightly lower (an abundance-matching estimate), although their stellar mass distribution is still highly uncertain.

An independent environment-based estimate provides tentative support for our abundance-matching predictions. By comparing the Mpc-scale overdensities around six LRDs at $z=4$--$5$ with those around ordinary galaxies, \citet{2025ApJ...988..246M} inferred a characteristic host stellar mass of $M_\star\sim 5\times10^7\,M_\odot$ (without directly inferring the stellar mass from the LRD host spectra). Mapping this stellar mass through standard stellar-to-halo mass relations gives $M_{\rm h}\sim4\times10^{10}\,M_\odot$. Both estimates agree well with our abundance-matching predictions of $M_\star\sim10^8\,M_\odot$ (or slightly lower) and $M_{\rm h}\sim10^{10}$--$6\times10^{10}\,M_\odot$.

Stellar mass can also be directly inferred based on LRD host spectra. Modeling of stacked JWST NIRSpec and MIRI observations with spectral decomposition between host stars and an LRD component yields $M_* \sim 10^{8.3}\,M_\odot$ \citep{2026arXiv260220247P}. Spatial decomposition gives extended stellar mass of $M_* \sim 10^{8.7}\, M_\odot$ for one resolved host and upper limits for others \citep{2025ApJ...983...60C}. Hybrid stellar+AGN decomposition fits yield a median $M_\star\sim 10^{8.8}\, M_\odot$ for 15 broad-line LRDs \citep{2025ApJ...986..126K}. These estimates remain highly model dependent: the principal uncertainties include how the unresolved UV and optical continua are divided between the compact LRD and ordinary stars, as well as the assumed star-formation history, dust attenuation, and metallicity. Moreover, UV-based measurements primarily constrain young, relatively unobscured stars and may miss a dust-embedded or older population. Current observations suggest a potentially wide range of stellar masses between $10^8$ and $10^9\,M_\odot$, although the stellar mass distribution remain uncertain. The lower end of this is roughly consistent with our abundance matching result of $\sim10^8M_\odot$, but the higher end suggests a tention. However, such a potential discrepancy between the actual stellar mass and that inferred from abundance matching can be reconciled if LRD hosts have undergone (or is undergoing) an intense star burst activity which brings them significantly above the average stellar-to-halo mass relation. If this is the case, LRD formation may be physically linked to intense star formation (e.g., the formation of a dense nuclear star cluster, see \S \ref{sec:formation_channels}).



\subsection{Spectroscopic signatures and nuclear environment}
\label{sec:spectral_environment}

\paragraph{Weak hard-photon emission.}
A hydrogen-burning SMS on the cool branch has neither a compact accretion flow nor a hot
corona capable of producing a luminous hard-photon component. Moreover,
at the effective temperatures predicted by our models,
$T_{\rm eff}\sim 7000\,{\rm K}$, the Wien tail of the photospheric
spectrum contains essentially no photons capable of ionizing hydrogen,
and still fewer photons above the He\,II or higher ionization edges.
The SMS interpretation therefore predicts intrinsically
weak X-ray emission and weak high-ionization lines such as He\,II and
[Ne\,V]. This is broadly consistent with the X-ray weakness of
most LRDs relative to ordinary broad-line AGNs and with the apparent
deficit of hard ionizing photons inferred from their emission-line
spectra
\citep{2024ApJ...974L..26Y,2025ApJ...995...24K,
2026A&A...708A.293T,
2026A&A...707A..52Z,2026ApJ..1003...10W}.

Neither signature is unique to the SMS model. An obscured or mildly
super-Eddington accreting black hole can also appear intrinsically or
observationally X-ray weak
\citep{2024A&A...691A..52K,2024ApJ...976...96P,
2025ApJ...988L..22I,2026NatAs..10..868L}. Moreover, dense gas
surrounding an accreting black hole can absorb and reprocess the
ionizing continuum before it reaches the lower-density line-emitting
gas \citep{2025ApJ...980L..27I,2026ApJ..1004..153K,
2026arXiv260118864S}. The resulting high-ionization line strengths
depend not only on the intrinsic spectrum, but also on the gas covering
fraction, density, column density, and optical depth
\citep{2026A&A...707A..75T,2026arXiv260317667M,2026arXiv260604711C}.
Nevertheless, persistent luminous X-ray emission or strong
high-ionization lines spatially associated with the unresolved red continuum
source would require an additional hard-photon component in the SMS
interpretation (e.g., an accreting compact object or shocks) and would disfavor a single-component SMS model.
An X-ray-luminous source with otherwise LRD-like properties may
represent such a transitional or multicomponent system
\citep{2026ApJ..1000L..18H, 2026NatAs.tmp..118F}.


\paragraph{A metal-rich nuclear environment.}
The formation channel considered here requires the gas immediately
surrounding the SMS to be substantially enriched, with 
metallicity $Z\gtrsim 10^{-2}$. In particular,
iron-group opacity is important for producing the inflated, cool
envelopes. This is a prediction for the compact SMS-forming reservoir instead of the
galaxy-scale metallicity. Consequently, the low oxygen abundances
inferred from narrow lines in some LRD hosts,
typically $Z_{\rm neb}\sim 0.05$--$0.2\,Z_\odot$ \citep{2026arXiv260631515N}, do not by themselves
contradict the model: these lines may originate in more extended,
lower-density gas that is chemically distinct from the enriched nuclear
component.

Recent ultra-deep rest-UV spectroscopy of four UV-bright LRDs at $z\simeq7$ reveals nitrogen-enhanced gas, electron densities reaching $n_e\sim10^{6}\,{\rm cm^{-3}}$, and neutral columns $N_{\rm HI}\gtrsim10^{22}\,{\rm cm^{-2}}$, with the UV-emitting regions inferred to lie within $\lesssim8$ pc of the nuclei \citep{2026arXiv260911094T}. These observations provide encouraging evidence for compact, dense, and chemically unusual nuclear reservoirs, although nitrogen enhancement does not by itself establish the total metallicity or iron-group abundance required by our models.

The cool SMS cannot itself provide enough ionizing photons to illuminate
this enriched gas. If massive hot stars form contemporaneously from the
same nuclear reservoir, however, their ionizing radiation may produce
nebular lines that probe its composition. The model then predicts that
the compact nebular component should reveal a higher enrichment than
the extended host, potentially including enhanced iron-group emission.
Permitted and forbidden Fe\,II emission has already been reported in
several individual LRDs
\citep{2025ApJ...994L...6T,2026A&A...707A..75T,
2026ApJ..1004..153K,2026MNRAS.545f2235J, 2026arXiv260911094T}.
The presence of these lines alone is not a metallicity measurement,
because their strengths also depend sensitively on density, column
density, and fluorescence, etc. A more decisive
test would require abundance measurements that isolate the compact,
high-density gas from the lower-density host. Demonstrating that this
nuclear component is itself strongly metal-poor, particularly in its
iron-group abundance, would pose a direct challenge to the enriched-SMS
formation channel.

\subsection{A chemical fossil record in nuclear star clusters}
\label{sec:chemical_fossil_record}

A potentially long-lived consequence of the SMS formation channel is
a fossil population of ordinary low-mass stars. If the enriched
nuclear gas required to form an inflated SMS also fragments into
lower-mass objects, stars with initial masses
$\lesssim0.8\,M_\odot$ can survive to the present day. Unlike the
short-lived SMS, their atmospheric abundances may retain information
about the chemical composition of the nuclear reservoir.

However, even in this case, the connection to present-day nuclear star clusters is
statistical. Subsequent mergers, disruptions, and repeated episodes of nuclear star
formation will mix and dilute the original population. Nevertheless,
nuclear star clusters are assembled through a combination of in-situ
star formation and the inspiral of dense stellar systems, and they
commonly contain old populations surrounding massive black holes
\citep{2020A&ARv..28....4N}. It is therefore plausible that some stars
formed in the immediate environments of LRD progenitors are retained
in the nuclear star clusters of their descendants.

The principal signature would be the joint occurrence of old age and
substantial metal enrichment. Stars formed at $z\simeq5$ should now
have ages of $\simeq12\,{\rm Gyr}$, while the coolest SMS models require
a nuclear gas metallicity of order $Z\simeq 10^{-2}$ (or above) with
a significant iron-group abundance. The model therefore motivates a search for very old stars
whose iron abundance is unexpectedly high for their age and whose
multi-element abundance patterns indicate formation in a common,
rapidly enriched nuclear environment.


The nuclear star cluster surrounding Sgr~A$^\ast$ provides the nearest
laboratory for this test. Its stellar population is predominantly old,
with a substantial fraction formed more than $10\,{\rm Gyr}$ ago
\citep{2020A&A...641A.102S}, and its metallicity distribution extends
from $[{\rm M/H}]<-1$ to $[{\rm M/H}]>+0.3$
\citep{2017MNRAS.464..194F}. High-resolution near-infrared spectra of
its brightest giants already yield abundances for $\alpha$ and iron-peak elements
\citep{2025ApJ...979..174R,2025ApJ...982L..14N}. These measurements
demonstrate the feasibility of chemical archaeology in the Galactic
center, but current samples remain small and are dominated by luminous
cool giants.

The sensitivity and adaptive-optics-assisted angular resolution of
next-generation 30--40m telescopes should extend Galactic-center
spectroscopy from the brightest giants toward solar-mass stars in the
crowded nuclear cluster \citep{2014CQGra..31x4007S}. Particularly
valuable targets for the ELT would be stars near the old main-sequence
turnoff and on the early subgiant branch.



\subsection{Time-dependent behavior}
\label{sec:time_dependent_behavior}


The models presented in this work are hydrostatic and therefore do not
predict the time-dependent behavior of the inflated envelope. The
proximity to the Eddington limit, near-sonic convection, and strong
opacity variations associated with the Fe bump and hydrogen
recombination can nevertheless lead to stochastic fluctuations, global
pulsations, or even episodic mass loss.


A useful reference timescale is the dynamical time evaluated at the
photosphere. Using $L_\star\approx L_{\rm Edd,T}\propto M_\star$ and
$L_\star=4\pi R_{\rm ph}^{2}\sigma_{\rm SB}T_{\rm eff}^{4}$ gives
\begin{equation}
\begin{split}
    t_{\rm dyn}
    &\equiv
    \left(\frac{R_{\rm ph}^{3}}{GM_\star}\right)^{1/2} \simeq
    2.5\,{\rm yr}\,\left(\frac{M_\star}{10^{6}\,M_\odot}\right)^{1/4}\\
    &
    \times \left(\frac{\kappa_{\rm T}}
               {0.34\,{\rm cm^{2}\,g^{-1}}}\right)^{-3/4}
    \left(\frac{T_{\rm eff}}{7000\,{\rm K}}\right)^{-3}.
\end{split}
\label{eq:sms_dynamical_timescale}
\end{equation}
The weak mass dependence follows from the Eddington scaling, while the
strong temperature dependence reflects the rapid increase of
photospheric radius toward lower $T_{\rm eff}$. For a source at
$z\simeq5$, the corresponding observed-frame time is
$(1+z)t_{\rm dyn}\sim15\,{\rm yr}$ for $T_{\rm eff}\sim 7000\rm\, K$.

Eq.~(\ref{eq:sms_dynamical_timescale}) should interpreted as the variability timescale --- for $T_{\rm eff}\sim 5000$--$7000\rm\, K$, we expect LRDs to show photometric variability on timescales of a few years to a decade in the rest frame. If the star undergoes global pulsation, then the period would be longer by the angular frequency a factor of $2\pi$, although the eigenfrequencies and growth rates depend on the nonadiabatic structure of the envelope. Nonadiabatic calculations of related quasi-star envelopes find unstable fundamental
radial modes with periods of $20$--$180\,{\rm yr}$ in the rest frame, longer than the surface dynamical time \citep{2026ApJ..1000L...4C}.

Pulsations may also produce shocks or episodic mass ejection without developing into a persistent wind.
Pulsational instabilities have been found in supergiant protostars and metal-rich
very massive stars \citep{2013MNRAS.431.3036I,2014arXiv1403.7241G}, and recent
time-dependent SMS calculations show that late strange-mode episodes
can eject compact circumstellar shells while leaving most of the
stellar mass intact \citep{2026ApJ..1006L..21N}. The accretion
histories and metallicities of those models differ
substantially from ours, so neither the ejected mass nor the
recurrence time is presently predictable for the cool SMS solutions.

Existing variability measurements are consistent with slow
photospheric evolution, but do not yet provide a distinctive test.
Multi-epoch imaging of several hundred LRDs finds little
population-wide variability, with tentative detections confined to a
small minority of objects \citep{2025ApJ...985..119Z}. The TWINKLE
program likewise detects no significant photometric, H$\alpha$-flux,
or line-profile variability among 18 LRDs over rest-frame baselines of
approximately $140$--$220\,{\rm d}$
\citep{2026arXiv260413000L}. These baselines are much shorter than the
fiducial dynamical time in
eq.~(\ref{eq:sms_dynamical_timescale}). Moreover, the absence of
rest-frame UV variability \citep{2025ApJ...983L..26T} is not a direct
constraint on the SMS in our picture, because the UV
continuum may be dominated by surrounding hot stars rather than the
unresolved SMS photosphere.

Evidence for variability on longer timescales is emerging. The multiply imaged LRD candidate R2211-RX1 shows intrinsic brightness and color differences of up to $0.7$ mag across lensing time delays spanning $130$ yr in the source frame ($z=4.3$), consistent with changes in the temperature of an extended photosphere \citep{2025arXiv251205180Z}. Additionally, J1717+3807 ($z=0.196$) exhibits coherent fading by $\simeq0.12$--$0.16$ mag in the WISE bands over 11.7 yr in the rest frame, together with weaker variability in the $i$ band (containing H$\alpha$) and no significant change in $r$ or $g$ (likely due to more contributions from host stars) \citep{2026arXiv260521574L}. These observations suggest that some LRDs likely vary preferentially in their optical or infrared components on decade-to-century timescales.


The most informative test will therefore be long-baseline monitoring
of the unresolved rest-frame optical component, and nearby LRDs may provide particularly useful targets. Correlated changes in optical luminosity, color temperature, and Balmer features could reveal photospheric pulsations or episodic shell
ejection. Time-dependent radiation-hydrodynamic
calculations performed for the metal-enriched envelopes studied
here may provide a robust prediction for the variability of the SMS models.

\subsection{A possible connection to little blue dots}
\label{sec:little_blue_dots}

JWST spectroscopic samples also contain compact broad-line sources
with blue rest-frame UV and optical continua, recently termed
``little blue dots'' (LBDs). LBDs and LRDs share several unusual
properties: both can exhibit broad Balmer lines, weak
X-ray emission, weak hot-dust emission, and little short-timescale
variability. The best-studied examples nevertheless differ in their
line-of-sight gas properties and ionization states. In particular,
prominent Balmer absorption is present in the red prototype but absent
in the blue prototype, while He\,II emission is substantially
stronger in the latter
\citep{2026arXiv260122214B}. This phenomenological
continuity raises the possibility that at least some LRDs and LBDs
represent related physical objects rather than entirely distinct
populations.

Our SMS calculations provide one possible interpretation in which the
red-to-blue sequence is intrinsic rather than caused by line-of-sight
obscuration \citep{2026arXiv260222386M}. At fixed stellar mass, the luminosity remains close to $L_{\rm Edd,T}$ and depends only weakly
on metallicity, whereas the photospheric temperature changes
substantially. At $Z\sim10^{-3}$--$10^{-4}$, the weaker opacity bump produces less
inflation, and our $10^{5}$--$10^{6}\,M_\odot$ models instead reach
$T_{\rm eff}\sim1.5\times10^{4}$--$6\times10^{4}\,{\rm K}$
(Fig.~\ref{fig:sms_hr_combined}). Lower-metallicity SMSs could
therefore appear as compact blue sources with similar luminosities to
LRDs. In this sense, some LBDs may be the lower-metallicity cousins of
LRDs.

This interpretation makes several qualitative predictions. At a given
luminosity, the nebular gas surrounding SMS-powered LBDs
should show lower metallicities or lower iron-group
abundances than the environments of LRDs. Their hotter photospheres
would produce many more hydrogen-ionizing photons than the cool LRD
models and could consequently generate stronger nebular and
high-ionization lines. In the absence of an
additional accreting compact object, however, the stellar continuum
should remain intrinsically X-ray weak.

The connection should not be applied indiscriminately to the entire
LBD population. Current samples suggest that blue compact broad-line
sources are several times more numerous than LRDs
\citep{2026arXiv260122214B}. If all of them were SMSs with lifetimes
and collapse outcomes comparable to those inferred for LRDs, their
remnants would substantially increase the cumulative number density
estimated in \S\ref{sec:remnant_demographics}, potentially weakening
the agreement with present-day SMBH demographics.
It remains possible that only a subset of LBDs belongs to the SMS sequence, with the broader
observational class containing ordinary accreting black holes or compact star-forming systems.

An important alternative is that LRDs and LBDs contain the same
accreting black-hole engine and differ primarily because of
inclination or the amount of dense gas around the broad-line region
\citep{2026arXiv260122214B,2026arXiv260222386M}. These interpretations would
predict different population-level correlations.

\begin{table*}[t!]
\centering
\caption{Principal predictions of the metal-enriched SMS model and
their observational counterparts. The entries progress from direct
model results to conditional or more speculative consequences.
}
\label{tab:sms_observation_comparison}
\begingroup
\footnotesize
\setlength{\tabcolsep}{4pt}
\renewcommand{\arraystretch}{1.2}
\begin{tabular}{@{}p{0.31\textwidth}p{0.32\textwidth}
                    p{0.30\textwidth}@{}}
\toprule
SMS-model prediction &
Observational comparison &
Caveats \\
\midrule

\textbf{Cool photosphere.}
Metal-rich SMSs with $M\gtrsim10^{4}\,M_\odot$ reach
$T_{\rm eff}\sim7000\,{\rm K}$. &
LRD continua imply model-dependent color temperatures of
$\sim\!3000$--$9000\,{\rm K}$
\citep{2025arXiv251121820D}. &
Non-LTE effects may strongly modify the opacity, leading to a
different continuum shape and Balmer features than
predicted by our LTE models. \\

\textbf{Maximum luminosity.}
The GR instability gives
$L_{\rm max}\simeq(0.5$--$2)\times10^{44}\,{\rm erg\,s^{-1}}$
for nonrotating to moderately rotating SMSs, rising to
$\sim5\times10^{44}\,{\rm erg\,s^{-1}}$ near the mass-shedding limit. &
The LRD luminosity function steeply declines above
$\lambda L_{5100}\simeq10^{44}\mbox{--}3\times10^{44}\,{\rm erg\,s^{-1}}$
\citep{2025arXiv250902662M}. &
The prediction is bolometric; a precise comparison requires the LRD
bolometric correction. Matching the observed scale requires rapidly rotating SMS cores, and a detailed comparison
depends on the (uncertain) distribution of core rotation rates. \\

\textbf{Demographics and hosts.}
A $\sim1\,{\rm Myr}$ lifetime gives
$n_{\rm BH,LRD}\sim 10^{-2}\,{\rm cMpc^{-3}}$ and predicts
$M_{\rm h}\sim10^{10}$--$6\times10^{10}\,M_\odot$ and
$M_\star\sim10^{8}\,M_\odot$ (or slightly lower) at $z\sim5$. &
The remnant density is consistent with the local SMBH abundance.
Clustering and SED estimates broadly favor
$M_{\rm h}\sim\,$a few$\times10^{10}\,M_\odot$ and
$M_\star\sim10^{8}$--$10^{9}\,M_\odot$
\citep{2025ApJ...988..246M,2026arXiv260220247P}. &
Conditional on most LRDs producing massive black holes; black hole mergers,
occupation fractions, wandering SMBHs, and host SED decomposition remain
uncertain. LRD hosts may have elevated stellar masses compared to the average stellar-to-halo mass relation. \\

\textbf{X-ray weakness and nuclear environment.}
A cool SMS is intrinsically X-ray weak and requires a compact,
metal-rich ($Z\gtrsim10^{-2}$) formation site. &
Most LRDs are X-ray weak, but a small fraction of them have detected X-ray emission \citep{2026ApJ..1000L..18H, 2026NatAs.tmp..118F}. Some LRDs contain compact, dense, nitrogen-enhanced gas and Fe\,II emission
\citep{2025ApJ...995...24K,2026arXiv260911094T}. &
Obscuration can also suppress AGN X-rays, and the observed lines do
not directly measure the total metallicity or iron-group abundance. \\

\textbf{Chemical fossil record.}
Co-formed low-mass stars may survive as an ancient, metal-rich
population in present-day nuclear star clusters. & The Galactic nuclear cluster is predominantly old {\catcode`\&=12 \citep[$\gtrsim10\,{\rm Gyr}$,][]{2020A&A...641A.102S}} and metal-rich, extending to super-solar \citep{2017MNRAS.464..194F}. &
Co-formation with the SMS is an assumption; it is not yet established
whether the metal-rich stars are specifically old enough
($\gtrsim12\,{\rm Gyr}$) to have formed during the LRD epoch.\\

\textbf{Variability.}
The dynamical time of the outer envelope is a few years in the rest frame, or
$\sim15\,{\rm yr}$ at $z\sim5$.  &
LRDs show little short-baseline variability, while a lensed candidate
varies across a $\sim130\,{\rm yr}$ rest-frame baseline
\citep{2026arXiv260413000L,2025arXiv251205180Z}. &
The hydrostatic models do not predict pulsation periods, amplitudes,
or mass-loss episodes. \\

\textbf{Little blue dots (LBDs).}
Lower-metallicity SMSs remain hotter,
$T_{\rm eff}\sim10^{4}$--$6\times10^{4}\,{\rm K}$, and may
appear as LBDs. &
LRDs and LBDs share several properties, while LBDs can have bluer
continua and stronger He\,II emission
\citep{2026arXiv260122214B}. &
Only a subset of observed LBDs can belong to this lower-metallicity sequence without
overproducing SMBHs in the present-day universe. \\

\textbf{Collapse transient.}
Ejecta-envelope interaction may produce a
$L\sim10^{47}\,{\rm erg\,s^{-1}}$ rest-UV transient lasting
$\sim1$ year in observer's frame for $z\sim5$. &
No secure counterpart is known; the fiducial event would be optically
bright ($r\sim20$ mag) with an all-sky rate of a few per year if all LRDs are rapidly rotating SMSs. &
Highly speculative; the signal strongly depends on the collapse energy release; the fiducial event assumes a rapidly rotating SMS core. \\

\bottomrule
\end{tabular}
\endgroup
\end{table*}

\subsection{A possible terminal collapse transient}
\label{sec:collapse_transient}

A more speculative prediction concerns the terminal transient produced
when the core of the SMS encounters the GR instability. Depending on the progenitor
mass, composition, and rotation, previous collapse calculations have
obtained explosion or outflow energies spanning
$E_{\rm exp}\sim10^{54}$--$10^{57}\,{\rm erg}$
\citep{1986ApJ...307..675F,2012ApJ...749...37M,
2014ApJ...790..162C,2023MNRAS.523.1629N,
2025ApJ...981..119F}. If fast ejecta are launched from the collapsing
core, they must propagate through the inflated envelope at
$R_{\rm env}\sim10^{16}\,{\rm cm}$. As the envelope is likely much
less massive than the core ejecta, this interaction resembles
supernova ejecta colliding with a low-mass circumstellar shell, even
though the interacting material is formally part of the progenitor.

The characteristic optical depth of an envelope with mass
$M_{\rm env}$ is
\begin{equation}
    \tau_{\rm env}
    \sim
    \frac{\kappa_{\rm env} M_{\rm env}}
         {4\pi R_{\rm env}^{2}}
    \simeq
    16
    \frac{\kappa_{\rm env}}{0.1\,{\rm cm^{2}\,g^{-1}}}
    \frac{M_{\rm env}}{100\,M_\odot}
    \left(\frac{R_{\rm env}}{10^{16}\,{\rm cm}}\right)^{-2},
\label{eq:collapse_envelope_optical_depth}
\end{equation}
where we have taken a fiducial value for the envelope mass $M_{\rm env}\sim 100M_\odot$, appropriate for the cool envelopes in the $Z=10^{-2}$ models. The photon-diffusion and shock-crossing times then satisfy
\begin{equation}
    \frac{t_{\rm diff}}{t_{\rm sh}}
    \sim
    \tau_{\rm env}\frac{v_{\rm sh}}{c},
    \quad
    t_{\rm sh}\sim\frac{R_{\rm env}}{v_{\rm sh}}.
\label{eq:collapse_diffusion_ratio}
\end{equation}
For $v_{\rm sh}\sim0.1c$, the condition for photons to escape before
substantial expansion is $\tau_{\rm env}\lesssim c/v_{\rm sh}\sim10$.
The fiducial envelope roughly satisfies this condition (to within a factor of $\sim$2): depending on the
effective opacity $\kappa_{\rm env}$, $t_{\rm diff}$ is comparable to the shock-propagation time $t_{\rm sh}$. We expect the radiation deposited by the shock to rapidly escape without suffering significant adiabatic losses.

If the shock energy is efficiently thermalized, the radiated energy is
limited approximately by the kinetic energy transferred to the
low-mass envelope,
\begin{equation}
    E_{\rm sh}
    \sim
    \frac{1}{2}M_{\rm env}v_{\rm sh}^{2}
    \sim 9\times 10^{53}\,{\rm erg}
    \frac{M_{\rm env}}{100\,M_\odot}
    \left(\frac{v_{\rm sh}}{0.1c}\right)^2,
\label{eq:collapse_transient_energy}
\end{equation}
which is comparable to or smaller than the explosion energies from the collapsing core.
Since $t_{\rm diff}\sim t_{\rm sh}$, the duration of the emission is roughly given by the photon diffusion time
\begin{equation}
    t_{\rm diff}
    \sim
    60\,{\rm d}{M_{\rm env}\over 100M_\odot} \lrb{R_{\rm env}\over 10^{16}\mr{\,cm}}^{-1}.
\label{eq:collapse_envelope_photon_diffusion_time}
\end{equation}
This gives a characteristic luminosity
\begin{equation}
    L_{\rm sh}
    \sim
    \frac{E_{\rm sh}}{t_{\rm diff}}
    \simeq
    2\times 10^{47}\,{\rm erg\,s^{-1}}
    \left(\frac{v_{\rm sh}}{0.1c}\right)^2
    \frac{R_{\rm env}}{10^{16}\,{\rm cm}}.
\label{eq:collapse_transient_luminosity}
\end{equation}
The emission should initially emerge primarily in the rest-frame UV, with effective temperature $\sim 4\times10^4\rm\, K$ for our fiducial parameters. At $z\simeq5$, the duration is stretched to $t_{\rm obs}=(1+z)t_{\rm sh}\sim1\,{\rm yr}$, while the rest-frame UV continuum is shifted into the observed optical bands. GR collapse of an inflated SMS may therefore produce an exceptionally luminous optical transient lasting about a year in the observer frame. Each transient has peak $r$-band apparent magnitude of $\sim20\rm\, mag$ for our fiducial parameters and $z=5$, and the all-sky rate is of the order a few per year (if all LRDs are rapidly rotating SMSs). However, if the explosion energy is much smaller, $E_{\rm exp}\ll 10^{54}\rm\, erg$ (e.g., for a slowly rotating SMS core), then the shock will rapidly slow down within the envelope and nearly all of $E_{\rm exp}$ will be radiated on the envelope's diffusion timescale, so the luminosity will be given by $L_{\rm sh}\sim E_{\rm exp}/t_{\rm diff}$ --- a dimmer and redder transient will be produced. The detailed light curve and color evolution of such transients will require radiation-hydrodynamic calculations of shock propagation through the inflated envelope. Another complication is that the UV emission from the collapsing transient may be obscured by dust along the line of sight, and the absorbed energy will be re-radiated at longer wavelengths and on longer timescales.

\section{Discussion}
\label{sec:discussion}

\subsection{Atmosphere and line-formation uncertainties}
\label{sec:atmosphere_line_formation}


Our calculations determine the global structure of the inflated
envelope using LTE Rosseland-mean opacities and an approximate
photospheric boundary condition. They do not solve the
frequency-dependent, non-LTE transfer problem required to predict the
continuum shape, the Balmer break, or individual line profiles.
A photosphere with $T_{\rm eff}\sim7000\,{\rm K}$ produces very few photons above the hydrogen Lyman edge. Broad Balmer
emission therefore does not follow directly from steady photoionization
by the cool photospheric continuum. Nevertheless, two mechanisms may
produce Balmer lines without requiring the bulk of the atmosphere to be
hot.

\paragraph{(1) Ly$\alpha$ trapping and Balmer-continuum reprocessing.}
In LTE, the excited-state population of approximately neutral hydrogen
is strongly suppressed near $T\sim 7000\rm\, K$,
\begin{equation}
    \left(\frac{n_2}{n_1}\right)_{\rm LTE}
    =
    4\exp\left(-\frac{10.2\,{\rm eV}}{k_{\rm B}T}\right)
    \simeq 2\times10^{-7}.
\label{eq:n2_n1_LTE}
\end{equation}
In a dense atmosphere with a very large Ly$\alpha$ optical depth,
however, repeated resonant absorption and re-emission can maintain an
$n=2$ population far above its LTE value \citep[e.g.,][]{2025ApJ...984..175D}.
Ly$\alpha$ trapping has been invoked to explain the substantial $n=2$
columns inferred from Balmer line absorption in LRDs and other dense
nuclear environments \citep{2024ApJ...963..129M,2024MNRAS.535..853J}.
Other studies likewise associate the Balmer line absorption and break with
dense gas, although through different excitation physics
\citep{2025ApJ...980L..27I,2025A&A...701A.168D}.

Once a significant $n=2$ population is present, collisional excitation from the $n=2$ becomes efficient. For instance, the $n=2\rightarrow 3$ transition only has an energy barrier of $1.9\,{\rm eV}$. Photons shortward of the Balmer edge can also photoionize hydrogen from the first excited state, for which the threshold energy is only $3.4\,{\rm eV}$. Such photons are abundant in a $7000\,{\rm K}$ radiation field. Recombination into higher excited
states also produces Balmer line photons through the subsequent cascade.
Thus, Balmer continuum absorption and Balmer-line production are physically coupled: the same photoionization-recombination cycle that generates a Balmer break can redistribute part of the absorbed continuum energy into Balmer lines. This provides a possible explanation for the coexistence of a strong Balmer break, Balmer absorption, and Balmer emission in the same source.

An elevated $n=2$ population directly increases the Balmer bound-free opacity. This may produce a modest increase in the Rosseland-mean opacity and the photospheric radius. We therefore regard the $T_{\rm eff}\sim7000\,{\rm K}$ endpoint of our LTE cool-branch models as uncertain. Forward LTE calculations of low-density,
optically thick atmospheres independently show that LRD-like red
continua and Balmer breaks can be produced at
$T_{\rm eff}\simeq4000$--$6000\,{\rm K}$
\citep{2025ApJ...994..113L,2026arXiv260302317L}.
These calculations establish the plausibility of cooler
atmospheres, but demonstrating such a shift in our SMS solutions will
require frequency-dependent non-LTE transfer coupled
self-consistently to the global structure.


If the continuum absorption opacity is smaller than that of electron scattering (even with a contribution from the Paschen continuum that depends on the $n=3$ population), Balmer line photons may undergo many electron scatterings before escaping. For the thermal speed adopted here, $v_{\rm e}\sim\sqrt{3\kB T/\me}\simeq
560\mr{\,km\,s^{-1}}\,(T/7000\mr{\,K})^{1/2}$, based on the random walk picture in frequency space, the characteristic linewidth from electron scattering is
\begin{equation}
    \Delta v_{\rm es}\sim \tau_{\rm s} v_{\rm e}\sim 2800\mr{\, km/s}\, {\tau_{\rm s}\over 5} \lrb{T\over 7000\mr{\,K}}^{1/2},
\end{equation}
where $\tau_{\rm s}$ is the scattering optical depth from the Balmer
thermalization layer to the surface. For comparison, the broad H$\alpha$ components in the EIGER/FRESCO sample have $\Delta v_{\rm FWHM}\sim1200$--$3700\,
{\rm km\,s^{-1}}$, while an exceptionally bright RUBIES LRD reaches
$\sim4000\,{\rm km\,s^{-1}}$
\citep{2024ApJ...963..129M,2025ApJ...984..121W}.
Thus, scattering depths of several to ten could in principle produce
wings on the observed velocity scale
\citep{2026MNRAS.545f2131C,2026arXiv260118864S}.


The emergent balance between Balmer absorption and emission line depends on the density profile, velocity gradient, Ly$\alpha$ escape probability, and the coupling of the
$2s$ and $2p$ levels. Existing non-LTE atmosphere calculations
demonstrate that Balmer emission and absorption can coexist in some
SMS-like atmospheres, but other calculations find tension between
producing a pronounced Balmer break and strong emission lines
simultaneously \citep{2026ApJ...998..124N,2026arXiv260512141M}.
The effectiveness of this mechanism in the metal-enriched, low-density
envelopes studied here therefore remains to be established.

\paragraph{(2) Shock-powered ionization.}
A second possibility is local ionization by shocks near the
photosphere. In our cool solutions, radiative diffusion suppresses
convection in the outer layers near the surface (see Fig.~\ref{fig:limiter_envelope}). Our effective
Eddington limiter represents the mean influence of the resulting
inhomogeneity but does not capture its time-dependent dynamics.
Multidimensional radiation-hydrodynamic simulations of related
radiation-dominated stellar envelopes find near-sonic turbulence,
strong density fluctuations, and shocks when radiative diffusion
becomes faster than convective transport
\citep{2015ApJ...813...74J,2018Natur.561..498J}.

For illustration, the post-shock temperature of a strong
gas-pressure-dominated shock is
\begin{equation}
    T_{\rm sh}
    \simeq
    \frac{3\mu m_{\rm p}v_{\rm sh}^{2}}{16k_{\rm B}}
    \simeq
    1.4\times10^{5}\,{\rm K}
    \frac{\mu}{0.6}
    \left(\frac{v_{\rm sh}}{100\,{\rm km\,s^{-1}}}\right)^2.
\label{eq:photospheric_shock_temperature}
\end{equation}
Trans-sonic shocks with velocities of tens to hundreds of
${\rm km\,s^{-1}}$ could therefore heat and partially ionize localized
regions even while the mean photospheric temperature remains near
$7000\,{\rm K}$. As the shocked gas cools, recombination can produce
Balmer emission. Repeated shocks could maintain a small ionized fraction with a significant emission measure. The estimate
in eq.~(\ref{eq:photospheric_shock_temperature}) is only illustrative,
because radiation pressure, photon diffusion, and non-equilibrium
ionization will modify the shock structure. A related warm,
collisionally excited outer layer has been proposed as a source of
Balmer emission and scattering wings in LRDs
\citep{2026A&A...707A..75T}.

However, near-photospheric shocks do not directly produce Balmer line widths of thousands of ${\rm km\,s^{-1}}$. Electron scattering through an extended partially ionized atmosphere can broaden the emergent profiles \citep{2026MNRAS.545f2131C,2026arXiv260118864S}.

A predictive calculation will therefore require a non-LTE atmosphere coupled to time-dependent, multidimensional radiation hydrodynamics. Such a calculation must
simultaneously determine the Ly$\alpha$ radiation field, hydrogen
departure coefficients, continuum thermalization depth, and line escape probabilities. Until then, the cool continuum is a robust consequence of the inflated-envelope
models, whereas the strengths, decrements, and profiles of the Balmer
lines should be regarded as plausible but not yet quantitative
predictions of the SMS interpretation.

\subsection{Possible formation channels}
\label{sec:formation_channels}

The present calculations describe the structure and evolution of an
SMS after it has assembled, but do not determine how such an object
forms. Much of the early literature on SMSs and direct-collapse black
holes focused on primordial atomic-cooling halos, in which molecular
hydrogen cooling is suppressed and the nearly metal-free gas can
collapse without extensive fragmentation
\citep[e.g.,][]{2003ApJ...596...34B}. The environmental requirements of
this channel do not naturally match the observed LRD population. The
LRD abundance peaks near $z\sim5$, substantially later than the first
pristine halos \citep{2026arXiv260700084K}, and their host galaxies are chemically enriched,
although often still subsolar
\citep{2026arXiv260631515N}. More importantly, the
coolest SMS envelopes in this work require a local absolute metal mass
fraction of order $Z\sim10^{-2}$, far above that assumed in primordial
direct-collapse models.

At such metallicities, efficient metal and dust cooling makes the
monolithic collapse of a single cloud into an SMS unlikely: the gas
should instead fragment and form many stars. Radiation-hydrodynamic
calculations indeed find a transition from central SMS growth toward
dense-cluster formation as the metallicity and resulting fragmentation
increase \citep{2025MNRAS.539.2561C}. We suggest that this fragmentation
need not terminate the formation pathway. It may instead first create the
dense stellar system from which an SMS is subsequently assembled.

A plausible channel is therefore the formation of a compact,
metal-enriched nuclear star cluster followed by runaway stellar collisions. Such collisions can transfer mass from already-formed stars to a central merger product, avoiding the
requirement that $\gtrsim10^{6}\,M_\odot$ of enriched gas remain in a
single fragment. Collisional runaway is well established as a possible
growth mechanism in sufficiently compact young clusters
\citep{2004Natur.428..724P,2006MNRAS.368..121F,
2006MNRAS.368..141F}, and related stellar-bombardment models have explored its extension toward SMS masses
\citep{2020ApJ...892...36T,2026ApJ...999..110N}.


Dense nuclear gas and stellar environment may itself promote the retention of the metals needed for SMS formation \citep{2024ApJ...976..166P}. One potential physical mechanism is that, at sufficiently high gas density, individual supernova remnants become radiative and fade into the ambient medium before occupying a large enough volume to overlap. The remnants then deposit momentum locally rather than combining into a coherent superbubble capable of removing the nuclear gas. Even if a coherent superbubble successfully forms, for a sufficiently high stellar density, the stars may inject enough gas in the interior of the superbubble from supernova ejecta and stellar mass loss, which then enhances the cooling rate inside the superbubble.


Cosmological simulations by \citet{2020MNRAS.493.4315M} have produced a bound cluster with \(M_\star\simeq1.4\times10^7\,M_\odot\) in a galaxy of halo mass $7.6\times10^{10}M_\odot$ near \(z\simeq5\), demonstrating that the required cluster mass scale can be assembled during the LRD epoch. Although their subgrid mechanical-feedback prescription accounts for cooling and the terminal momentum of SNe, it does not explicitly resolve the fading and overlap of individual remnants in dense gas; the resulting metal-retention efficiency in this ultra-dense regime therefore remains uncertain.


Establishing whether a metal-enriched dense nuclear cluster can realize the formation of SMSs near $Z\sim10^{-2}$ requires calculations of its gas supply, star formation, relaxation and collision rates, and mass loss. These questions are beyond the scope of this work and we treat the formation rate of metal-rich SMSs as a major uncertainty of the model.

We also note recent simulations by \citet{2026arXiv260818249C} showed that SMSs of $\sim10^5M_\odot$ may form in self-gravitating massive AGN accretion disks. Although such a scenario is inconsistent with the environmental constraints for the observed LRDs, since the AGN disk is highly metal enriched with $Z\gtrsim$a few$\times 10^{-2}$, we expect the outer envelopes of these SMSs to be highly inflated and that they will show similar properties as LRDs with red continuum and broad Balmer lines.

\subsection{Future work}
\label{sec:future_work}

Three developments are particularly important for turning the SMS
interpretation into a more predictive model.

First, multidimensional
radiation-hydrodynamic calculations are needed to determine the
structure of the Fe opacity-inflated envelope, especially its formally
super-Eddington outer layers. Such calculations could establish
whether porosity, turbulent convection, and shocks maintain a
quasi-hydrostatic inflated envelope or instead drive a wind and
episodic mass loss. They would also replace the effective-Eddington
limiter adopted here with a self-consistent description of radiative
transport through an inhomogeneous medium.

Second, the envelope solutions should be coupled to
frequency-dependent, non-LTE atmosphere calculations. Following the
hydrogen level populations and Ly$\alpha$ radiation field is essential
for predicting the color temperature, Balmer break, and the
strengths and profiles of the Balmer lines. Synthetic spectra computed
in this way would permit direct comparison with LRD observations and
determine whether the cool photospheric continuum and broad Balmer
features can arise from a single SMS atmosphere.

Finally, cosmological galaxy formation simulations are needed to connect enriched gas
inflow, nuclear-cluster formation, and runaway stellar collisions to
the observed LRD population. These calculations should determine the
rate and environmental dependence of SMS formation, the metallicity
and halo-mass distributions of their hosts, and the fraction of systems
that reach the GR-instability threshold. Following the subsequent
mergers and dynamical evolution of the remnants would then provide a
self-consistent prediction for the present-day SMBH mass function,
occupation fraction, and population of wandering massive black holes.

\section{Conclusions}
\label{sec:conclusions}

We have explored the hypothesis that at least a subset of little red
dots (LRDs) are thermally relaxed, hydrogen-burning supermassive stars (SMSs) rather
than accreting black holes. We constructed stellar models consisting
of a radiation-pressure-dominated, CNO-burning convective core, a thin
radiative interior, and an outer envelope with a self-consistent
Rosseland-mean opacity, non-adiabatic convective transport, and a
photospheric closure condition. The luminosity is fixed by the
convective-core boundary and remains close to the Thomson-scattering
Eddington luminosity,
$L_\star\approx (1-\beta_c)L_{\rm Edd,T}$. The observed luminosity of an
LRD therefore maps almost directly onto the mass of the SMS.

The principal result is that metal opacity can produce a
strongly inflated, low-mass envelope around an otherwise
compact SMS. As the temperature decreases through the Fe-opacity
bump, the envelope evolves toward a low-density state regulated near
the local Eddington limit (by the $\Gamma\approx 1$ attractor). The resulting convective envelope can extend over
two orders of magnitude in radius while containing only a small fraction,
$\Delta m_{\rm env}/M\lesssim10^{-4}$, of the stellar mass. Hydrogen
recombination ultimately reduces the opacity and radiative
acceleration, allowing the inflated structure to terminate at a bound,
cool photosphere. In our grid of $10^{4}$--$10^{6}\,M_\odot$ models,
the degree of inflation increases strongly with metallicity. Models
with a metal mass fraction $Z\simeq10^{-2}$ reach
$T_{\rm eff}\sim7000\,{\rm K}$ and photospheric radii comparable to
those inferred for LRDs. The effective temperature of this cool branch is insensitive (at a few percent level) to the adopted effective-Eddington limiter and
remains present when the iron-group abundance is reduced by
$10^{-0.5}$ relative to the solar pattern. Lower-metallicity models
remain hotter and more compact, suggesting that some ``little blue dots'' \citep{2026arXiv260122214B} could be lower-metallicity members of the same SMS sequence.

The onset of the general-relativistic instability introduces a maximum
mass and hence a maximum luminosity for the hydrogen-burning SMS
population. For an initial composition $X=0.7$ and
$Z=10^{-2}$, we find threshold masses of
$M_{\rm th}\simeq3.2\times10^{5}\,M_\odot$ for a nonrotating SMS and
$M_{\rm th}\simeq1.2\times10^{6}\,M_\odot$ for a moderately rotating
SMS core ($1/\sqrt{2}$ of the break-up angular frequency). These
thresholds correspond to maximum luminosities of
$4.8\times10^{43}$ and $1.8\times10^{44}\,{\rm erg\,s^{-1}}$,
respectively --- comparable to the luminosity scale of $10^{44}$--$3\times10^{44}\rm\, erg\,s^{-1}$ above which the LRD
luminosity function rapidly declines \citep{2025arXiv250902662M}. SMS cores rotating at the
mass-shedding limit reach $L_{\rm max}\sim5\times10^{44}\,{\rm erg\,s^{-1}}$, roughly an order of
magnitude above the nonrotating threshold.

Hydrogen depletion can also cause an initially stable SMS to cross the
general-relativistic  instability boundary during its main-sequence evolution. The time
required to reach this point is approximately $\tau_{\rm SMS}\sim 1\rm\, Myr$. The cancellation
between the available nuclear fuel and the Eddington luminosity makes
this lifetime nearly independent of stellar mass. The subsequent
nonlinear outcome remains uncertain: relativistic collapse may produce
a black hole retaining most of the stellar mass, whereas sufficiently
rapid nuclear burning can instead cause pulsations or complete
thermonuclear disruption in parts of parameter space
\citep{2012ApJ...749...37M,2025ApJ...981..119F}. Our demographic
results adopt the fiducial black-hole-forming outcome.

Under the assumptions that each observed LRD represents one SMS and
that each SMS leaves a massive black-hole remnant, the measured LRD
abundance \citep{2026arXiv260700084K} implies a cumulative remnant density $n_{\rm dead}(z=0)
    \simeq
    4\times10^{-2}\,{\rm cMpc^{-3}}
    \left(\tau_{\rm SMS}/1\,{\rm Myr}\right)^{-1}$.
Allowing for a factor of a few reduction from mergers among
the descendants gives $n_{\rm BH,LRD}\sim 10^{-2}\,{\rm cMpc^{-3}}$ for a
$1\,{\rm Myr}$ lifetime, comparable to the estimated present-day abundance of
SMBHs above $10^{6}\,M_\odot$
\citep{2004MNRAS.351..169M}. A lifetime of $10\,{\rm Myr}$ would
instead underproduce the local SMBH population. This agreement between the
nuclear-burning lifetime and SMBH demographics is one of the strongest
quantitative results of the SMS interpretation. Reversing the mean
growth histories of the corresponding present-day halos suggests
characteristic abundance-matching LRD host halo masses of
$M_{\rm h}\sim10^{10}$--$6\times10^{10}\,M_\odot$ and stellar masses of $M_*\sim 10^8M_\odot$ (or slightly lower) at $z\sim5$.

Our model is then compared with accretion-powered alternatives. Classical ``quasi-star'' models require a  massive envelope to survive or re-establish hydrostatic equilibrium around a newborn black hole, while ``black hole star'' models begin with an already massive black hole surrounded by a dense reprocessing cocoon \citep{2026ApJ...996...48B,2025arXiv250316596N}. Both may reproduce important features of individual LRDs, but their LRD-visible lifetimes must be set by the envelope survival time or accretion duty cycle. If these populations are to produce a substantial fraction of present-day SMBHs, the demographic comparison favors an LRD-phase duration of order 1 Myr (cumulative);
substantially longer phases would produce too few present-day SMBHs.
``Black hole star'' models must assume the initial seed and
additionally invoke a black hole mass function or a gas-supply distribution
that reproduces the observed LRD upper luminosity cutoff. In the SMS
interpretation, the lifetime, maximum luminosity, and remnant mass instead arise from successive stages of a single stellar evolutionary sequence in a predictable framework.

Our model also makes several qualitative observational predictions.
An isolated cool SMS should be intrinsically weak in X-rays and in
high-ionization lines. Its formation requires a strongly enriched
gas reservoir likely in a dense nuclear star cluster, potentially producing a compact nebular component with higher metallicity (and in particular higher iron-group abundance) than the extended host galaxy. Low-mass stars formed from the same gas could survive as an old, metal-rich fossil population in present-day nuclear star clusters.

LRD spectra and broad Balmer lines are not yet predictions of the present
models. Ly$\alpha$ trapping may enhance the hydrogen $n=2$ population
\citep{2024MNRAS.535..853J, 2025ApJ...984..175D}
and couple Balmer-continuum absorption to recombination-line emission,
while trans-sonic shocks near the photosphere may provide an additional
source of local ionization. Establishing whether either mechanism
reproduces the observed Balmer breaks, Balmer line emission and absorption, and line profiles
requires frequency-dependent, non-LTE atmosphere calculations, which likely needs to be coupled
to multidimensional radiation hydrodynamics. The variability of the
inflated envelope and the possible transient generated when core
ejecta shocks the outer envelope remain more speculative consequences.

Finally, this work does not demonstrate how a
$10^{5}$--$10^{6}\,M_\odot$ star forms at the required metallicity.
Because metal cooling promotes fragmentation, we favor a pathway in
which enriched gas first forms a compact nuclear star cluster and
runaway stellar collisions subsequently assemble the central SMS
\citep{2025MNRAS.539.2561C,2004Natur.428..724P}.

The same extreme nuclear conditions may also help produce the required metal-rich reservoir. We suggest that sufficiently high stellar density and intra-cluster gas density may prohibit the formation or rapid expansion of a volume-filling superbubble. This could strongly suppress the coherent large-scale coupling of supernova momentum and may allow a high fraction of the newly synthesized metals to be retained and recycled \citep{2024ApJ...976..166P}. 
Dense-gas metal retention could therefore be a decisive ingredient enabling the formation of the enriched SMSs required by the LRD interpretation, although this picture should be tested with multidimensional simulations of supernova feedback.

Multidimensional SMS envelope calculations, non-LTE synthetic
spectra, and cosmological simulations of enriched nuclear-cluster
formation and remnant evolution will therefore provide the decisive
tests to our model. Subject to these uncertainties, metal-enriched SMSs offer a
single framework connecting LRDs to the formation of a substantial fraction of the present-day
SMBH population.









\section*{Acknowledgment}

The author is indebted to many conversations along the way with Chris McKee, Liang Dai, and Dan Stark who provided persistent encouragement for this work to be carried out. The author also thanks Xiaohui Fan, Enrico Ramirez-Ruiz, Eliot Quataert, Howard Huang, Masaru Shibata, Yilun Ma, Feige Wang, Kohei Inayoshi, Luis Ho, Massimo Pascale, Sterl Phinney, Jenny Greene, Carl Fields, Savannah Cary, Ryan Chornock, Raffaella Margutti, Dan Kasen, Harley Katz, Shri Kulkarni, Eugene Chiang, Daichi Tsuna whose helpful comments along the way helped steer the direction of exploration. W.L.'s research is supported by a Sloan Research Fellowship (Award Number FG-2026-79505) from the Alfred P. Sloan Foundation and by an LSST Scialog Early Science grant from the Research Corporation for Science Advancement.

\bibliographystyle{mnras}
\bibliography{ref.bib}

\appendix

\section{A. Opacity model and validation}
 \label{app:opacity_model}

\subsection{A1. Hydrogen and helium continuum opacity}
\label{app:HHe_opacity}

We calculate the metallicity-independent H/He contribution to the Rosseland-mean opacity directly from the monochromatic continuum opacity. The calculation assumes LTE, an ideal gas with hydrogen and helium mass fractions \(X=0.7\) and \(Y=0.3\), and Saha ionization equilibrium. Metals are omitted here because their contribution is treated separately. This is motivated by the fact that, at relatively low temperatures $T\lesssim 3\times10^4\rm\, K$ where H/He recombination plays an important role, the metal contribution to the Rosseland-mean opacity becomes subdominant.

\begin{figure*}
    \centering
    \includegraphics[width=\textwidth]{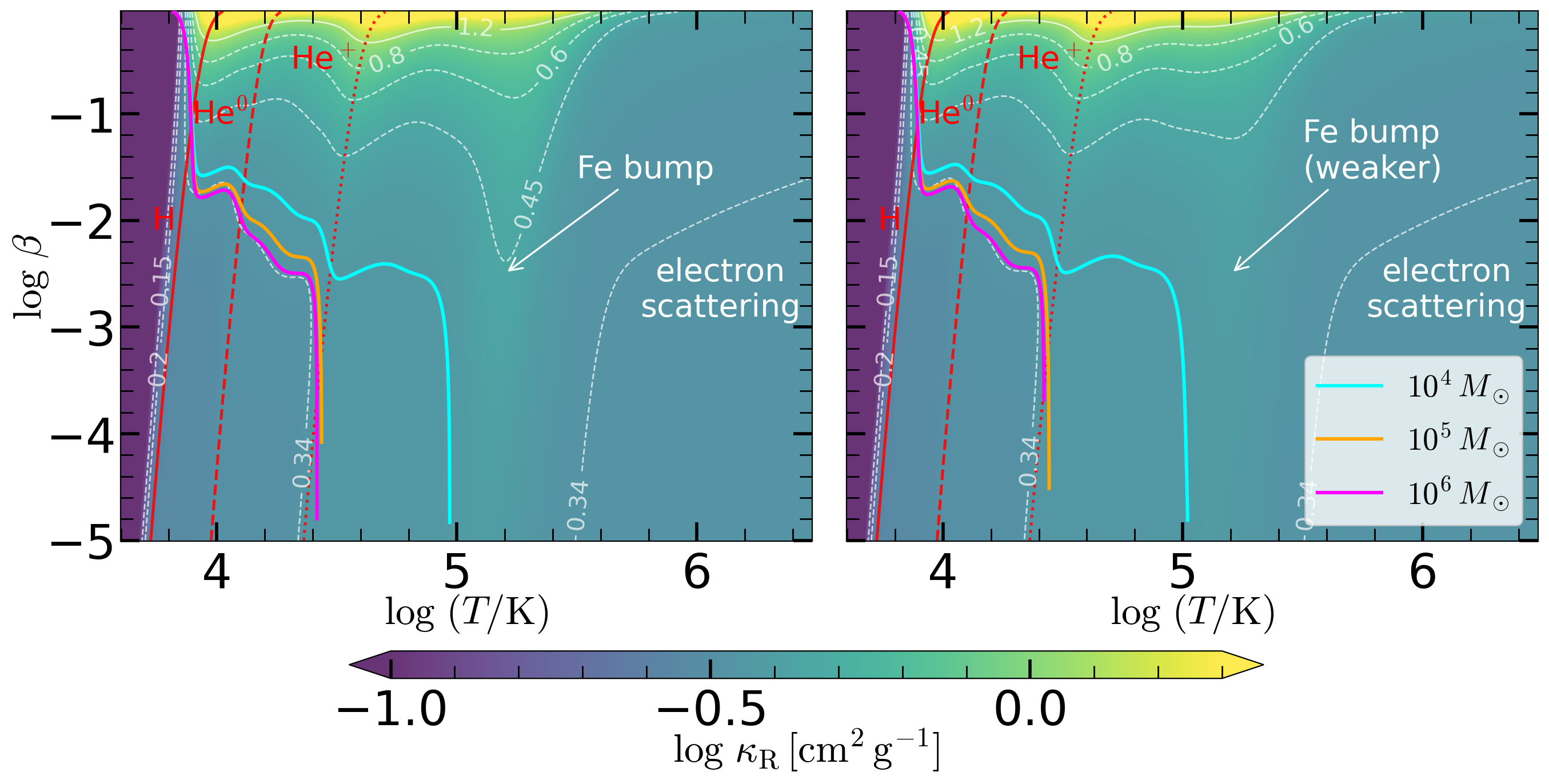}
    \caption{The total Rosseland-mean opacity model
    $\kappa_{\rm R}(T,\beta)$ at $Z=10^{-2.5}$ (Z-2.5), for the
    solar Fe-group abundance pattern ($f_{\rm Fe}=1$; left) and the reduced
    Fe-group abundance pattern ($f_{\rm Fe}=10^{-0.5}$; right). White dashed contours trace constant $\kappa_{\rm R}$ (labeled in ${\rm cm^2\,g^{-1}}$); red curves mark the
    Saha ionization-front boundaries for ${\rm H\to H^+}$
    (solid), ${\rm He^0\to He^+}$ (dashed), and ${\rm He^+ \to He^{++}}$
    (dotted), each for an ionization fraction of $y=0.9$. The iron opacity bump and
    the electron-scattering region at high $T$ are labeled. Colored curves show the photospheric closure
    curves $\beta_{\rm ph}(T_{\rm ph})$ (\S\ref{sec:hse_envelopes}) for three
    stellar masses, $M=10^{4},10^{5},10^{6}\,M_{\odot}$, illustrating
    how the photospheric state at fixed metallicity depends on mass and the weak dependence on the Fe abundance.
    }
    \label{fig:kapR_tot_z3}
\end{figure*}

We denote by \(y_{\rm H}\) the ionized fraction of hydrogen, by
\(y_{\rm He^0}\) the fraction of neutral helium that has undergone the first
ionization, and by \(y_{\rm He^+}\) the fraction of singly
ionized helium that has undergone the second ionization. It is useful to
define
\begin{equation}
    \Phi(T,\chi)
    \equiv
    \frac{(2\pi\me\kB T)^{3/2}}{h^3}
    \exp\left(-\frac{\chi}{\kB T}\right).
\end{equation}
For hydrogen with ionization energy $\chi_{\rm H}=\mr{Ry}=13.6\,{\rm eV}$, the Saha equation becomes
\begin{equation}
    \frac{y_{\rm H}^2}{1-y_{\rm H}}=Q_{\rm H},
    \ 
    Q_{\rm H}=
    \frac{\mproton}{X\rho}\Phi(T,\chi_{\rm H}),
\end{equation}
whose solution is 
\begin{equation}
    y_{\rm H}
    = \frac{2}{1+\sqrt{1+4/Q_{\rm H}}}.
    \label{eq:app_saha_H}
\end{equation}

The two helium ionization stages \({\rm He^0\rightarrow He^+}\) and
\({\rm He^+\rightarrow He^{++}}\) are treated sequentially. This
approximation is accurate at the low envelope densities of interest
because the two transitions have little overlap in
temperature. For the first stage with $\chi_{\rm He^0}=24.6\,{\rm eV}$, the electron density includes the hydrogen ionization fraction, and we define
\begin{equation}
    R_{\rm He^0}=\frac{4Xy_{\rm H}}{Y},
    \ 
    Q_{\rm He^0}
    = \frac{16\mproton}{Y\rho}
    \Phi(T,\chi_{\rm He^0}).
\end{equation}
The factor of 4 in \(Q_{\rm He^0}\) relative to the hydrogenic Saha
coefficient follows from the statistical weights of the
\({\rm He^0}\) and \({\rm He^+}\) ground states. For the second stage ($\chi_{\rm He^+}=54.4\,{\rm eV}$), H and He are taken to be at least singly ionized, and we define
\begin{equation}
    R_{\rm He^+}=1+\frac{4X}{Y},
    \ 
    Q_{\rm He^+}
    =
    \frac{4\mproton}{Y\rho}
    \Phi(T,\chi_{\rm He^+}).
\end{equation}
The helium ionization fractions are then given by
\begin{equation}
    y_j
    =
    \frac{2Q_j}
    {R_j+Q_j+\sqrt{(R_j+Q_j)^2+4Q_j}},
    \label{eq:app_saha_He}
\end{equation}
for $j={\rm He^0}$ or ${\rm He^+}$.
The mean molecular weight (ignoring metals) is
\begin{equation}
    \mu^{-1}
    =
    X(1+y_{\rm H})
    +\frac{Y}{4}
    \left[1+y_{\rm He^0}(1+y_{\rm He^+})\right].
    \label{eq:app_mu_HHe}
\end{equation}
This expression approaches \(\mu=1.29\) for a neutral mixture and
\(\mu=0.615\) for a fully ionized mixture.

At fixed \((T,\beta)\), we solve the
following implicit equation for density $\rho$,
\begin{equation}
    \frac{\rho}{\mu(\rho,T)}
    =
    \frac{\mproton}{\kB T}
    \frac{\beta}{1-\beta}
    \frac{\arad T^4}{3}.
    \label{eq:app_rho_beta_T}
\end{equation}
The solution is bracketed by the neutral and fully ionized limits of
\(\mu\). The resulting \(\rho(T,\beta)\), ionization fractions, and mean
molecular weight are tabulated and interpolated when constructing the
opacity grid.

The monochromatic H/He opacity is written as
\begin{equation}
    \kappa_\nu^{\rm H+He}
    =
    \kappa_{\rm es}
    +\kappa_{\rm bf,\nu}
    +\kappa_{\rm ff,\nu}
    +\kappa_{{\rm H^-},\nu}.
    \label{eq:app_kappa_HHe_nu}
\end{equation}
The electron-scattering contribution follows from the number of free
electrons,
\begin{equation}
    \kappa_{\rm es}
    =
    \frac{\sigma_{\rm T}}{\mproton}
    \left[
        Xy_{\rm H}
        +\frac{Y}{4}y_{\rm He^0}(1+y_{\rm He^+})
    \right].
    \label{eq:app_kappa_es}
\end{equation}
Because \(y_{\rm He^+}\) only becomes appreciable after
\(y_{\rm He^0}\approx 1\), this is equivalent to adding the two helium
ionization stages separately.

For hydrogen and hydrogenic \({\rm He^+}\), the bound-free cross-section
from principal quantum number \(n\) is approximated by
\begin{equation}
    \sigma_{{\rm bf},n}^{(Z)}(\nu)
    =
    \frac{16n}{3\sqrt{3}Z^2}
    \frac{he^2}{\me c\,{\rm Ry}}
    \left(\frac{\nu}{\nu_{n,Z}}\right)^{-3},
    \label{eq:app_sigma_bf}
\end{equation}
with $h\nu_{n,Z}=Z^2{\rm Ry}/n^2$ and for \(\nu\geq\nu_{n,Z}\), and $\sigma_{{\rm bf},n}=0$ below the threshold. Here \(Z=1\) for
hydrogen and \(Z=2\) for \({\rm He^+}\). Note that the charge number here should not be confused with the metal mass fraction. The level population fractions are computed from Boltzmann equilibrium,
\begin{align}
    f_{n,{\rm H^0}}
    &=
    n^2(1-y_{\rm H})
    \exp\left[
        -\frac{{\rm Ry}(1-n^{-2})}{\kB T}
    \right],\\
    f_{n,{\rm He^+}}
    &=
    n^2y_{\rm He^0}(1-y_{\rm He^+})
    \exp\left[
        -\frac{4{\rm Ry}(1-n^{-2})}{\kB T}
    \right].
\end{align}
We retain levels from $n=1$ up to \(n_{\max}=10\).

Neutral helium is not hydrogenic. Its ground-state photoionization
cross-section is therefore taken from the analytic fits of
\citet{1996ApJ...465..487V}. Excited configurations with \(n\geq2\) are
treated approximately as hydrogenic states with an effective charge \(Z_{\rm eff}=1\),
because the inner \(1s\) electron approximately screens one unit of nuclear
charge. Their populations are approximated by
\begin{equation}
    f_{n\geq 2,{\rm He^0}}
    =
    4n^2(1-y_{\rm He^0})
    \exp\left[
        -\frac{\chi_{\rm He^0}-{\rm Ry}/n^2}{\kB T}
    \right],
\end{equation}
where the factor of 4 accounts approximately for the singlet and triplet. Stimulated emission is included in every bound-free term
through the factor \(1-\exp[-h\nu/(\kB T)]\).

The free-free opacity is evaluated using the Kramers form
\begin{equation}
    \kappa_{\rm ff,\nu}
    =
    C_{\rm ff}T^{-1/2}\nu^{-3}
    \left(1-\exp\left[-\frac{h\nu}{\kB T}\right]\right)
    \frac{1}{\rho}
    \sum_i Z_i^2 n_i n_{\rm e},
    \label{eq:app_kappa_ff}
\end{equation}
with \(i={\rm H^+}\), \({\rm He^+}\), and
\({\rm He^{++}}\), and
\begin{equation}
    C_{\rm ff}
    =
    \frac{4\sqrt{\pi}e^6}{3\sqrt{3}\me^2hc}
    \left(\frac{2\me}{\kB}\right)^{1/2}.
\end{equation}
The bound-free and free-free Gaunt factors are set to unity.

We also include continuum absorption by the negative hydrogen ion. Its LTE
number density is
\begin{equation}
    n_{\rm H^-}
    =
    \frac{y_{\rm H}(1-y_{\rm H})}{4}
    \left(\frac{X\rho}{\mproton}\right)^2
    \frac{h^3}{(2\pi\me\kB T)^{3/2}}
    \exp\left(\frac{\chi_{\rm H^-}}{\kB T}\right),
\end{equation}
with $\chi_{\rm H^-}/\kB=8760\,{\rm K}$.
The bound-free photodetachment cross-section is taken from
\citet{1979MNRAS.187P..59W}, while the free-free absorption coefficient is
taken from \citet{1988A&A...193..189J}. Although included for completeness,
\({\rm H^-}\) makes a negligible contribution at the low densities of the
SMS photospheric layers.

\begin{figure*}
    \centering
    \includegraphics[width=0.85\textwidth]{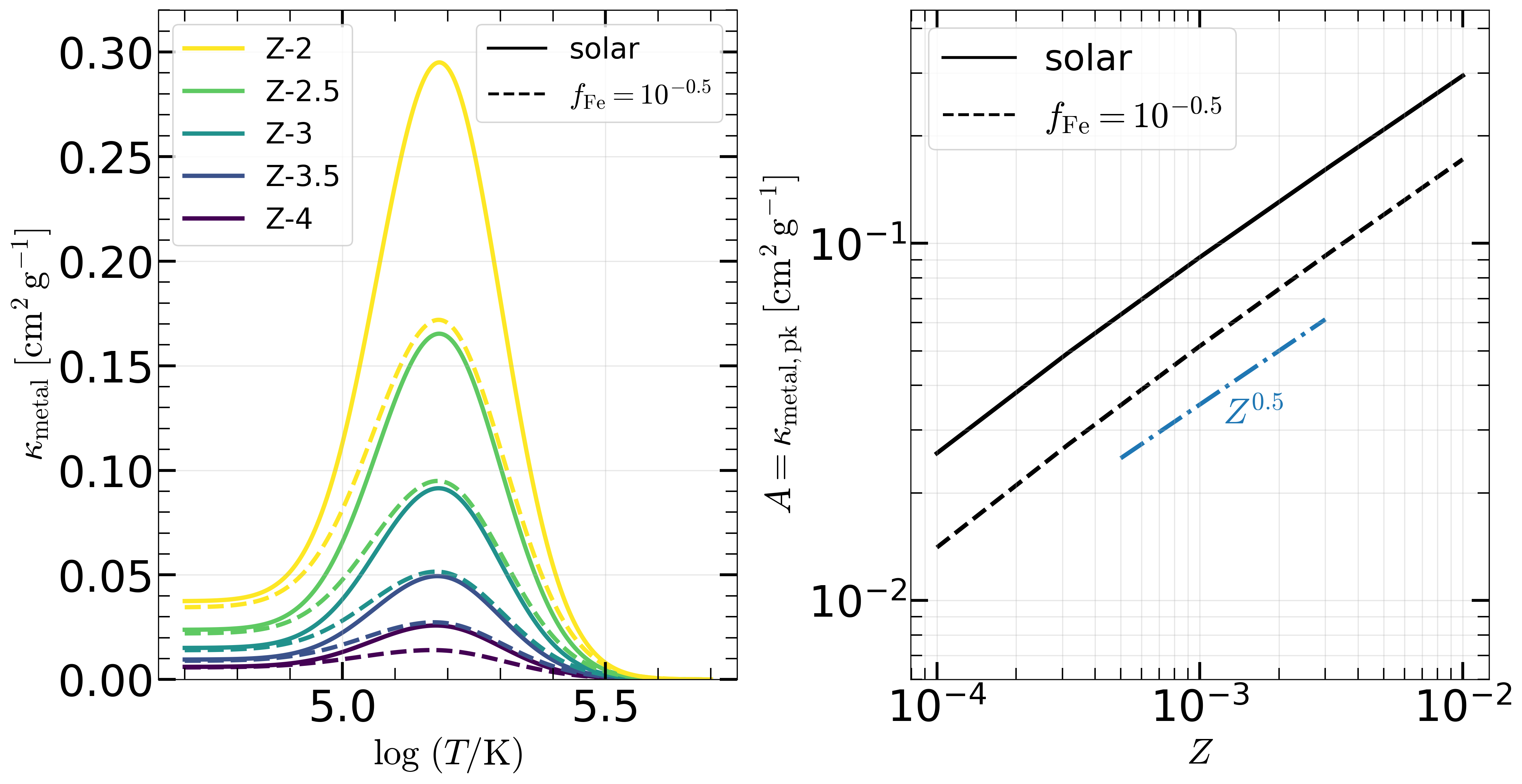}
    \caption{
        The metal contribution to the Rosseland-mean opacity $\kappa_{\rm metal}(T,\beta, Z)$ (\S\ref{sec:opacity}) at different Fe abundances, $f_{\rm Fe}=1$ (solid) and $10^{-0.5}$ (dashed), across five metallicities (Z-4 to Z-2, color). All cases are for a fixed $\beta=10^{-3}$. The bump peak amplitude increases with metallicity roughly as $\kappa_{\rm metal,pk}\propto Z^{0.5}$.
    }
    \label{fig:kap_metal_fem}
\end{figure*}

Finally, the monochromatic contributions in
eq.~\eqref{eq:app_kappa_HHe_nu} are used to compute the
Rosseland-mean:
\begin{equation}
    \frac{1}{\kappa_{\rm H+He}}
    =
    \frac{\pi}{4\sigma_{\rm SB}T^3}
    \int_0^\infty
    \frac{1}{\kappa_\nu^{\rm H+He}}
    \frac{\partial B_\nu}{\partial T}\,{\rm d}\nu.
    \label{eq:app_kappa_R_HHe}
\end{equation}

The principal approximations are LTE/Saha equilibrium, unity Gaunt factors, and the neglect of bound-bound transitions. Molecular hydrogen may
become important below approximately \(5000\,{\rm K}\), so the coolest
edge of the opacity table should be interpreted cautiously. The contribution from bound-bound opacity (mainly Balmer lines) should be minor for the low-density envelopes (unless the $n=2$ or $3$ levels have high non-LTE departure coefficients as discussed in \S \ref{sec:atmosphere_line_formation}).
In the low-density regime relevant to the inflated SMS envelopes, direct evaluation shows that electron scattering dominates through most of the ionized layers, while bound-free absorption becomes at most comparable near hydrogen and helium
recombination temperatures.

\subsection{A2. Metal opacity and the Fe-opacity bump}
\label{app:metal_opacity}

We obtain the metal contribution to the opacity from Rosseland-mean tables
generated through the LANL TOPS interface to the OPLIB opacity
calculations \citep{2016ApJ...817..116C, 2024ApJ...968...56F}. We generate a metal-free table for $X=0.7, Y=0.3, Z=0$ on the same
\((\log T,\log\rho)\) grid. The metal contribution is then isolated by
subtracting the two raw tables
\begin{equation}
    \kappa_{\rm metal}(Z,f_{\rm Fe})
    =
    \kappa_{\rm R}^{\rm TOPS}(Z,f_{\rm Fe})
    -
    \kappa_{\rm R}^{\rm TOPS}(Z=0),
    \label{eq:app_kappa_metal_extract}
\end{equation}
before constructing an interpolating spline.
Unlike the H/He term derived in Appendix~\ref{app:HHe_opacity},
\(\kappa_{\rm metal}\) includes the bound-bound transitions of the metal
ions and therefore captures the iron-group opacity bump.

\begin{figure*}
    \centering
    \includegraphics[width=0.95\textwidth]{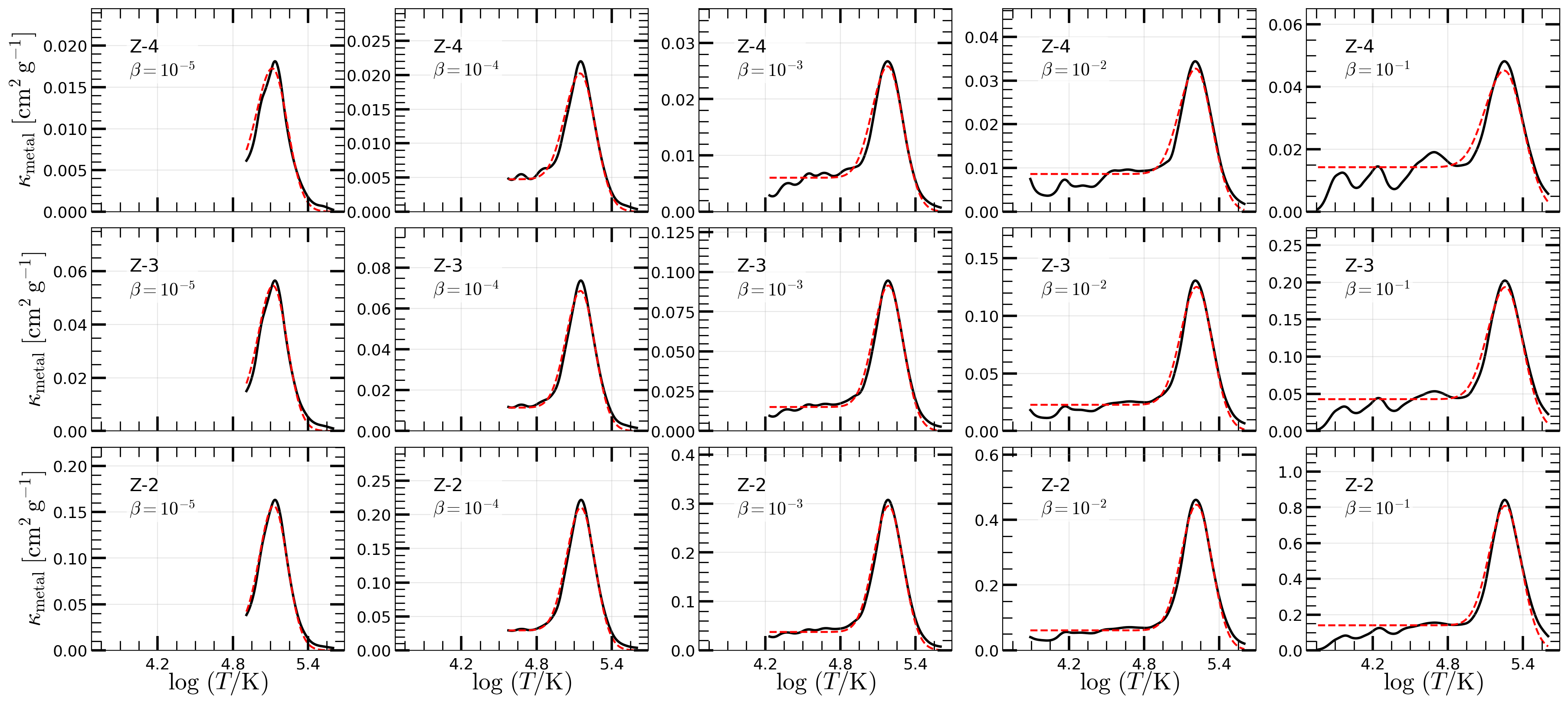}
    \caption{The Gaussian$+$plateau model fit (red dashed,
    \S\ref{sec:opacity}) against the raw metal opacity extracted from the TOPS tables
    ($\kappa_{\rm metal}=\kappa_{\rm R}(Z)-\kappa_{\rm R}(Z=0)$; black
    solid), for three representative metallicities (rows, Z-4, Z-3, Z-2)
    and a grid of gas-pressure fractions (columns,
    $\beta=10^{-5}$ to $10^{-1}$). The fit reproduces the Fe bump amplitude, width,
    and location closely at every $(\beta,Z)$, while the raw table curves show
    residual low-temperature structure that the smooth
    fitting model does not capture. The total opacity $\kappa_{\rm R}$ at such low temperatures is dominated by H and He contributions, so the residual has negligible effect on our results.
    }
    \label{fig:kap_metal_fit}
\end{figure*}

The extracted opacity consists primarily of a broad iron-group bump near
\(\log T\simeq5.2\)--\(5.3\), together with a plateau at lower temperatures.
We represent this structure using the Gaussian+plateau model in eq.~\eqref{eq:kappa_metal_fit},
The parameter \(A\) describes the amplitude of the iron bump, \(T_0\) is its central
temperature, and \(\sigma\) is its logarithmic width. The logistic factor
\(H\) switches on the plateau \(\kappa_{\rm p}\) below the bump while suppressing
it above the bump. Consequently,
\(\kappa_{\rm metal}\rightarrow\kappa_{\rm p}\) well below \(T_0\),
\(\kappa_{\rm metal}=A\) at \(T=T_0\), and the fitted metal contribution
rapidly declines at $T>T_0$.

We determine the four parameters $(A, \log T_0, \sigma, \kappa_{\rm p})$ using a two-stage fitting procedure. In the first stage, we locate the iron-bump maximum over \(5.0\leq\log T\leq5.7\) and fit a three-parameter Gaussian ($A, \log T_0, \sigma$) in a restricted
window around the peak.

In the second stage, we fix \(\log T_0\) and \(\sigma\) from the first stage,
and refit \(A\) and \(\kappa_{\rm p}\) 
over the wider interval
\begin{equation}
    \log T_{\min}(\beta,Z,f_{\rm Fe})
    \leq \log T \leq 5.7.
\end{equation}
At high gas-pressure fractions, the raw metal opacity develops a second bump near \(\log T\simeq4.7\)--\(4.8\) (likely due to CNO), which cannot be represented by our single-bump model (no significant effect on the envelope solutions). We therefore choose \(\log T_{\min}\) adaptively: starting below the iron bump, we identify the opacity minimum and exclude temperatures at which the opacity rises by more than a factor
of 1.5 above this minimum toward the secondary feature. If no such rise is
present, we adopt \(\log T_{\min}=4.5\).


To obtain a smooth opacity function in \(\beta\), we fit each of the 4 model
parameters as a polynomial in
\begin{equation}
    x
    \equiv
    \log\left(\frac{P_{\rm gas}}{P_{\rm rad}}\right)
    =
    \log\left(\frac{\beta}{1-\beta}\right).
    \label{eq:app_x_beta}
\end{equation}
At fixed temperature, \(P_{\rm gas}/P_{\rm rad}\) is approximately
proportional to density, making \(x\) a more physically natural fitting
coordinate than \(\log\beta\).

\begin{figure*}
    \centering
    \includegraphics[width=\textwidth]{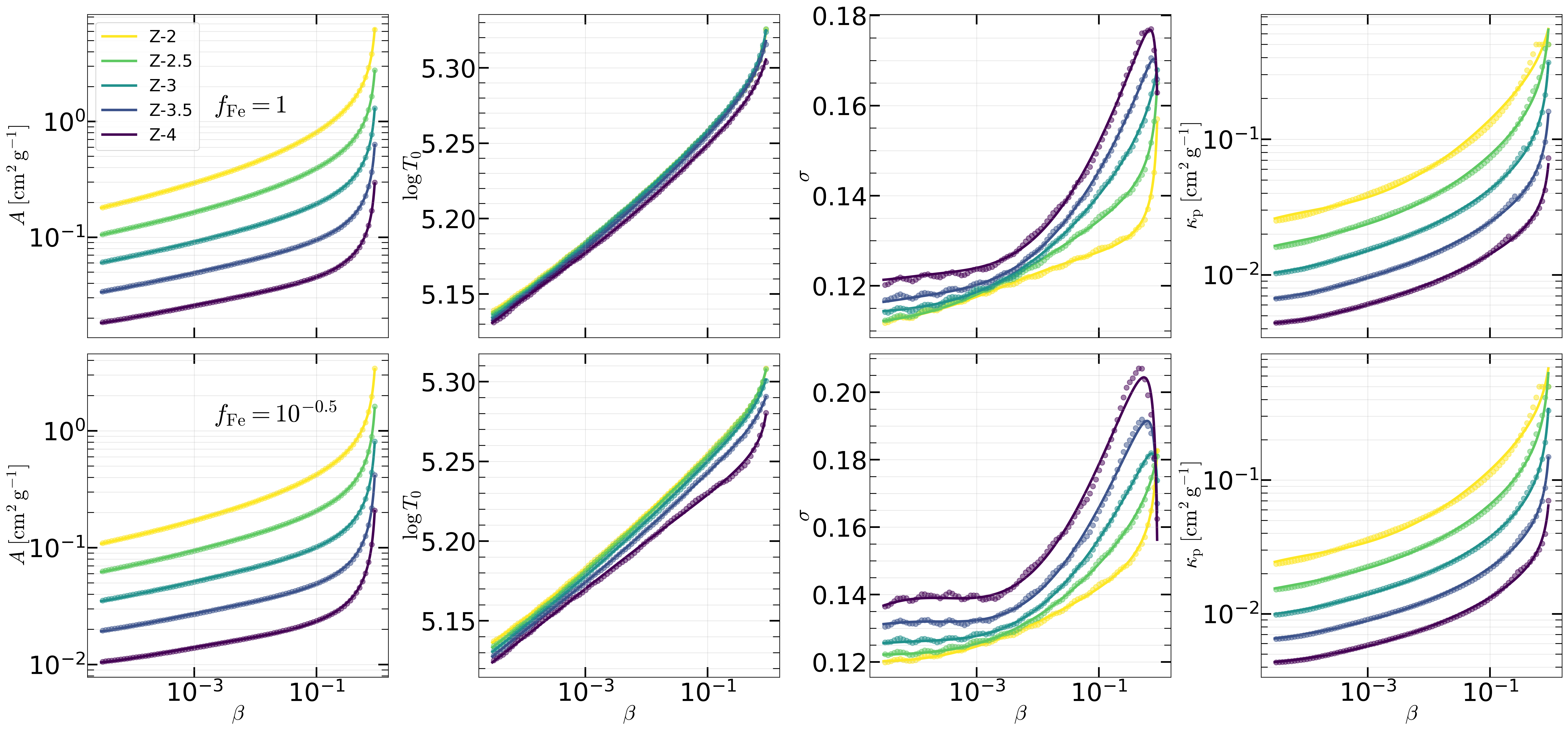}
    \caption{The Gaussian+plateau model parameters
    $(A,\log T_0,\sigma,\kappa_{\rm p})$ (\S\ref{sec:opacity}) versus $\beta$, for
    the metallicity grid, at the solar Fe-group abundance pattern
    ($f_{\rm Fe}=1$; top row) and the reduced Fe-group abundance pattern
    ($f_{\rm Fe}=10^{-0.5}$; bottom row). Dots show independent
    per-$\beta$ fits to the raw TOPS-derived $\kappa_{\rm metal}(
    T,\beta)$ data; curves show the adopted degree-4 (degree-5 for $\sigma$) polynomial fit in
    $x=\log [\beta/(1-\beta)]$, restricted to the domain it is
    used over, $\beta\in[3\times10^{-5},0.9]$. The two agree to within a few percent everywhere, for both abundance patterns.
    }
    \label{fig:kap_metal_params}
\end{figure*}

For each metallicity and Fe-group abundance pattern, the fitted parameters
are written as
\begin{equation}
    p(x)=\sum_{i=0}^{i_{\rm max}}c_{i}x^i,
    \ 
    p\in
    \left\{
        \log A,\,
        \log T_0,\,
        \sigma,\,
        \log\kappa_{\rm p}
    \right\}.
    \label{eq:app_kappa_metal_poly}
\end{equation}
We use \(i_{\rm max}=4\) for \(\log A\), \(\log T_0\), and
\(\log\kappa_{\rm p}\), and $i_{\rm max}=5$ for \(\sigma\) which shows more complex structure (caused by CNO opacity bump interference). The resulting best-fit curves closely follow the independently fitted values at each \(\beta\), as shown
in Fig.~\ref{fig:kap_metal_params}. The polynomial coefficients are
listed in Table~\ref{tab:kap_metal_coeffs}.

We construct fits for five metallicities,
\(Z=10^{-4},10^{-3.5},10^{-3},10^{-2.5}\), and \(10^{-2}\), and the solar abundance pattern and another pattern with reduced Fe-group abundances --- the mass fractions of Cr, Mn, Fe, Co, and Ni are reduced by \(f_{\rm Fe}=10^{-0.5}\), with the removed metal mass redistributed among the remaining metals so that the total \(Z\) is unchanged. As shown in Fig.~\ref{fig:kap_metal_fem}, reducing \(f_{\rm Fe}\) primarily lowers
the opacity bump amplitude, while its location and shape remain nearly unchanged.
Across the metallicity range used in this work, the Fe-bump peak
amplitude is well described by
\begin{equation}
    \kappa_{\rm metal,pk}\propto Z^{\simeq0.5}.
\end{equation}
This sublinear scaling arises because the Fe-opacity bump is produced by
a forest of strong, collisionally broadened lines whose cores are already
saturated relative to the electron-scattering background. For a saturated
Lorentzian line, the effective Rosseland width is set by where its damping
wings fall to the scattering floor and scales with the line-center opacity $\kappa_0$ as \(\Delta\nu_{\rm eff}\propto\kappa_0^{1/2}\); since the line-center opacity
\(\kappa_0\propto Z\), this naturally gives
\(\kappa_{\rm metal,pk}\propto Z^{1/2}\).

The parameter fits are calibrated over \(3\times10^{-5}\leq\beta\leq0.9\) (relevant for our SMS models), but extrapolation down to $10^{-5}$ shows excellent agreement.
The metal-opacity fit itself is intended primarily for
\(\log T\gtrsim4.5\), encompassing the iron bump and the plateau. At lower
temperatures, the TOPS density floor and metal recombination make a direct
extrapolation uncertain. In the final opacity prescription, this
low-temperature contribution is tapered by the hydrogen ionization
fraction \(y_{\rm H}\), as in eq.~\eqref{eq:kappaR_decomp}; the total
opacity there is dominated by the explicitly calculated H/He contributions,
so the envelope solutions are insensitive to the detailed form of this
taper. Fig.~\ref{fig:kap_metal_fit} compares the fitted metal opacity
with the directly extracted TOPS values and shows that the iron-bump
amplitude, location, and width are reproduced to within a few percent
throughout the domain relevant to the envelope calculations.

\begin{table*}
    \centering
    \footnotesize
    \caption{Coefficients of the degree-4 (degree-5 for $\sigma$)
    polynomial fits to the Gaussian+plateau model parameters of
    eq.~\eqref{eq:kappa_metal_fit}, as a function of
    $x=\log\left[\beta/(1-\beta)\right]$, for the solar
    ($f_{\rm Fe}=1$) and reduced ($f_{\rm Fe}=10^{-0.5}$) Fe-group
    abundance patterns across the metallicity grid. $c_{5}$ applies only to $\sigma$
    (degree 5). Fig.~\ref{fig:kap_metal_params} shows the resulting fits
    against the raw per-$\beta$ values.
    }
    \label{tab:kap_metal_coeffs}
    \begin{tabular}{llrrrrrr}
        \hline
        Metallicity & Parameter & $c_5$ & $c_4$ & $c_3$ & $c_2$ & $c_1$ & $c_0$ \\
        \hline
        Z-4 & $\log A$ & N/A & 3.103e-3 & 3.657e-2 & 1.566e-1 & 3.925e-1 & -1.084 \\
        Z-4 & $\log T_0$ & N/A & -2.220e-4 & -2.161e-3 & -5.675e-3 & 3.157e-2 & 5.283 \\
        Z-4 & $\sigma$ & -1.470e-4 & -2.099e-3 & -1.030e-2 & -1.649e-2 & 1.519e-2 & 1.740e-1 \\
        Z-4 & $\log \kappa_{\rm p}$ & N/A & 9.010e-4 & 9.599e-3 & 6.157e-2 & 3.383e-1 & -1.573 \\
        \hline
        Z-3.5 & $\log A$ & N/A & 2.607e-3 & 3.174e-2 & 1.433e-1 & 4.010e-1 & -7.472e-1 \\
        Z-3.5 & $\log T_0$ & N/A & -1.850e-4 & -1.850e-3 & -4.747e-3 & 3.434e-2 & 5.291 \\
        Z-3.5 & $\sigma$ & -8.100e-5 & -1.202e-3 & -6.056e-3 & -9.032e-3 & 1.647e-2 & 1.651e-1 \\
        Z-3.5 & $\log \kappa_{\rm p}$ & N/A & 9.300e-4 & 1.175e-2 & 8.052e-2 & 4.059e-1 & -1.287 \\
        \hline
        Z-3 & $\log A$ & N/A & 1.990e-3 & 2.554e-2 & 1.258e-1 & 4.084e-1 & -4.177e-1 \\
        Z-3 & $\log T_0$ & N/A & -9.200e-5 & -1.039e-3 & -2.668e-3 & 3.598e-2 & 5.294 \\
        Z-3 & $\sigma$ & -1.700e-5 & -2.950e-4 & -1.678e-3 & -1.571e-3 & 1.620e-2 & 1.558e-1 \\
        Z-3 & $\log \kappa_{\rm p}$ & N/A & 8.180e-4 & 1.259e-2 & 9.704e-2 & 4.764e-1 & -9.924e-1 \\
        \hline
        Z-2.5 & $\log A$ & N/A & 1.433e-3 & 1.990e-2 & 1.106e-1 & 4.233e-1 & -8.611e-2 \\
        Z-2.5 & $\log T_0$ & N/A & -4.200e-5 & -6.340e-4 & -1.753e-3 & 3.575e-2 & 5.293 \\
        Z-2.5 & $\sigma$ & 3.200e-5 & 4.100e-4 & 1.823e-3 & 4.325e-3 & 1.413e-2 & 1.453e-1 \\
        Z-2.5 & $\log \kappa_{\rm p}$ & N/A & -1.055e-3 & -4.426e-3 & 5.651e-2 & 4.877e-1 & -7.041e-1 \\
        \hline
        Z-2 & $\log A$ & N/A & 8.940e-4 & 1.472e-2 & 9.940e-2 & 4.540e-1 & 2.624e-1 \\
        Z-2 & $\log T_0$ & N/A & -5.300e-5 & -7.420e-4 & -1.989e-3 & 3.538e-2 & 5.292 \\
        Z-2 & $\sigma$ & 5.300e-5 & 7.390e-4 & 3.635e-3 & 7.729e-3 & 1.168e-2 & 1.342e-1 \\
        Z-2 & $\log \kappa_{\rm p}$ & N/A & -4.129e-3 & -3.685e-2 & -5.188e-2 & 3.834e-1 & -4.701e-1 \\
        \hline
        Z-4\_lowFe & $\log A$ & N/A & 3.866e-3 & 4.564e-2 & 1.961e-1 & 4.525e-1 & -1.340 \\
        Z-4\_lowFe & $\log T_0$ & N/A & 2.600e-5 & 1.470e-4 & -4.640e-4 & 2.622e-2 & 5.255 \\
        Z-4\_lowFe & $\sigma$ & -2.300e-4 & -3.610e-3 & -1.939e-2 & -3.707e-2 & 6.161e-3 & 2.042e-1 \\
        Z-4\_lowFe & $\log \kappa_{\rm p}$ & N/A & 4.160e-4 & 6.486e-3 & 6.157e-2 & 3.545e-1 & -1.590 \\
        \hline
        Z-3.5\_lowFe & $\log A$ & N/A & 3.549e-3 & 4.210e-2 & 1.823e-1 & 4.456e-1 & -1.018 \\
        Z-3.5\_lowFe & $\log T_0$ & N/A & -1.660e-4 & -1.795e-3 & -5.916e-3 & 2.657e-2 & 5.272 \\
        Z-3.5\_lowFe & $\sigma$ & -1.790e-4 & -2.707e-3 & -1.408e-2 & -2.504e-2 & 1.228e-2 & 1.902e-1 \\
        Z-3.5\_lowFe & $\log \kappa_{\rm p}$ & N/A & 5.290e-4 & 9.513e-3 & 8.315e-2 & 4.216e-1 & -1.319 \\
        \hline
        Z-3\_lowFe & $\log A$ & N/A & 2.952e-3 & 3.576e-2 & 1.609e-1 & 4.376e-1 & -6.970e-1 \\
        Z-3\_lowFe & $\log T_0$ & N/A & -1.570e-4 & -1.776e-3 & -5.900e-3 & 2.844e-2 & 5.281 \\
        Z-3\_lowFe & $\sigma$ & -9.400e-5 & -1.470e-3 & -7.812e-3 & -1.295e-2 & 1.612e-2 & 1.780e-1 \\
        Z-3\_lowFe & $\log \kappa_{\rm p}$ & N/A & 7.890e-4 & 1.357e-2 & 1.065e-1 & 4.889e-1 & -1.044 \\
        \hline
        Z-2.5\_lowFe & $\log A$ & N/A & 2.285e-3 & 2.878e-2 & 1.392e-1 & 4.364e-1 & -3.742e-1 \\
        Z-2.5\_lowFe & $\log T_0$ & N/A & -9.000e-5 & -1.207e-3 & -4.480e-3 & 2.954e-2 & 5.285 \\
        Z-2.5\_lowFe & $\sigma$ & -1.400e-5 & -2.920e-4 & -1.818e-3 & -1.623e-3 & 1.858e-2 & 1.669e-1 \\
        Z-2.5\_lowFe & $\log \kappa_{\rm p}$ & N/A & -2.780e-4 & 4.055e-3 & 8.589e-2 & 5.099e-1 & -7.691e-1 \\
        \hline
        Z-2\_lowFe & $\log A$ & N/A & 1.668e-3 & 2.260e-2 & 1.227e-1 & 4.520e-1 & -3.764e-2 \\
        Z-2\_lowFe & $\log T_0$ & N/A & -7.900e-5 & -1.163e-3 & -4.545e-3 & 2.874e-2 & 5.285 \\
        Z-2\_lowFe & $\sigma$ & 4.000e-5 & 5.340e-4 & 2.571e-3 & 6.878e-3 & 1.903e-2 & 1.557e-1 \\
        Z-2\_lowFe & $\log \kappa_{\rm p}$ & N/A & -3.232e-3 & -2.653e-2 & -1.342e-2 & 4.220e-1 & -5.302e-1 \\
        \hline
    \end{tabular}
\end{table*}

 \subsection{A3. Combined opacity model and validation}

\begin{figure*}
    \centering
    \includegraphics[width=0.85\textwidth]{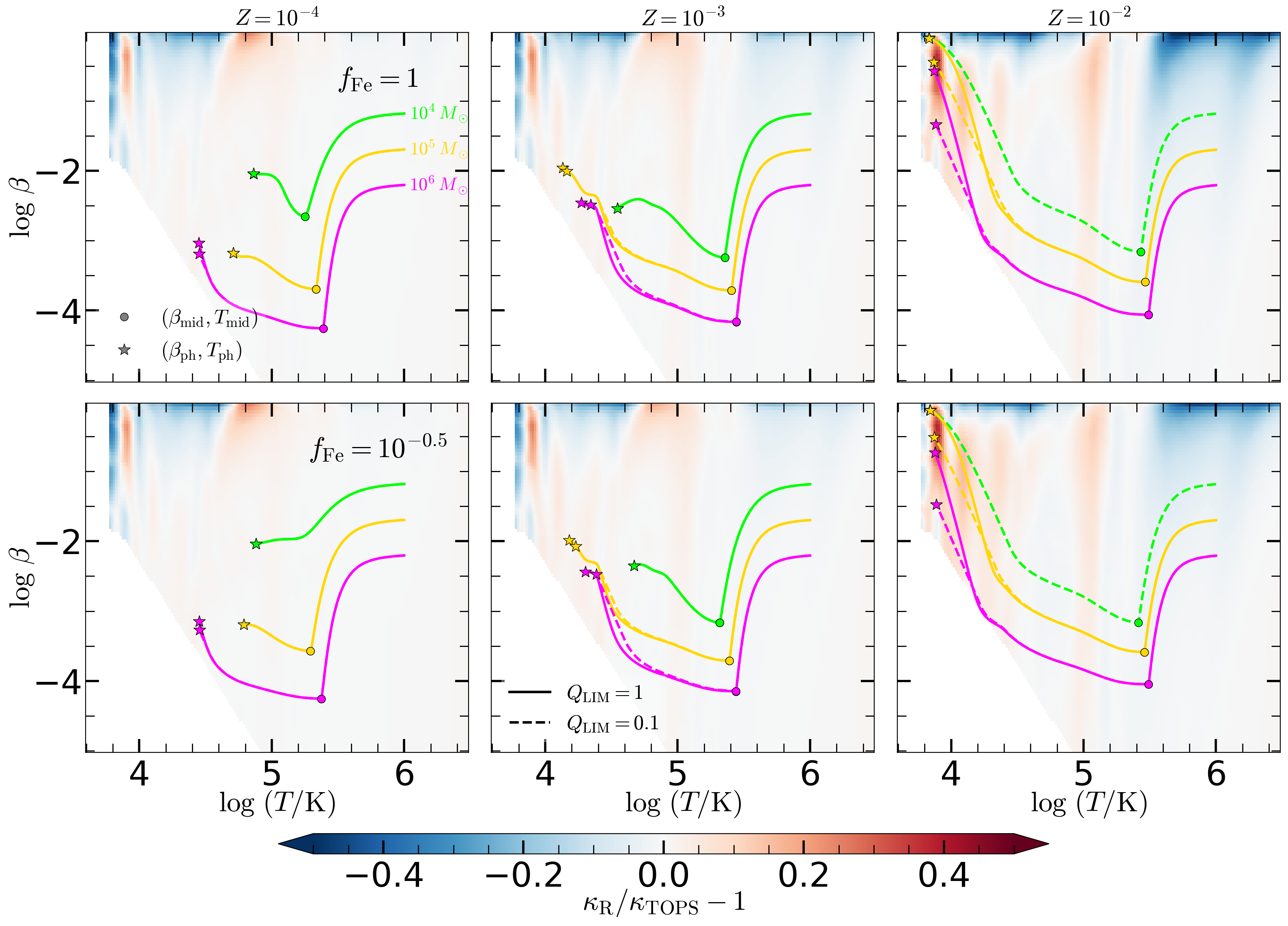}
    \caption{Fractional residual $\kappa_{\rm R}/\kappa_{\rm TOPS}-1$
    between the opacity model of this work and the raw TOPS
    tables, over the $(\log T,\log \beta)$ plane, for three representative
    metallicities ($Z=10^{-4},10^{-3},10^{-2}$; columns) and two Fe-group
    abundance patterns ($f_{\rm Fe}=1$, top row; $f_{\rm Fe}=10^{-0.5}$,
    bottom row). Colored curves show the converged envelope solutions
    (\S\ref{sec:hse_envelopes}) for $M=10^{4},10^{5},10^{6}\,M_\odot$
    (green, gold, magenta) at effective-Eddington limiters
    $Q_{\rm LIM}=1$ (solid) and $0.1$ (dashed), tracing the $(\log
    T,\log \beta)$ path for each model from its base to its
    photosphere; circles mark the inner transition-line crossing
    $(\beta_{\rm mid},T_{\rm mid})$ that anchors the two-point shooting
    method and stars mark the resulting photospheric state
    $(\beta_{\rm ph},T_{\rm ph})$. The two
    $M=10^{4}\,M_\odot$, $Q_{\rm LIM}=1$ cases at $Z=10^{-2}$ (both Fe
    patterns) have no converged photospheric solution and are omitted. The
    fractional opacity residual stays small ($\lesssim$10\%) along most parts of the tracks in the deep interior of the envelope; for $Z=-2$ solutions that reaches $\beta\gtrsim 0.1$ near the cool photosphere at $T\lesssim 10^4\rm\, K$, the maximum opacity residual reaches up to $\sim30\%$, likely due to larger errors in our model or inaccurate interpolation of the TOPS table near the grid boundary.
    }
    \label{fig:kap_tot_comp_residual}
\end{figure*}

The full Rosseland-mean opacity $\kappa_{\rm R}(T,\beta,Z)$ used throughout this work combines the H/He continuum opacity of \S\ref{app:HHe_opacity} with the metal-opacity fit of \S\ref{app:metal_opacity}, following the taper prescription of eq.~\eqref{eq:kappaR_decomp}. While
Fig.~\ref{fig:kap_metal_fit} validated the metal-opacity fit against the
raw TOPS-derived $\kappa_{\rm metal}$ in isolation, Fig.~\ref{fig:kap_tot_comp_residual}
tests the fully assembled model $\kappa_{\rm R}$ directly against the raw
TOPS tables over the entire $(\log T,\beta)$ domain used in this paper.
The fractional residual is below the 10\% level
across the region actually sampled by the converged envelope solutions
(\S\ref{sec:hse_envelopes}). An exception is for the higher density ($\beta\gtrsim 0.5$) regions near the cool photosphere at $T\lesssim 10^4\rm\, K$ for the $Z\simeq 10^{-2}$ models, where our approximate treatment for He$^0$ bound-free opacity and collisionally broadened bound-bound transitions of hydrogen may play moderate roles (causing a mismatch up to $\sim 30\%$). This region is close to the boundary of the (relatively coarse) TOPS grid, so the spline interpolation may also become inaccurate. Non-LTE effects (\S \ref{sec:atmosphere_line_formation}) may lead to even large (order unity) mismatch that is not captured by our model. However, in the deep interior of the envelope at $T\gtrsim10^4\rm\, K$, our opacity model is adequate for the SMS structure calculations presented here, and hence the our solutions with inflated envelopes are robust.

\section{B. Non-adiabatic mixing-length model}
\label{app:mlt}


We use a local, non-adiabatic mixing-length model similar to that of
\citet{1965ApJ...142..841H} to determine the division of the stellar
luminosity between radiative diffusion and convection in the envelope at $r > r_{\rm base}$. At each point in a
convectively unstable layer, the closure determines the superadiabaticity
\(\nabla-\nabla_{\rm ad}\), the convective luminosity, and the convective
velocity while allowing rising fluid elements to exchange heat radiatively
with their surroundings.

Consider a fluid element displaced outward by one mixing length defined based on the pressure scale height,
\begin{equation}
    \ell=\alpha_{\rm MLT}H,
    \quad
    H\equiv\left|\frac{\d r}{\d\ln P}\right|
      =\frac{P}{\rho g},
    \quad
    g=\frac{GM}{r^{2}},
\label{eq:app_mlt_length}
\end{equation}
where \(\alpha_{\rm MLT}\) is a dimensionless parameter of order unity. In this work, we adopt $\alpha_{\rm MLT} = 1$ for all cases. We also define the local sound speed
\begin{equation}
    c_{\rm s}^{2}\equiv gH={P/\rho}.
\label{eq:app_mlt_cs}
\end{equation}
The fluid element is assumed to remain in pressure equilibrium with the ambient medium during
its displacement. If its evolution were adiabatic, its fractional
temperature excess after traveling a distance \(\ell\) would be
\((\ell/H)(\nabla-\nabla_{\rm ad})\). Radiative leakage reduces this
contrast, which we describe using a convective-efficiency parameter
\(0\leq\eta\leq1\):
\begin{equation}
    \frac{\Delta T}{T}
    =
    \frac{\ell}{H}\,
    \eta\left(\nabla-\nabla_{\rm ad}\right)
    =
    \alpha_{\rm MLT}\eta
    \left(\nabla-\nabla_{\rm ad}\right).
\label{eq:app_mlt_temperature_excess}
\end{equation}
The limits \(\eta\rightarrow1\) and \(\eta\rightarrow0\) correspond,
respectively, to nearly adiabatic motion and rapid thermal equilibration
with the surroundings.

For a gas-radiation mixture at fixed total pressure,
\begin{equation}
    \left(\frac{\partial\ln\rho}{\partial\ln T}\right)_{P}
    =
    -\frac{4-3\beta}{\beta}
    =
    -\frac{4}{\beta}\left(1-\frac{3\beta}{4}\right).
\label{eq:app_mlt_density_derivative}
\end{equation}
The temperature excess therefore makes a rising element underdense. Its
buoyant acceleration is
\begin{equation}
    a_{\rm buo}
    =
    -g\frac{\Delta\rho}{\rho}
    =
    \frac{4g}{\beta}
    \left(1-\frac{3\beta}{4}\right)
    \frac{\Delta T}{T}.
\label{eq:app_mlt_buoyancy}
\end{equation}
The velocity acquired over the mixing length is
\begin{equation}
    v_{\rm c}
    \simeq
    \alpha_{\rm MLT}c_{\rm s}
    \left(\frac{\eta}{\beta}\right)^{1/2}
    \left(1-\frac{3\beta}{4}\right)^{1/2}
    \left(\nabla-\nabla_{\rm ad}\right)^{1/2}.
\label{eq:app_mlt_velocity}
\end{equation}
As usual in MLT, the coefficient in this estimate depends on the
assumed acceleration and mixing history and should be regarded as
accurate to order unity.

The specific heat at constant pressure for the same gas-radiation
equation of state is
\begin{equation}
    c_{P}
    =
    \frac{k_{\rm B}}{\mu m_{\rm p}}
    \frac{16-12\beta-\tfrac{3}{2}\beta^{2}}{\beta^{2}}
    \simeq
    \frac{16k_{\rm B}}{\mu m_{\rm p}\beta^{2}}
    \left(1-\frac{3\beta}{4}\right),
\label{eq:app_mlt_cp}
\end{equation}
where the last expression applies for \(\beta\ll1\). The enthalpy carried
by the rising elements then gives the convective flux
\begin{align}
    F_{\rm conv}
    &\simeq
    \frac{1}{2}\rho c_{P}\Delta T\,v_{\rm c}
    \nonumber\\
    &\simeq
    8\alpha_{\rm MLT}^{2}\rho c_{\rm s}^{3}
    \left(\frac{\eta}{\beta}\right)^{3/2}
    \left(1-\frac{3\beta}{4}\right)^{3/2}
    \left(\nabla-\nabla_{\rm ad}\right)^{3/2}
    \nonumber\\
    &=8\alpha_{\rm MLT}^{-1}\rho v_{\rm c}^{3},
\label{eq:app_mlt_convective_flux}
\end{align}
and hence
\begin{equation}
    L_{\rm conv}=4\pi r^{2}F_{\rm conv}.
\label{eq:app_mlt_convective_luminosity}
\end{equation}

To determine \(\eta\), we compare the time required for radiative diffusion
to remove the temperature excess of a fluid element with its advection
time $t_{\rm dy}$. Approximating the element as an optically thick sphere with radius
of order \(\ell/2\), its thermal-leakage time is
\begin{equation}
    t_{\rm th}
    \simeq
    \frac{\rho^{2}c_{P}\kappa_{\rm T}\ell^{2}}
         {24\arad cT^{3}},
    \ 
    t_{\rm dy}\simeq\frac{\ell}{v_{\rm c}}.
\label{eq:app_mlt_timescales}
\end{equation}
Their ratio can be written
\begin{equation}
    \gamma
    \equiv
    \frac{t_{\rm th}}{t_{\rm dy}}
    \simeq
    \frac{2\alpha_{\rm MLT}}{9\beta}
    \frac{1-3\beta/4}{1-\beta}
    \tau_{H}\frac{v_{\rm c}}{c},
    \ 
    \tau_{H}\equiv\rho\kappa_{\rm T}H.
\label{eq:app_mlt_gamma}
\end{equation}
The factor \(\beta^{-1}\) reflects the large heat capacity of a
radiation-pressure-dominated fluid: eliminating a small temperature
contrast requires removing substantially more heat than the instantaneous
radiation-energy contrast alone would suggest. Balancing the generation
of \(\Delta T\) by adiabatic displacement against radiative damping gives \citep{1965ApJ...142..841H}
\begin{equation}
    \eta=\frac{\gamma}{1+\gamma}.
\label{eq:app_mlt_eta}
\end{equation}
Thus, \(\gamma\gg1\) describes efficient convection with trapped radiation,
whereas \(\gamma\ll1\) describes rapid leakage and inefficient convection.

Combining eqs.~\eqref{eq:app_mlt_velocity},
\eqref{eq:app_mlt_gamma}, and \eqref{eq:app_mlt_eta} gives a direct
relation between the convective efficiency and the superadiabaticity,
\begin{equation}
    \gamma(\gamma+1)
    =
    \mc{K}_{\rm MLT}
    \left(\nabla-\nabla_{\rm ad}\right),
\label{eq:app_mlt_gamma_nabla}
\end{equation}
where
\begin{equation}
    \mc{K}_{\rm MLT}
    =
    \frac{4\alpha_{\rm MLT}^{4}}{81\beta^{3}}
    \frac{(1-3\beta/4)^{3}}{(1-\beta)^{2}}
    \left(\tau_{H}\frac{c_{\rm s}}{c}\right)^{2}.
\label{eq:app_mlt_K}
\end{equation}
The local radiative luminosity associated with the actual temperature
gradient is
\begin{equation}
    L_{\rm rad}
    =
    4(1-\beta)L_{\rm Edd}\nabla.
\label{eq:app_mlt_Lrad}
\end{equation}
Substituting eqs.~\eqref{eq:app_mlt_convective_flux} and
\eqref{eq:app_mlt_gamma_nabla} into
\(L_{\star}=L_{\rm rad}+L_{\rm conv}\) reduces the local MLT closure to
\begin{equation}
    A\gamma^{3}+B\gamma(\gamma+1)=C,
\label{eq:app_mlt_cubic}
\end{equation}
with
\begin{align}
    A
    &=
    \frac{729(4\pi r^{2}\rho c^{3})}{\alpha_{\rm MLT}^{4}}
    \left[
    \frac{\beta(1-\beta)}{1-3\beta/4}
    \right]^{3}
    \tau_{H}^{-3},
    \nonumber\\
    B
    &=
    \frac{81L_{\rm Edd}}{\alpha_{\rm MLT}^{4}}
    \left[
    \frac{\beta(1-\beta)}{1-3\beta/4}
    \right]^{3}
    \left(\frac{c}{c_{\rm s}\tau_{H}}\right)^{2},
    \nonumber\\
    C
    &=
    L_{\star}
    -
    4(1-\beta)L_{\rm Edd}\nabla_{\rm ad}.
\label{eq:app_mlt_coefficients}
\end{align}
Note that $C$ here should not be confused with that in eq.~(\ref{eq:C_const}).
In a convectively unstable layer,
\(C/L_{\rm Edd}=\Gamma_{\star}-\Gamma_{\rm conv}>0\), while \(A\) and
\(B\) are also positive. The left-hand side of
eq.~\eqref{eq:app_mlt_cubic} is monotonic for \(\gamma>0\), so the
equation has a unique positive root. Once this root is obtained, we
calculate $\nabla-\nabla_{\rm ad}$ and $\eta$,
followed by \(L_{\rm rad}\), \(L_{\rm conv}=L_{\star}-L_{\rm rad}\), and
\begin{equation}
    v_{\rm c}
    =
    \left(
    \frac{\alpha_{\rm MLT}L_{\rm conv}}
         {32\pi r^{2}\rho}
    \right)^{1/3}.
\label{eq:app_mlt_velocity_from_luminosity}
\end{equation}
When the \(A\gamma^{3}\) term dominates,
\(\gamma\simeq(C/A)^{1/3}\); in the rapid-leakage limit where the linear
part of the \(B\) term dominates, \(\gamma\simeq C/B\).

To see why the deep convective envelope remains nearly
adiabatic, we re-arrange eq.~\eqref{eq:app_mlt_velocity} as
\begin{equation}
    \nabla-\nabla_{\rm ad}
    =
    \frac{\beta}{\alpha_{\rm MLT}^{2}\eta(1-3\beta/4)}
    \left(\frac{v_{\rm c}}{c_{\rm s}}\right)^{2}.
\label{eq:app_mlt_small_superadiabaticity}
\end{equation}
Efficient and subsonic convection implies
\(\nabla-\nabla_{\rm ad}\ll\beta\). In this regime,
\(\Gamma\simeq1-\beta+4(\nabla-\nabla_{\rm ad})\), and therefore
\(\d\beta/\d\ln P\simeq-4(\nabla-\nabla_{\rm ad})\): the gas-pressure
fraction changes only slowly through the convective envelope. We terminate a trial envelope integration if \(v_{\rm c}\geq c_{\rm s}\), because the MLT picture would break down.


While the full Rosseland-mean opacity \(\kappa_{\rm R}(T,\beta,Z)\) is
used to calculate \(L_{\rm Edd}\), \(\Gamma_{\star}\), and the convective
instability criterion, here eqs.
\eqref{eq:app_mlt_timescales}--\eqref{eq:app_mlt_gamma} use the Thomson
opacity \(\kappa_{\rm T}\) to estimate radiative diffusion time within a fluid
element. This is an approximation in our adopted MLT closure.
Replacing \(\kappa_{\rm T}\) by a self-consistent local transport opacity
for the fluid element would be a natural refinement of the model.

\section{C. Numerical construction}
\label{app:shooting}

\subsection{C1. Deep-interior shooting method}
\label{app:deep_interior_shooting}


For fixed central temperature \(T_c\), metallicity \(Z\), and composition,
the only shooting parameter is the central density \(\rho_c\). We integrate
the convective core outward until radius $r_{\rm cc}$ where the Schwarzschild condition
\(\nabla_{\rm rad}=\nabla_{\rm ad}\) is encountered. The luminosity is then
held fixed at \(L_\star=L(r_{\rm cc})\), and the radiative envelope is
continued to \(T=T_{\rm base}=10^{6}\,\mathrm{K}\).
Both regions are integrated
with an implicit solver suitable for the stiff equations.

At \(r_{\rm base}\), we impose the boundary condition \(P_{\rm rad}=\Gamma P\) under the assumption of a constant opacity $\kappa_{\rm T}$. Using the residual function $f(\rho_c)$ in eq.~\eqref{eq:f_rhoc_residual}, we determine the physical solution by adjusting \(\rho_c\) until \(f(\rho_c)=0\).

The interval in \(\rho_c\) over which a trial integration reaches
\(T_{\rm base}\) can be extremely narrow. We therefore scan \(\rho_c\)
logarithmically and classify each trial according to whether it reaches
\(T_{\rm base}\) or terminates earlier on an invalid branch. The search
interval is iteratively narrowed until two successful trials bracket a
sign change in \(f(\rho_c)\), after which a bisection solver determines
\(\rho_c\).


\subsection{C2. Transition line and photospheric-boundary curve}
\label{app:boundary_curves}

The transition line defined by eq.~\eqref{eq:transition_line_def} is not necessarily single-valued in either \(T\) or \(\beta\), owing to the non-monotonic temperature dependence of the opacity. At fixed \(\beta\), the equation
\begin{equation}
    g(\beta,T)=1-\Gamma_\star(\beta,T)-\beta=0
\end{equation}
may have multiple roots associated with the Fe-opacity bump and the hydrogen and helium recombination region. We therefore scan the full envelope temperature range,
\(4\times10^{3}\,{\rm K}\leq T\leq T_{\rm base}\), at each value of \(\beta\), bracket all sign changes of \(g\), and refine each root individually. Roots in adjacent \(\beta\) columns are then connected by continuity in the
\((\log T,\log\beta)\) plane. Where a curve folds in \(\beta\), its points are ordered by arclength rather than by either coordinate alone. This procedure retains all connected segments without selecting a branch in advance. The resulting segments can be identified physically as an inner branch associated with the Fe bump, an outer branch associated with recombination\footnote{If extended our $\beta$ range for the opacity model down to $\ll 10^{-5}$, the inner and outer transition-line branches would be connected into a continuous curve.}, or a merged branch connecting the two at sufficiently low metallicities.

The photospheric-boundary curve is constructed from the integral closure condition in eq.~\eqref{eq:photospheric_closure_exact}. For each trial pair
\((T_{\rm ph},\beta_{\rm ph})\), the radiative equation
\(d\beta/d\ln P=1-\Gamma_\star-\beta\) is integrated between
\(P_{\rm ph}\) and \(P_0\), simultaneously evaluating the opacity integral. Using eq.~\eqref{eq:Tph_def} to eliminate \(r_{\rm ph}\), we define the residual
\begin{equation}
    \mathcal{R}_{\rm ph}(\beta_{\rm ph},T_{\rm ph})
    \equiv
    \int_{P_0}^{P_{\rm ph}}
    \kappa_{\rm R}\!\left[\beta(P),P\right]\,dP
    -
    \frac{8\pi GM\sigma_{\rm SB}T_{\rm ph}^{4}}
         {3L_\star},
\label{eq:photospheric_boundary_residual}
\end{equation}
which then gives $\beta_{\rm ph}(T_{\rm ph})$ from \(\mathcal{R}_{\rm ph}=0\).

Across the model grid, numerical scans show that
\(\mathcal{R}_{\rm ph}\) has at most one sign-changing root in
\(\beta_{\rm ph}\) at fixed \(T_{\rm ph}\). We therefore tabulate the curve directly as the single-valued function \(\beta_{\rm ph}(T_{\rm ph})\), using a bisection solver at each temperature. Temperature intervals for which no root exists are retained as gaps between disconnected curve segments. The local expression in eq.~\eqref{eq:photospheric_closure_local} provides a useful analytic explanation for the proximity of this curve to the outer transition line, but the numerical envelope solutions use the integral closure condition above.

\subsection{C3. Two-point envelope shooting}
\label{app:two_point_shooting}

As discussed in \S\ref{sec:two_point_shooting}, direct shooting from \(r_{\rm base}\) to the photosphere is impractical because the envelope trajectory is extremely sensitive to the base perturbation \(f_\beta\) (eq. \ref{eq:f_beta_perturbation}). We therefore replace \(f_\beta\) with the gas-pressure fraction at the inner transition line, \(\beta_{\rm mid}\), as the shooting coordinate. This requires a two-stage algorithm.

In Stage A, we integrate each trial \(f_\beta\) from \(r_{\rm base}\) to its first crossing of the inner transition line,
\begin{equation}
    1-\Gamma_\star(T_{\rm mid},\beta_{\rm mid})
    -\beta_{\rm mid}=0,
\end{equation}
and record \((\beta_{\rm mid},r_{\rm mid})\) for each $f_\beta$. Although \(\beta_{\rm mid}(f_\beta)\) is extremely sensitive, the crossing quantities \(r_{\rm mid}\), \(T_{\rm mid}\) vary smoothly with \(\beta_{\rm mid}\); we tabulate and interpolate them as functions of \(\beta_{\rm mid}\).

In Stage B, a trial \(\beta_{\rm mid}\) with its interpolated state \((r_{\rm mid},T_{\rm mid})\) is integrated outward until \(T(r_{\rm stop})=T_{\rm eff}(r_{\rm stop})\) (a candidate photosphere), yielding \((r_{\rm stop}, T_{\rm stop},\beta_{\rm stop})\). The signed displacement from the candidate photospheric point to the photospheric-boundary curve of \S\ref{app:boundary_curves} is
\begin{equation}
    s(\beta_{\rm mid})
    \equiv
    \log \beta_{\rm stop}
    -
    \log \beta_{\rm ph,c}(T_{\rm stop}),
\label{eq:two_point_shooting_residual}
\end{equation}
with \(s>0\) (\(s<0\)) on the gas-pressure-rich (poor) side of the boundary curve. We scan in \(\beta_{\rm mid}\), bracket sign changes in \(s\), and accept a root once
\begin{equation}
    |s|<\max\!\left(s_0,2\sigma_s\right), \ s_0 = 10^{-2},
\label{eq:shooting_noise_tolerance}
\end{equation}
where $\sigma_s$ (usually $\lesssim 0.1$) is the local residual noise of the $s(\beta_{\rm mid})$ sequence, estimated from its small-scale, non-monotonic reversals near the bracket. The fixed tolerance $s_0$ is reached directly when the integration resolves $s$ to better precision. In some cases, the extreme sensitivity of $\beta_{\rm mid}(f_\beta)$ hits the floating-point precision in our integration, so bisection stalls once $\sigma_s$ exceeds $s_0$. However, these solutions are not treated as lower quality, because they are sufficiently close to the converged solution for our purpose in this work.


In some cases, no sign-changing bracket can be resolved, and we accept an \emph{approximate} solution when the closest sampled models with $|s|<0.5$ have the asymptotically flat predictions for $T_{\rm eff}(s\rightarrow0)$, i.e.,
\begin{equation}
    \frac{|\Delta T_{\rm eff}(s\rightarrow0)|}{T_{\rm eff}}<0.03,
\end{equation}
which shows that \(T_{\rm eff}\) is insensitive to the unresolved residual.

Otherwise, if the search terminates at the boundary of the opacity domain (e.g. \(\beta_{\rm ph}\) approaching the tabulated limit \(0.9\) while remaining on one side of the closure curve) or nearby trials establish a robust monotonic trend toward a root outside the sampled range, the closest sampled model is instead reported as a one-sided bound on \(T_{\rm eff}\), not an extrapolated solution.

For cases with an accepted solution, we recover a base perturbation consistent with the converged or closest-to-converge \(\beta_{\rm mid, final}\) by solving \(\beta_{\rm mid}(f_\beta)=\beta_{\rm mid,final}\) within the first-stage bracket, requiring agreement to within $1\%$ (tighter tolerances would approach floating-point precision in \(f_\beta\)).

\section{D. Core structure and GR instability scaling}

\subsection{D1. Constant-\texorpdfstring{\(\beta\)}{beta} core structure and binding energy}
\label{app:constant_beta_core}

The deep-interior solutions have an approximately constant gas-pressure
fraction, \(\beta\equiv P_{\rm gas}/P\approx \beta_c\), from the center to
\(r_{\rm base}\). It is therefore useful to consider the idealized
Eddington model in which \(\beta\) is exactly constant
\citep{1964RvMP...36..545F,1984ApJ...280..825B}. From the gas and
radiation equations of state,
\begin{equation}
    \frac{\beta}{1-\beta}
    =
    \frac{P_{\rm gas}}{P_{\rm rad}}
    =
    \frac{3\rho k_{\rm B}}
    {a_{\rm rad}\mu m_{\rm p}T^3},
\end{equation}
a constant \(\beta\) implies the polytropic relation
\begin{equation}
    P=K_\beta\rho^{4/3},
    \ 
    K_\beta=
    \left[\frac{3(1-\beta)}{a_{\rm rad}}\right]^{1/3}
    \left(\frac{k_{\rm B}}
    {\mu m_{\rm p}\beta}\right)^{4/3}.
\label{eq:constant_beta_polytrope}
\end{equation}
The equilibrium structure is therefore an \(n=3\) polytrope. Note that the
polytropic exponent \(4/3\), related to radial stratification, should not be confused with the adiabatic
index of the fluid, \(\gamma_{\rm ad}=4/3+\beta/6+\mathcal{O}(\beta^2)\) (eq. \ref{eq:adiabatic_index}),
which governs radial stability.

For a Newtonian \(n=3\) polytrope with vanishing surface
pressure, the gravitational potential energy is
\begin{equation}
    \Omega
    =
    -\frac{3}{5-n}\frac{GM^2}{R}
    =
    -\frac{3}{2}\frac{GM^2}{R}.
\label{eq:n3_gravitational_energy}
\end{equation}
The internal energy density of radiation and non-relativistic
monatomic gas is
\begin{equation}
    u
    =
    3P_{\rm rad}+\frac{3}{2}P_{\rm gas}
    =
    3P\left(1-\frac{\beta}{2}\right).
\end{equation}
The scalar virial theorem gives
\(\Omega=-3\int P d V\), so the total internal energy is
\begin{equation}
    U = \int u\, d V
    = -\left(1-\frac{\beta}{2}\right)\Omega.
\end{equation}
The total binding energy is therefore
\begin{equation}
    E_{\rm bind}
    =
    U+\Omega
    =
    -\frac{3\beta}{4}\frac{GM^2}{R},
\label{eq:constant_beta_binding_energy}
\end{equation}
These relations are exact within the constant-\(\beta\), Newtonian model.

The same model gives a direct relation between \(\beta\) and the stellar
mass. Let \(\theta(\xi)\) denote the \(n=3\) Lane--Emden solution and
define
\begin{equation}
    \phi(\xi)\equiv-\xi^2\frac{d\theta}{d\xi},
    \ 
    \phi_{\rm max}\equiv\phi(\xi_{\rm max})=2.01824,
\end{equation}
where \(\xi_{\rm max}=6.89685\) is the first zero of \(\theta\) (or density). The Lane--Emden
mass is then
\begin{align}
    M
    &=
    4\pi\phi_{\rm max}
    \left(\frac{K_\beta}{\pi G}\right)^{3/2}
    \nonumber\\
    &=
    \phi_{\rm max}
    \left(\frac{k_{\rm B}}
    {\mu m_{\rm p}\beta}\right)^2
    \left(\frac{48}
    {\pi a_{\rm rad}G^3}\right)^{1/2}
    (1-\beta)^{1/2}
    \nonumber\\
    &=
    4.77\times10^5\,M_\odot
    \left(\frac{\mu}{0.615}\right)^{-2}
    \left(\frac{\beta}{10^{-2}}\right)^{-2}
    (1-\beta)^{1/2}.
\label{eq:constant_beta_mass}
\end{align}
Neglecting the very weak factor \((1-\beta)^{1/2}\) in the limit of $\beta\ll 1$ and identifying
\(\beta\approx \beta_c\) gives the central gas-pressure fraction
\begin{equation}
    \beta_c
    \approx
    6.90\times10^{-3}
    \left(\frac{M}{10^6\,M_\odot}\right)^{-1/2}
    \left(\frac{\mu}{0.615}\right)^{-1}.
\label{eq:betac_mass_analytic}
\end{equation}


Our SMS core model is truncated at \(r_{\rm base}\), where
the pressure is small but nonzero. The virial theorem then contains the
surface term \(4\pi r_{\rm base}^3P_{\rm base}\). Across our model grid,
this term is less than \(10^{-5}GM^2/r_{\rm base}\) and is hence negligible.
This justifies the use of \(r_{\rm base}\) as the effective
polytropic radius.


\subsection{D2. Derivation of the GR-instability scaling}
\label{app:gr_instability}




\paragraph{First-order criterion.}
In our notation, the relativistic stability criterion of
\citet{2016ApJ...818..157S,2025ApJ...978...58S} can be written as
\begin{equation}
    \mathcal{F}
    \equiv
    \gamma_{\rm ad}-\frac{4}{3}
    -
    \left[
    \epsilon_{\rm GR}-\epsilon_{\rm GR}^{2}
    -
    \left(
    \frac{10}{3}-2\gamma_{\rm ad}
    -\epsilon_{\rm GR}-\eta
    \right)\eta
    \right],
\label{eq:GR_stability_functional}
\end{equation}
where \(\epsilon_{\rm GR}\) is defined in eq.~\eqref{eq:epsilonGR_def}, with \(\rho_c\) the central rest-mass density. Marginal stability corresponds to \(\mathcal{F}=0\), while \(\mathcal{F}<0\) denotes instability. For a radiation-dominated gas, the adiabatic index is given by eq.~(\ref{eq:adiabatic_index}). Expanding eq.~\eqref{eq:GR_stability_functional} jointly in the small quantities \((\beta_c,\epsilon_{\rm GR},\eta)\) gives
\begin{align}
    \mathcal{F}
    ={}&
    \underbrace{
    \frac{\beta_c}{6}
    -\epsilon_{\rm GR}
    +\frac{2}{3}\eta
    }_{\mathcal{O}(1)}
    \nonumber\\
    &+
    \underbrace{
    \frac{\beta_c^2}{48}
    +\epsilon_{\rm GR}^2
    -\frac{\beta_c\eta}{3}
    -\epsilon_{\rm GR}\eta
    -\eta^2
    }_{\mathcal{O}(2)}
    +\mathcal{O}(3).
\label{eq:GR_stability_expansion}
\end{align}
Retaining all terms consistently through first order therefore recovers the criterion (eq.~\ref{eq:epsilonGR_marginal_stability}) used in the main text. At \(\eta_{\rm max}\simeq0.009\), the largest second-order correction is purely rotational \(\eta_{\rm max}^2=8.1\times10^{-5}\), only \(1.4\%\) of the leading rotational term \(2\eta_{\rm max}/3\).

\paragraph{Composition homology.}
Near the GR instability threshold (eq.~\ref{eq:epsilonGR_marginal_stability}), the nuclear burning is dominated by the CNO cycle. Locally approximating its temperature dependence by a power law gives
\begin{equation}
    \epsilon_{\rm CNO}
    \propto
    \rho_c XZ_{\rm CNO}T_c^\nu,
    \ 
    \frac{L_{\rm Edd,T}}{M}
    \propto(1+X)^{-1}.
\label{eq:CNO_Eddington_homology}
\end{equation}
Combining thermal equilibrium,
\(\epsilon_{\rm CNO}\sim L_\star/M\), with the radiation-dominated SMS homology relation
\(\rho_c\propto T_c^3M^{-1/2}\) yields
\begin{equation}
    T_c^{\nu+3}
    \propto
    \mathcal{A}^{-1}M^{1/2},
\label{eq:Tc_composition_homology}
\end{equation}
where \(\mathcal{A}(X,Z)\) is defined in
eq.~\eqref{eq:AB_def_main}.

The two quantities in the stability condition scale as
\begin{equation}
    \epsilon_{\rm GR}
    \propto
    \frac{T_c^4}{\rho_c}
    \propto
    T_cM^{1/2},
    \ 
    \beta_c
    \propto
    \frac{\rho_c}{\mu T_c^3}
    \propto
    \mathcal{B}M^{-1/2},
\label{eq:epsilon_beta_homology}
\end{equation}
where \(\mathcal{B}(X)\) is defined in eq.~\eqref{eq:AB_def_main}. It is important to notice that, at a fixed mass, $\epsilon_{\rm GR}$ is linearly proportional to the central temperature, whereas $\beta_c$ only depends on the composition via the mean molecular weight or $\mc{B}$. At lower CNO mass fraction, the core temperature needs to be higher to sustain the same luminosity, and this leads to larger compactness $\epsilon_{\rm GR}$ and hence lower threshold mass for GR instability\footnote{When hydrogen is fully exhausted, the mean molecular weight increases to $\mu=4/3$ and the onset of helium burning also increase $T_c$ significantly; these two effects combined reduce the threshold mass by about an order of magnitude compared to zero-age main sequence.}.

For a rotating SMS core with \(\eta>0\), we apply the rotational term in the stability
condition to the nonrotating deep-interior sequences; rotational changes
to \(T_c\), \(\rho_c\), and \(\beta_c\) are neglected. Published rotating
SMS models suggest that these changes remain at the percent level even
near mass shedding \citep{2025ApJ...978...58S}; a more precise treatment
would require self-consistent rotating equilibrium sequences.

To quantify the relative importance of gas-pressure support in a rotating SMS core, we define
\begin{equation}
    \phi_\beta
    \equiv
    \frac{\beta_c/6}
         {\beta_c/6+(2/3)\eta}
    =
    \frac{\beta_c}{\beta_c+4\eta}.
\label{eq:phi_beta_GR}
\end{equation}
At fixed rotation parameter \(\eta\), logarithmic differentiation of the marginal-stability condition gives
\(d\ln\epsilon_{\rm GR}=\phi_\beta\,d\ln\beta_c\). Eqs.
\eqref{eq:Tc_composition_homology} and
\eqref{eq:epsilon_beta_homology} therefore imply
\begin{align}
    (\nu+3)\,d\ln T_c
    &=
    -d\ln\mathcal{A}
    +\frac{1}{2}d\ln M_{\rm th},
    \nonumber\\
    d\ln T_c+\frac{1}{2}d\ln M_{\rm th}
    &=
    \phi_\beta
    \left(
    d\ln\mathcal{B}
    -\frac{1}{2}d\ln M_{\rm th}
    \right).
\label{eq:GR_homology_differentials}
\end{align}
Solving these relations gives the local composition scaling $M_{\rm th}
    \propto \mathcal{A}^{p} \mathcal{B}^{q}$ with
\begin{equation}
    p(\eta)
    =
    \frac{2}{(\nu+3)(1+\phi_\beta)+1},
    \
    q(\eta)
    = (\nu+3)\phi_\beta p.
\label{eq:GR_pq_homology}
\end{equation}
For a nonrotating star, \(\phi_\beta=1\), and hence
\begin{equation}
    p(0)=\frac{1}{\nu+7/2},
    \
    q(0)=1-\frac{p(0)}{2}.
\label{eq:GR_pq_nonrotating}
\end{equation}
As \(\eta\) increases, $q$ drops due to the decreasing contribution of gas pressure to the marginal-stability boundary.

\paragraph{Calibration to the numerical sequences.}
For each adopted \(\eta\), we locate \(M_{\rm th}\) by finding the sign change of $ \epsilon_{\rm GR}(M)-\beta_c(M)/6 -2\eta/3$ along each deep-interior sequence and interpolating in \(\log M\). At
\(X=X_0\), one has \(\mathcal{B}=1\) and
\(\mathcal{A}\simeq Z/Z_0\). A power-law fit to the resulting
\(M_{\rm th}(Z)\) values therefore directly determines
\(M_0(\eta)\) and \(p(\eta)\) in
eq.~\eqref{eq:Mth_composition_fit}. The nonrotating fitted slope
\(p(0)=0.0653\) corresponds through
eq.~\eqref{eq:GR_pq_nonrotating} to an effective local nuclear burning temperature exponent
\begin{equation}
    \nu=p(0)^{-1}-\frac{7}{2}\simeq11.8.
\end{equation}
For each rotating sequence, \(\phi_\beta\) is evaluated at the directly
computed threshold at the reference composition \((X_0=0.7,Z_0=10^{-2})\). The resulting values are
approximately \(1\), \(0.260\), and \(0.099\) for
\(\eta=0\), \(\eta_{\rm max}/2\), and \(\eta_{\rm max}\), respectively,
giving \(q\simeq0.967\), \(0.377\), and \(0.170\). This procedure gives the coefficients reported in
Table~\ref{tab:Mth_fit}; the numerical sequences supply the normalization and metallicity dependence, while the local homology calculation supplies the additional mean-molecular-weight dependence. Although this homology scaling is derived locally about the reference composition ($X_0=0.7, Z_0=10^{-2}$), it reproduces the numerical threshold grid over \(0.05\leq X\leq0.7\) and \(10^{-4}\leq Z\leq10^{-2}\) to within a few percent.

\section{F. Complete model grids}
\label{app:model_tables}

The stellar models are constructed in two stages separated at
$T_{\rm base}=10^{6}\,{\rm K}$. The deep-interior solution determines
the central structure, luminosity, and conditions at $r_{\rm base}$ and
is independent of the Fe-group abundance and the effective-Eddington
limiter. The outer-envelope calculation then depends additionally on
$f_{\rm Fe}$ and $Q_{\rm LIM}$. Tables~\ref{tab:deep_interior_models}
and~\ref{tab:outer_envelope_models} provide the corresponding results below and above $r_{\rm base}$, respectively.


\begingroup
\LongTables

\setlength{\aboverulesep}{0.8ex}
\setlength{\belowrulesep}{0.8ex}

\makeatletter
\renewcommand{\tablehead}[1]{%
  \@table@not@headedfalse
  \kill
  \caption{\\\@tablecaption}\\
  \toprule
  #1\hskip\tabcolsep\\
  \midrule
  \endfirsthead
  \caption[]{--- \emph{Continued}}\\
  \toprule
  #1\hskip\tabcolsep\\
  \midrule
  \endhead
  \bottomrule
  \endfoot
}
\renewcommand{\enddata}{}
\makeatother

\begin{deluxetable*}{ccccccc}
\tablecaption{Deep-interior model grid for $X=0.7$, $f_{\rm cno}=0.7$. $T_c$
and $\rho_c$ are central temperature and density, while $r_{\rm base}$,
$\Gamma_{\rm base}$, and $L_\star$ are evaluated where the deep-interior
integration reaches $T_{\rm base}=10^6\,{\rm K}$. The tabulated $L_\star$
applies to all outer-envelope solutions with the same mass and metallicity $(M,Z)$. Here $M$
is only the shooting target; the converged masses differ from it by
$\lesssim 0.6\%$, which is why the luminosities at the same mass differ slightly.
\label{tab:deep_interior_models}}
\tablehead{
\colhead{$Z$} &
\colhead{$M$} &
\colhead{$T_c$} &
\colhead{$\rho_c$} &
\colhead{$r_{\rm base}$} &
\colhead{$1-\Gamma_{\rm base}$} &
\colhead{$L_\star$} \\
&
\colhead{[$M_\odot$]} &
\colhead{[$10^7\,{\rm K}$]} &
\colhead{[${\rm g\,cm^{-3}}$]} &
\colhead{[${\rm AU}$]} &
&
\colhead{[$10^{44}\,{\rm erg\,s^{-1}}$]}
}
\startdata
$10^{-4}$   & $10^{4}$ & 7.50 & 0.570  & 0.465 & 0.0725  & 0.0137 \\
$10^{-4}$   & $10^{5}$ & 8.17 & 0.226  & 1.42  & 0.0223  & 0.144  \\
$10^{-4}$   & $10^{6}$ & 8.87 & 0.0909 & 4.20  & 0.00696 & 1.47   \\
\specialrule{0.4pt}{2.5pt}{2.5pt}
$10^{-3.5}$ & $10^{4}$ & 6.94 & 0.453  & 0.500 & 0.0727  & 0.0136 \\
$10^{-3.5}$ & $10^{5}$ & 7.55 & 0.179  & 1.53  & 0.0223  & 0.144  \\
$10^{-3.5}$ & $10^{6}$ & 8.18 & 0.0714 & 4.54  & 0.00695 & 1.47   \\
\specialrule{0.4pt}{2.5pt}{2.5pt}
$10^{-3}$   & $10^{4}$ & 6.44 & 0.359  & 0.541 & 0.0722  & 0.0138 \\
$10^{-3}$   & $10^{5}$ & 6.98 & 0.142  & 1.65  & 0.0223  & 0.145  \\
$10^{-3}$   & $10^{6}$ & 7.56 & 0.0563 & 4.90  & 0.00695 & 1.47   \\
\specialrule{0.4pt}{2.5pt}{2.5pt}
$10^{-2.5}$ & $10^{4}$ & 5.97 & 0.288  & 0.580 & 0.0723  & 0.0138 \\
$10^{-2.5}$ & $10^{5}$ & 6.47 & 0.113  & 1.77  & 0.0223  & 0.144  \\
$10^{-2.5}$ & $10^{6}$ & 7.00 & 0.0446 & 5.27  & 0.00695 & 1.47   \\
\specialrule{0.4pt}{2.5pt}{2.5pt}
$10^{-2}$   & $10^{4}$ & 5.55 & 0.232  & 0.618 & 0.0727  & 0.0136 \\
$10^{-2}$   & $10^{5}$ & 6.01 & 0.0903 & 1.90  & 0.0223  & 0.144  \\
$10^{-2}$   & $10^{6}$ & 6.49 & 0.0355 & 5.67  & 0.00695 & 1.47   \\
\enddata
\end{deluxetable*}

\begin{deluxetable*}{ccccccc}
\tablecaption{Outer-envelope effective temperatures
$T_{\rm eff}/10^3\,{\rm K}$. The two columns within each mass group
correspond to the labeled values of $f_{\rm Fe}$. The two upper limits mark models
for which no converged photospheric solution was found and report the
nearest valid trial. A dagger marks an approximate
solution, for which no root satisfying the nominal tolerance could be
bracketed and the tabulated value is instead the nearest sampled trial
(Appendix~\ref{app:two_point_shooting}).
\label{tab:outer_envelope_models}}
\tablehead{
\colhead{} &
\multicolumn{2}{c}{$M=10^4\,M_\odot$} &
\multicolumn{2}{c}{$M=10^5\,M_\odot$} &
\multicolumn{2}{c}{$M=10^6\,M_\odot$} \\
\specialrule{0.4pt}{2.5pt}{2.5pt}
\colhead{$Z$} &
\colhead{$1$} & \colhead{$10^{-0.5}$} &
\colhead{$1$} & \colhead{$10^{-0.5}$} &
\colhead{$1$} & \colhead{$10^{-0.5}$}
}
\startdata
\multicolumn{7}{c}{$Q_{\rm LIM}=1$} \\
$10^{-4}$   & 72.8 & 76.1 & 51.2 & 61.5 & 28.2 & 28.3 \\
$10^{-3.5}$ & 56.7 & 67.8 & 21.6 & 24.8 & 25.7 & 26.4 \\
$10^{-3}$   & 35.1 & 47.0 & 13.6 & 15.2 & 18.8 & 20.2 \\
$10^{-2.5}$ & 7.05 & 9.98 & 7.61 & 9.25$^\dagger$ & 12.7$^\dagger$ & 13.9$^\dagger$ \\
$10^{-2}$   & $<11.4$ & $<9.53$ & 6.79 & 6.94 & 7.45 & 7.56 \\
\specialrule{0.4pt}{2.5pt}{2.5pt}
\multicolumn{7}{c}{$Q_{\rm LIM}=0.1$} \\
$10^{-4}$   & 72.8 & 76.1 & 51.2 & 61.5 & 28.3 & 28.4 \\
$10^{-3.5}$ & 56.8 & 67.8 & 25.4 & 27.4 & 26.9 & 27.2 \\
$10^{-3}$   & 35.1 & 47.0 & 14.5 & 16.9 & 22.0 & 24.3 \\
$10^{-2.5}$ & 7.38 & 10.1 & 7.75 & 9.26 & 12.8$^\dagger$ & 14.3 \\
$10^{-2}$   & 6.82 & 6.91 & 7.41 & 7.47 & 7.68 & 7.75 \\
\enddata
\end{deluxetable*}

\endgroup

\end{document}